# Multi-task deep-learning optimization of trade-off properties for superior-performance Fe-based soft magnetic alloys

Kang-Yuan Li(李康源)[1], Mao-Zhi Li(李茂枝)[1,*], Wei-Hua Wang(汪卫华)[2,3,4]

[1]*School of Physics and Key Laboratory of Quantum State Construction and Manipulation (Ministry of Education), Renmin University of China; Beijing, 100872, China*
[2]*Songshan Lake Materials Laboratory, Dongguan, Guangdong 523808, China*
[3]*Institute of Physics, Chinese Academy of Sciences, 100190 Beijing, China*
[4]*Center of Materials Science and Optoelectronics Engineering, University of Chinese Academy of Sciences, Beijing 100049, China*

**Abstract**

Fe-based amorphous alloys are promising soft magnetic materials for developing next-generation devices with high frequency and efficiency. However, optimization of Fe-based alloys with ultra-high saturation magnetic flux density ($B_s$), ultra-low coercivity ($H_c$), and good glass-forming ability is a notorious problem, owing to the vast composition space and complex trade-offs of these properties. Thus, conventional design methods encounter great challenges. Here we develop a generative multi-task deep learning (GMTDL) to achieve simultaneous optimization of compositions and trade-off properties. The GMTDL can sufficiently exploit and share the knowledge of datasets across different tasks, despite the limitation and imbalance of these datasets. Therefore, it exhibits superior performance in prediction of alloys with multiple targeted properties, outperforming previous machine learning-based design strategies. Moreover, the GMTDL can also tailor compositions, providing an efficient way to regulate properties and generate desired candidates for further experimental processing. The validity and reliability of GMTDL are rigorously tested by benchmarking with Fe-based alloys reported very recently. Moreover, some new alloys with ultra-high $B_s$ and ultra-low $H_c$ are predicted. The optimal content windows of key elements and their synergistic effects are also unraveled for practical guidance. Thus, our study establishes an effective and reliable paradigm for simultaneous prediction and optimization of high-performance materials with multiple properties.

*maozhili@ruc.edu.cn

## 1. Introduction

With the development of new infrastructure fields such as 5G communication, new energy vehicles, artificial intelligence, and intelligent manufacturing, it has been increasingly desirable to develop next-generation electrical equipment and electronic devices with high frequency, miniaturization, and high efficiency [1-2]. With increasing frequency, however, most of the electrical energy is consumed by magnetic components [3-4]. Thus, energy conversion efficiency associated with soft magnetic materials becomes a key factor in the development of the high-frequency devices [5]. So far, silicon steels are the most widely used soft magnetic materials, yet they suffer from high core loss, low energy conversion efficiency, and thereby serious heating problems of devices, when used at high frequencies such as megahertz [6-7]. While ferrites exhibit high permeability and low loss due to high electrical resistivity, their low saturation magnetization is detrimental to the miniaturization of high-frequency devices [8]. Fe-based amorphous alloys have emerged as a promising new class of soft magnetic materials for high-frequency applications [9]. Compared to silicon steel, these alloys possess lower coercivity ($H_c$) and higher resistivity, which contribute to reduced core losses. Although the saturation magnetic flux density ($B_s$) of Fe-based amorphous alloys is relatively low, it can be improved by tuning the content of ferromagnetic elements and some other metallic and metalloid elements [10,11]. Moreover, the formation of nanocrystals within the amorphous matrix induced by heat treatment can further enhance the $B_s$ [12,13]. Therefore, great efforts have been devoted to designing high-performance Fe-based alloys with high $B_s$, low $H_c$, and good glass-forming ability (GFA) [11,14].

However, optimization and regulation of Fe-based alloys with ultra-high $B_s$, ultra-low $H_c$, and good GFA have been a notorious problem [11,14]. While the increase of ferromagnetic elements favors the $B_s$, the GFA of these alloys rapidly decreases, leading to poor thermal stability and harsh and uncontrollable crystallization process. To enhance the GFA, some metalloid elements, such as B, P, Si, and C are often added, which mostly deteriorates the $B_s$. Meanwhile, minor addition of some metallic elements may be useful for controllable nanocrystalline structure [15]. Moreover, there is also a trade-off between the $B_s$ and $H_c$, that is, enhancing $B_s$ leads to a deterioration of $H_c$ [16]. Thus, it has been a long-standing challenge to balance the trade-offs among the $B_s$, $H_c$ and GFA for optimizing the superior-performance Fe-based amorphous alloys, since this is essentially a high-dimensional search and optimization in a vast and very complicated composition space integrating multiple properties.

So far, machine learning (ML) methods have been applied to accelerate the search for high-performance alloys [17-18]. However, such methods require manual input of alloy compositions. As alloy complexity increases, the number of potential compositions grows exponentially, rendering manual input impractical [19]. Very recently, rapid progress in generative approaches and autonomous laboratories (A-Lab) provides new efficient strategies for addressing high-dimensional optimization problems and has significantly accelerated new material design [20]. Nevertheless, most existing A-Labs focus on single-objective design [21-22], while practical applications mostly require simultaneous fulfillment of multiple performance objectives. This greatly increases the complexity of searching the performance space and finding the optimal solution.

To achieve multi-objective design, multiple single-task models are often developed for each single property prediction, and a step-by-step property screening is performed in the design procedure [23-26]. However, due to the diversity and complexity of alloy compositions and the scarcity of high-quality datasets, single-task models often face overfitting issues when applied to small datasets, which limits their practical performance. Recently, multi-task learning, a transfer learning strategy that shares information across multiple related tasks, has been shown to achieve multi-objective design, significantly enhancing model performance, including data augmentation, attention focusing, feature selection, and overfitting prevention [27-30]. For instance, Wu et al. developed a multi-task learning model to successfully predict multiple properties of quantum many-body systems from short-range correlation data [30].

Unlike the crystalline materials, however, amorphous alloys are in non-equilibrium states, thereby the disordered atomic structures are metastable, which may not be used as a criterion for materials design. Moreover, the discovered Fe-based amorphous alloy compositions are much fewer than the crystalline materials. Another serious problem is that the existing data of multiple properties of amorphous alloys are significantly imbalanced, posing a greater challenge for multiple objective design of amorphous alloys.

To address these challenges, we developed a generative multi-task deep learning (GMTDL) framework by integrating a multi-task Wasserstein autoencoder (MTWAE) and the non-dominated sorting genetic algorithm III (NSGA-III), and achieved efficient multi-objective inverse design of high-performance Fe-based amorphous alloys. Our framework effectively mitigates the challenge of the limited and imbalanced data by leveraging shared information across related multiple properties, significantly outperforming conventional single-task models in prediction accuracy. The validity and reliability of the framework are rigorously examined and benchmarked against recently reported Fe-based alloys which are not included in the datasets. Furthermore, by employing SHapley Additive exPlanations (SHAP) and statistical analysis, we provided comprehensive understanding for the intricate effects and synergistic interactions of some key elements on soft magnetic properties and GFA. Optimal content window of these elements is also revealed for practical guidance of materials design. Finally, we developed an online platform for the GMTDL framework to achieve convenient and efficient material design needs.

## 2. Methods

### 2.1. Generative multi-task deep learning framework

Figure 1 illustrates the GMTDL framework, which integrates the multi-task Wasserstein autoencoder (MTWAE) [31] and the non-dominated sorting genetic algorithm III (NSGA-III) [32]. Here the MTWAE model, combining the distributional matching capabilities of WAE with multi-task learning, maps alloy compositions to a latent space and captures the non-linear relationship between compositions and multiple properties, simultaneously optimizing multiple properties of materials. The NSGA-III-based evolutionary algorithm can efficiently explore the latent space to identify alloy compositions with novel multiple properties. Thus, the GMTDL framework fully leverages the advantages of WAE-based generative model and multi-objective optimization algorithm to achieve efficient optimization and regulation of Fe-based alloys with multiple desired properties, i.e., high saturation magnetic flux density ($B_\mathrm{s}$), low coercivity ($H_\mathrm{c}$), and good GFA.

**2.2. Data collection and processing**

We constructed the datasets for three properties of Fe-based metallic glasses (MGs), i.e., saturation magnetic flux density $B_\mathrm{s}$ of 574 compositions, coercivity $H_\mathrm{c}$ of 383 compositions, and critical casting diameter $D_\mathrm{c}$ of 311 compositions (see Fig. 1(a)). All data were rigorously collected from peer-reviewed experimental measurements (see Supplementary Table S1–S3). Among the MG compositions in these datasets, 30 elements are involved (see Supplementary Fig. S1). To input the elemental composition of each alloy into the multi-task learning, a composition descriptor is constructed as a 30-dimensional vector of $X = [x_1, x_2, \cdots, x_{30}]^T$ based on the molar ratios of 30 elements in an alloy, subject to the constraint of $\sum_{i=1}^{30} x_i = 1$. Here, $x_i$ denotes the molar ratio of element *i* in an alloy. If an alloy does not contain element $i$, $x_i = 0$ (see Supplementary Materials I for more details of data processing).

Considering that the different units and numerical levels among the dataset of $B_\mathrm{s}$, $H_\mathrm{c}$ and $D_\mathrm{c}$ may result in big gaps of gradient among each task, we standardized the raw datasets for easier model training. Note that the raw experimental datasets should be standardized in a linear transformation to guarantee that distribution shapes are not changed. Here Z-score transformation of $\hat{y}_i^{\mathrm{exp}} = \left(y_i^{\mathrm{exp}} - \bar{y}^{\mathrm{exp}}\right)/\sigma$ was employed that can always transform a raw data distribution to a distribution with a mean of 0 and a standard deviation of 1. Here $\hat{y}_i^{\mathrm{exp}}$ represents the transformed property values and $y_i^{\mathrm{exp}}$ refers to the original values, while $\bar{y}^{\mathrm{exp}}$ and $\sigma$ are the average value and standard deviation of a dataset, respectively. Note that $\bar{y}^{\mathrm{exp}}$ and $\sigma$ must be calculated on training datasets, then two parameters are used to the transformation of the test datasets. In our work, the $B_\mathrm{s}$, $H_\mathrm{c}$, and $D_\mathrm{c}$ datasets were randomly divided into two subsets: 80% for training and 20% for testing.

### 2.3. Multi-task Wasserstein autoencoder (MTWAE) model

To fully leverage the limited experimental data and directly generate novel alloy compositions with multiple desired properties, we developed the MTWAE model by extending the standard WAE model to a multi-task architecture. The MTWAE model consists of five distinct neural networks: an encoder of $q_{\phi}(Z|X)$, a decoder of $p_{\theta}(\hat{X}|Z)$, and three separate predictors of $f_{\omega_p}(y_p|Z)$ for each target property $y_p$ ($p \in \{B_s, H_c, D_c\}$). Here $Z$ denotes a latent space representation. $\phi$, $\theta$, and $\omega_p$ represent the parameters of the encoder network, the decoder network, and the predictor networks associated with property $y_p$, respectively. All network components are based on multi-layer perceptrons (MLPs). Here a softmax activation function was employed in the output layer in the decoder, which guarantees that all decoded molar fractions are non-negative and satisfied with the constraint of $\sum x_i = 1$ ($i = 1, \ldots, 30$) without any additional post-processing. (see Supplementary Materials II for detailed architecture of the MTWAE model).

The MTWAE model was trained to fit the distribution of the dataset of both alloy compositions and properties shown in Fig. 1(b). The model training is essentially an optimization of a total loss function, $L_{total} = \sum_{task} w_{task} L_{task}$, a sum of weighted individual task losses. Here $w_{task}$ denotes the weight of each task loss $L_{task}$ for balancing its contribution to the total loss, and $L_{task} = L_{recon} + L_{pro} + L_{MMD}$. $L_{recon}$, $L_{pro}$, and $L_{MMD}$ are reconstruction loss, prediction loss, and maximum mean discrepancy loss, respectively [33]. (see Supplementary Materials II for details of the MTWAE model training).

To address the significant imbalance in the size of these datasets, we employed a dual strategy of resampling and inverse sample size weighting. This approach effectively balances the contribution of each task during training, leading to improved overall performance (see Supplementary Figs. S3-S5). Furthermore, we optimized the latent space dimensionality, identifying $k = 8$ as the optimal dimension that balances reconstruction fidelity and prediction accuracy (see Supplementary Figs. S6). The final optimized model demonstrates strong predictive performance on the test dataset, achieving coefficients of determination $R^2$ ≈0.88 for $B_s$, 0.49 for $\ln(H_c)$ and 0.63 for $D_c$, respectively (see Supplementary Figs. S7). Note that the $R^2$ value for $\ln(H_c)$ is lower than those for $B_s$ and $D_c$. This is mainly because that coercivity is an extrinsic property and strongly affected by microstructural features and processing conditions beyond composition. In addition, the $H_c$ dataset is relatively smaller, providing less training information. Although incorporating processing- or microstructure-related descriptors may further improve the prediction of $H_c$, the acquisition of such information remains a significant challenge in this field. Nevertheless, the MTWAE model still outperforms the single-task ML models in $H_c$ prediction (see Fig. 2(f)).

### 2.4. Multi-objective evolutionary optimization

In the evolutionary optimization shown in Fig. 1(c), our framework employs the NSGA-III to effectively maintain diversity in solutions and uniformly approximate complex Pareto fronts by leveraging a reference-point-based approach. In this procedure, candidate solutions in the latent space are iteratively evolved via genetic operations (selection, crossover, and mutation). Property predictions by the predictors in the MTWAE model guide the evolution towards latent representations associated with optimal combinations of multiple properties, i.e., high $B_s$, low $H_c$, and high $D_c$, effectively managing trade-offs among these properties. Upon reaching a predefined criterion (e.g., maximum generations or target property thresholds), these optimal latent representations are decoded into explicit alloy compositions by the decoder in the MTWAE model (see Fig. 1d). To realize efficient multi-objective evolutionary optimization, we utilized the Genetic and Evolutionary Algorithm Toolbox for Python with High Performance (GEATPY) [34] (see Supplementary Materials III for implementation details).

## 3. Results

### 3.1. Generative exploration and diversity analysis of generated alloy compositions

First, we evaluated the ability of the MTWAE model to generate diverse alloy compositions with desired multiple properties simultaneously, which has been argued to be a prerequisite for addressing any inverse materials design task [35]. Here we chose $k$=8 as the optimal dimensionality of the latent space, and trained the MTWAE model on the entire training datasets containing $B_s$, $H_c$ and $D_c$ data. We randomly generated 3000 Fe-based alloy compositions and predicted corresponding $B_s$, $H_c$ and $D_c$ values by the trained MTWAE model.

Fig. 2(a-c) shows the distribution of the $B_s$, $\ln(H_c)$, and $D_c$ values for the existing and generated alloys, respectively. It is found that the property distribution of the generated alloys (blue) overlaps mostly with the distribution of the existing alloys (red). However, the density of the generated alloys with $B_s$>1.6T or $H_c$<1A/m is lower than that of the existing ones. This may be due to the random generation of alloys, which is difficult to search for alloys with extremely high $B_s$ or small $H_c$. Among 3000 generated alloys, 221 alloys are found to simultaneously satisfy the criteria of $B_s$>1.5(T), $\ln(H_c)$<1.5(A/m), and $D_c$>1mm, as shown in Fig. 2(a-c), denoted as target alloys (green). These results highlight the capability of the model in generating Fe-based MG compositions with high performance of soft magnetic properties.

To analyze the diversity of the generated alloy compositions, we reduced the high-dimensional latent space to 2 dimensions by using the t-distributed stochastic neighbor embedding (t-SNE) method [36]. Fig. 2(d) visualizes the composition distribution of the existing and generated alloys in the 2D latent space, respectively. It can be seen that the generated alloys (blue) not only cover the existing ones (red), but also extend to the nearby region. This indicates that our MTWAE model can capture the composition distribution in the training dataset and generate new alloy compositions in a reliable way. This also indicates that the alloy compositions generated by the MTWAE model are as diverse as the existing ones.

Moreover, the MTWAE model can generate a substantial amount of unique and new alloy compositions. Here two alloy compositions are considered similar if they contain the same elements and corresponding content differences are less than 1 at.%. Thus, a generated alloy composition can be regarded unique if it is not similar to any other alloy compositions generated by the same method, and an alloy composition is new if it is not similar to any alloy compositions in the $B_s$ dataset. As shown in Fig. 2(e), among 1000 generated alloy compositions, the percentage that are both unique and new is as high as 98%. The percentage still reaches 64%, after 100000 alloy compositions are generated. It is also found that 99% of the generated alloy compositions are new (see Fig. 2(e)). This demonstrates that the MTWAE model can generate diverse alloy compositions without significant saturation even at a large scale, and almost all of these alloy compositions are new, compared to the training dataset. These results illustrate the ability of the MTWAE model to generate unique and new alloy compositions.

Next, we demonstrated the property prediction ability of the MTWAE model by benchmarking with several traditional ML models, such as support vector machine (SVM), K-nearest neighbors (KNN), random forest (RF), and extreme gradient boosting (XGBoost) (see Supplementary Table S4 for the optimal hyperparameters of these traditional ML models). The prediction performance was systematically evaluated by the 5-fold cross-validation. As shown in Fig. 2(f), the MTWAE model significantly outperforms the traditional ML models in the prediction of $H_c$ and $D_c$. In the $B_s$ prediction task, its performance is comparable to the best-performing XGBoost model, despite a smaller weight assigned to the $B_s$ predictor in the training of the MTWAE model (see Methods for more details). These results indicate that through multi-task learning, the MTWAE model exploits the shared knowledge of datasets across different tasks and benefits from the combined datasets, which enhances the prediction accuracy and generalization capability, particularly when each dataset is quite limited [37-38]. For the traditional ML models, however, they are trained independently for each single property prediction, and thereby fail to share the knowledge of dataset across different tasks, resulting in prediction accuracy reduction and a higher risk of overfitting [27].

Therefore, the MTWAE model couples high generative diversity with superior prediction accuracy and thus establishes a practical foundation for reliable inverse design. Next, we tried to understand how the MTWAE model captures and organizes the complex relationships between alloy compositions and multiple properties. Here, we characterized the distribution of compositions with a specific property in the latent space by employing the principal component analysis (PCA) [39] to reduce the 8-dimensional latent space of the model to 2 dimensions. As shown in Fig. 2(g), the alloy compositions with high $B_s$ values (red dots) are mainly distributed on the left side of the latent space, while the compositions with low $B_s$ values (blue dots) are located on the right side. This indicates that the MTWAE model is able to organize the alloy compositions effectively in the latent space according to the $B_s$ properties. Different from the distribution of $B_s$, compositions with high $D_c$ values locate on the right side of the latent space, whereas those with low values are on the left side, as shown in Fig. 2(i). The difference in the distributions of $B_s$ and $D_c$ in the latent space indicates that the model successfully captures the trade-off between $B_s$ and $D_c$, and assigns clear regional divisions in the latent space for different properties. Notably, the distribution of alloy compositions with $H_c$ values in the latent space shows a complex feature (see Fig. 2(h)). Compositions with high $H_c$ values (red dots) are located at the end of the latent space, while those with low $H_c$ values locate in the central region, demonstrating a complex relationship between $H_c$ property and alloy compositions.

The above results reveal the significant advantages of the MTWAE model in multi-task learning: by sharing knowledge between different tasks of $B_s$, $\ln(H_c)$, and $D_c$, the MTWAE model is able to effectively express and organize the relationships between multiple properties and alloy compositions in a unified latent space, which results in the complementary and collaborative knowledge across tasks. In addition, the latent space unifying multiple properties allows to explore the alloy composition space with multiple targeted properties more efficiently in the inverse design framework, accelerating the discovery of high-performance alloy materials.

### 3.2. Efficient multi-objective inverse design of Fe-based amorphous alloys

In the above section, we established a robust MTWAE model which is capable of accurately predicting $B_s$, $H_c$, and $D_c$ while simultaneously generating corresponding alloy compositions. To realize efficient multi-objective inverse design for Fe-based MGs with high performance of soft magnetic properties, we incorporated the NSGA-III algorithm into the MTWAE model. Here NSGA-III is used to efficiently search for latent composition representations in latent space obtained by the MTWAE model, and offer multi-objective optimization for the trade-offs between the $B_s$~$D_c$, $B_s$~$H_c$, and $H_c$~$D_c$ property pairs. In the optimization, $B_s$ and $D_c$ are maximized, while $\ln(H_c)$ is minimized. The searched latent composition representations are subsequently decoded into corresponding alloy compositions using the decoder in the MTWAE model. In the multi-objective optimization, a balanced set of solutions, known as the Pareto optimal solutions is identified, which provide the best possible trade-offs among multiple properties.

Fig. 3(a) visualizes the Pareto optimal solutions in the diagram of $B_s$-$\ln(H_c)$-$D_c$ obtained by a population size of 200 across 500 generations. While the red circles (target MGs) represent the Pareto optimal solutions that meet all the target properties of $B_s$>1.5T, $\ln(H_c)$<1.5A/m, and $D_c$>1mm, the gray triangles (other MGs) represent the Pareto optimal solutions that do not meet at least one target property. The green, blue, and orange circles are the projections of the Pareto optimal solutions onto the $B_s$-$D_c$, $B_s$-$\ln(H_c)$, and $D_c$-$\ln(H_c)$ plot, respectively, representing the trade-offs between each pair of properties. The data exhibit clear trends: a negative $B_s$–$D_c$ relation (higher $B_s$ typically coincides with smaller $D_c$) and a positive $B_s$–$\ln(H_c)$ relation (higher $B_s$ accompanies higher $\ln(H_c)$). The Pareto optimal solutions are well spread, covering approximately $B_s$≈1.5~2.0T, $\ln(H_c)$≈-0.5~1.5A/m, and $D_c$≈2~6 mm, with a denser cluster near $B_s$≈1.6~1.8T, and $\ln(H_c)$≈0.3~1.0A/m. This indicates that the GMTDL framework achieves good trade-offs between the conflicting properties.

To illustrate the ability of the GMTDL framework to identify new alloy compositions with superior properties, we compared the Pareto optimal solutions (the candidate alloys) of each pair of properties with the experimental data in the datasets. As shown in Fig. 3(b–d), the Pareto optimal solutions extend the feasible performance space beyond the region covered by the experimental data, demonstrating that the GMTDL framework successfully identifies new alloy compositions and significantly pushes the achievable performance boundaries moving toward regions with better trade-offs between conflicting properties. Specifically, Fig. 3(b) showcases many candidate alloys that achieve satisfactory trade-off between high $B_s$ and low $\ln(H_c)$. In the experimental dataset, however, alloys with high $B_s$ (>1.5T) are typically accompanied by higher $\ln(H_c)$ (>1.5A/m). The trade-offs of $B_s$~$D_c$ and $D_c$~$H_c$ also show similar trend (see Figs. 3(c)&(d)). Therefore, our GMTDL framework successfully identifies new alloy compositions with more superior properties.

### 3.3. Model validation and discovery of high-performance Fe-based amorphous alloys

In this section, we rigorously validated our MTWAE model and assessed its generalizability and predictive power in two aspects. One is that we applied our MTWAE model to predicting the $B_s$ values for given Fe-based amorphous alloys. To demonstrate it, we applied the MTWAE model to 12 Fe-based alloys reported very recently that were not included in our datasets. As shown in Fig. 4(a), the MTWAE model exhibits remarkable accuracy in predicting $B_s$. For amorphous alloys of $Fe_{68.2}Co_{17.5}B_{13}Si_{0.5}Cu_{0.8}$ and $Fe_{79.7}Co_{6}B_{13}Si_{0.5}Cu_{0.8}$ fabricated by Guo *et al.*, the predicted $B_s$ values by the MTWAE model are 1.92T and 1.86T, respectively, with MAE≈0.04 T, very close to the experimentally measured $B_s$ values of 1.90T and 1.81T [40]. This confirms the predictive power of our model for Fe-based amorphous alloys with ultrahigh $B_s$.

For amorphous alloys of $Fe_{85}B_{12}Si_{2}V_{0.5}Cu_{0.5}$, $Fe_{76.5}Co_{8.5}B_{12}Si_{2}V_{0.5}Cu_{0.5}$, $Fe_{68}Co_{17}B_{12}Si_{2}V_{0.5}Cu_{0.5}$, and $Fe_{68.8}Co_{17.2}B_{11}Si_{2}V_{0.5}Cu_{0.5}$ synthesized by Li *et al.*, the predicted $B_s$ values are about 1.70~1.87T with MAE≈0.10T, quite close to the experimental values of 1.58~1.84T [16]. Moreover, our model reproduces the change tendency of the $B_s$ value with Co content, increasing first and then decreasing as Co content exceeds 17% [16]. These results indicate that our MTWAE model not only accurately predicts the $B_s$ values, but also correctly captures the essential Fe–Co exchange-coupling mechanism that governs $B_s$ [41-42].

Our MTWAE model can also estimate the $B_s$ values of the amorphous precursors for the Fe-based amorphous/nanocrystalline alloys. Very recently Yang, *et al*. developed some Fe-based amorphous/nanocrystalline alloys with $B_s$ exceeding 1.9T [43]. For the amorphous precursors of these Fe-based alloys, all the predicted $B_s$ values are larger than 1.7T, as shown in the right panel of Fig. 4(a). It is well known that the $B_s$ values of the amorphous precursors are smaller than those in the amorphous/nanocrystalline counterparts, because the fine grains smaller than 20 nm formed within the amorphous matrix after annealing possess higher saturation magnetization than the residual amorphous matrix, thereby further enhancing the $B_s$ values [43].

In addition to accurately predicting the $B_s$ of Fe-based alloys, the MTWAE model can simultaneously provide predictions for $H_c$ and $D_c$ (see Table 1). Note that here $D_c$ serves as an indicator of the GFA of alloys, and the predicted $D_c$ values reflect the potential GFA of alloys but do not necessarily correspond to the real critical diameter of a given alloy under current experimental conditions. The distribution of experimentally measured $D_c$ values is predominantly concentrated within the range of 1~3 mm, and the MTWAE model predictions align closely with this distribution (see Supplementary Fig. S2(c)), highlighting the remarkable reliability of the MTWAE model in predicting the GFA of Fe-based alloys. For the 12 alloy compositions listed in Table 1, the predicted $D_c$ values primarily range from 1.12 to 1.39 mm, all exceeding 1 mm, indicative of good GFA of these alloys. Although the critical diameters of the Fe-based alloys mentioned above were not measured by copper mold casting methods in experiments, their good GFA has been confirmed by the successful fabrication of Fe-based MG ribbons using melt spinning [16,40,43]. Furthermore, the predicted $H_c$ values are highly consistent with experimental findings; for instance, amorphous alloys prepared by Yang et al. typically exhibit coercivities above 10 A/m in their amorphous state [43], and correspondingly, the MTWAE model also predicts similar $H_c$ values for nearly all these alloys, confirming the accuracy of the MTWAE model in predicting coercivity for amorphous alloys.

The other aspect is that our GMTDL framework can also navigate the search in the vast compositional space to efficiently identify high-performance Fe-based amorphous alloys. With a population of 500 after 1000 generations, the GMTDL framework pinpoints a Pareto front of 95 optimal compositions that simultaneously satisfy the more stringent criteria, i.e., $B_s$ >1.75T, ln( $H_c$ )<1.5 A/m, and $D_c$ >1.0 mm. Some compositions with ultra-high $B_s$, ultra-low $H_c$, and good GFA are listed in Table 1 (see Supplementary Table S5 for all predicted ones). These newly discovered candidates exhibit much more superior properties, locating in a high-performance region of the property space that is sparsely populated by both alloys in the datasets (grey squares) and recently reported high-performance alloys (blue triangles) (see Fig. 4(b-d) and Supplementary Table S5). For instance, Fig. 4(b) reveals that most of these compositions exhibit substantially low $H_c$ (ln($H_c$)<1.5A/m), while maintaining high $B_s$ (>1.75T). Moreover, these candidate alloys also tend to possess larger $D_c$ than those in the dataset and recently reported Fe-based alloys, while retaining high $B_s$, as shown in Fig. 4(c). In addition, the newly discovered compositions simultaneously achieve both high $D_c$ and low ln($H_c$), as shown in Fig. 4(d).

To go beyond the property-space validation and the discovery of superior alloy compositions and assess whether the discovered candidates remain within empirically plausible thermodynamic–structural regimes or not, we further conducted a physics-guided plausibility check by projecting alloys in the datasets and the GMTDL candidates (see Supplementary Table S5) onto physically motivated descriptor planes. Here three physical descriptors were considered, i.e., atomic-size mismatch $\delta$, mixing enthalpy $\Delta H_{mix}$, and a theoretical saturation magnetic flux density $B_s^{cal}$ (see Supplementary Materials IV for definitions). While $\delta$ and $\Delta H_{mix}$ are related to the GFA [44], $B_s^{cal}$ corresponds to $B_s$ [45]. To visualize the distribution, we plotted the $\delta$–$\Delta H_{mix}$ map together with two magnetism-related projections, i.e., $B_s^{cal}$–$\delta$ and $B_s^{cal}$–$\Delta H_{mix}$. As shown in Fig. 5(a), the dataset-defined high-performance subsets occupy a compact region characterized by moderate $\delta$ and negative $\Delta H_{mix}$. Moreover, the GMTDL candidates largely fall within, or in proximity to the envelope spanned by these experimentally observed subsets. No obvious drift toward extreme outliers is observed in the composition–thermodynamics descriptor space. In addition, Fig. 5(b–c) indicate that the candidates preferentially populate the upper envelope of $B_s^{cal}$, consistent with the design objective of achieving higher predicted $B_s$ (e.g., more stringent target $B_s > 1.75$T used to define the highlighted candidates in Fig. 5). Moreover, Fig. 5(c) suggests a mild shift of candidates toward less negative $\Delta H_{mix}$ along the high-$B_s^{cal}$ ridge, which is consistent with the commonly observed trade-off between pushing saturation performance and maintaining strongly negative mixing enthalpy that favors glass formation of Fe-based alloys.

To further substantiate the validity of the GMTDL framework, we compared the predicted candidates (see Supplementary Table S5) with recently reported Fe-based alloys with high $B_s$ values (see Table 1). The GMTDL framework autonomously converges to the compositions that are close to the experimental exemplars in both compositions and properties. For instance, the candidate of $Fe_{68.47}B_{13.11}Si_{1.20}Co_{17.03}$ closely matches the reported alloy of $Fe_{68.2}B_{13}Si_{0.5}Co_{17.5}Cu_{0.8}$ in compositions with nearly identical total magnetic-element content, i.e., 85.50 vs 85.70 at.% of Fe+Co content, and exhibits comparable $B_s$ values (1.97 vs 1.90T). Likewise, compared to the reported $Fe_{68}Co_{17}B_{12}Si_2V_{0.5}Cu_{0.5}$ alloy, the candidate of $Fe_{67.13}Co_{17.44}B_{12.79}Si_{2.09}Nb_{0.43}$ alloy has a highly similar magnetic-element content (84.57 vs 85.00 at.%) and $B_s$ values (1.94T vs 1.84T), indicating that the framework rediscovers the same high-performance regions in composition space. These quantitative matches across composition, total magnetic-element content, and $B_s$ values provide firm evidence that the GMTDL framework can learn the governing composition-property relationship and reliably rediscover the experimentally validated regions of Fe-based alloys with ultra-high $B_s$. Thus, the framework generates large pools of reliable, high-performance alloy candidates, markedly reducing experimental trial-and-error and accelerating the discovery of next-generation soft magnetic materials. Note that alloys of number 13-20 in Table 1 as well as those in Supplementary Table 5 are predicted by the GMTDL framework and have not yet been experimentally tested.

## 4. Discussion

### 4.1. Comparison with previous ML-based Strategies

To demonstrate the search efficiency of our GMTDL framework for discovering compositions with target properties, i.e., $B_s$>1.5 T, ln($H_c$)<1.5 A/m, and $D_c$>1 mm, we conducted a comprehensive comparison with previous ML approaches in alloy design. Data-driven generative ML models represent one prominent approach [46-48]. While generative models alone can effectively capture data distribution, they often face great challenges in target property optimization [48]. ML models combined with genetic algorithms (GA+ML) constitute another approach [49-51], which has received increasing attention. Here we made a comparison between the GA+ML approach and our GMTDL framework by using the same datasets, target criteria, NSGA-III settings, and computing platform. We implemented GA+ML approach by using the NSGA-III algorithm. In this approach, a composition space needs to be defined. Here a composition space containing 6 most prevalent elements of Fe, B, Si, C, P, and Co was considered, yielding 26 alloy systems from ternary to 6-element compositions. Moreover, property predictions were employed by XGBoost for $B_{\mathrm{s}}$ and $D_{\mathrm{c}}$, and random forest for $H_{\mathrm{c}}$, respectively. With population sizes of 500 over 1,000 generations per alloy system, the GA+ML approach generates 13,000 samples total, of which 3,428 (26%) satisfies the performance criteria, as shown in Fig. 6(a). The average values are $B_{\mathrm{s}}$≈1.62T, ln($H_{\mathrm{c}}$)≈1.05A/m, and $D_{\mathrm{c}}$≈2.37mm, respectively (see Fig. 6(b)).

For the GMTDL framework, we trained ten MTWAE models with different parameter initializations to ensure robust statistical sampling. Each model was executed with a population of 500 over 1,000 generations, generating 5,000 alloy compositions total. Remarkably, 2,255 candidates (45%) satisfy the predefined criteria (see Fig. 6(a)), a dramatic improvement from the sub-1% success rate of standalone generative models [48]. The GMTDL generated samples exhibit superior average properties of $B_{\mathrm{s}}$≈1.65 T, ln($H_{\mathrm{c}}$)≈0.19A/m, and $D_{\mathrm{c}}$≈3.05mm (see Fig. 6(b)). These metrics substantially exceed target thresholds, with $B_{\mathrm{s}}$ surpassing requirements by 10%, ln($H_{\mathrm{c}}$) achieving an 87% improvement below threshold, and $D_{\mathrm{c}}$ tripling the minimum requirement.

Apparently, the GMTDL framework exhibits significant improvement in search efficiency compared to GA+ML approach, while delivering superior property distributions across all three metrics. This enhanced performance stems from a fundamental advantage: the GMTDL framework initiates searches from the distribution of existing samples, navigating design space proximate to this distribution. This approach generates novel yet chemically plausible designs, maintaining prediction reliability throughout optimization. On the contrary, the GA+ML method often encounters significant discrepancies between randomly generated samples and training data, undermining prediction reliability [52]. On the other hand, a representative GMTDL optimization run needs about 4.0s, whereas the GA+ML approach costs about 12.7-16.8s per alloy system, with an average time of 14.7s. Moreover, GA+ML must optimize 26 predefined alloy systems separately, and its cumulative search time was about 383s in our implementation. In contrast, GMTDL performs the search directly in a fixed 8-dimensional latent space and then decodes latent vectors into alloy compositions, leading to substantially lower search cost. Therefore, the GMTDL framework substantially outperforms traditional ML approaches in both efficiency and scalability for inverse design of high-performance alloys.

**4.2. Evolution of the Pareto front in the GMTDL framework**

To understand why the GMTDL framework can provide accurate and efficient prediction of high-performance Fe-based soft magnetic alloys, we tracked and visualized the evolution of the composition population across multiple generations in the evolutionary optimization procedure. We initialized the population with 200 individuals and examined 0, 10, and 200 generations, respectively. In Fig. 7(a-c), the t-SNE embeddings depict 3000 randomly generated compositions (blue heatmap) superimposed with Pareto optimal solutions (yellow points). These panels illustrate how the Pareto solutions transform from a random distribution in the initial generation (see Fig. 7(a)) to a more refined, higher-performance region of the design space after 200 Generations (see Fig. 7(c)). Specifically, in the generation 0, the randomly sampled latent variables yield a population with sparse Pareto optimal solutions, aligning with a nearly random composition distribution. Only a small subset of these individuals achieves desirable trade-off in $B_\mathrm{s}$, $H_\mathrm{c}$, and $D_\mathrm{c}$. In the generation 10 shown in Fig. 7(b), the Pareto solutions begin to grow, reflecting initial success in balancing the competing objectives and indicating that the algorithm is effectively exploring the latent space. In the generation 200 shown in Fig. 7(c), the Pareto-front solutions clearly separate from the main density cloud, converging toward less-populated regions that are less likely to arise from random sampling alone. This demonstrates the transformation from near-random exploration to targeted optimization, highlighting the capability of the GMTDL framework to uncover compositions with superior properties.

In addition, Fig. 7(d–f) compares the pairwise properties of $B_\mathrm{s}$~ln($H_\mathrm{c}$), $B_\mathrm{s}$~$D_\mathrm{c}$, and $D_\mathrm{c}$~ln($H_c$) across the generation 0, 10, and 200, which clearly shows how the Pareto-front solutions shift toward improved $B_\mathrm{s}$, reduced ln($H_c$), and enhanced $D_\mathrm{c}$ during the generations. These shifts confirm that the GMTDL framework successfully discovers the compositions with balanced trade-off properties. Consequently, the optimization process evolves from random sampling to a systematic, iterative search in regions of the composition space with superior multi-objective solutions. Such results further validate the efficiency and advantage of the GMTDL framework in identifying target compositions in a vast composition space.

We also evaluated whether the high-performance solutions collapse to a single local minimum or instead arise from multiple, compositionally distinct regions of the composition space by comparing representative candidates (yellow circles in Fig. 3(b-d)). Although several alloys achieve target properties, i.e., $B_s$>1.5T, ln($H_c$)<1.5A/m, and $D_c$>1mm, their chemistries differ markedly, indicating that they do not originate from the same local regions. For example, $Fe_{61.35}B_{23.38}Si_{1.00}Co_{8.98}Nb_{4.94}Y_{0.15}$ and $Fe_{72.28}B_{15.79}Si_{4.24}Co_{3.59}Nb_{1.90}Cu_{2.01}$ differ markedly in the content of Fe and B, while $Fe_{67.95}B_{21.62}Si_{1.19}Co_{7.58}Nb_{1.48}$ and $Fe_{77.13}B_{13.45}Si_{7.53}P_{0.19}Nb_{0.22}Zr_{1.21}Ga_{0.11}$ also exhibit significant differences in the content of Fe, B, and Co. To more clearly visualize the inherent compositional diversity of these candidate alloys, we plotted their compositions in a ternary Fe–B–Si diagram by focusing exclusively on three dominant alloying elements (see Supplementary Fig. S8). Despite containing additional alloying elements, this representation effectively highlights the broad compositional distribution of the candidate alloys. This demonstrates that our multi-objective design framework can discover a diverse range of novel candidate alloys.

Note that the present GMTDL framework is established on available literature datasets and evaluates candidates mainly through learned composition-property relationships. Consequently, possible effects associated with real processing conditions, such as composition deviation, cooling-rate variation, and microstructural heterogeneity, were not yet considered in the current study. Nevertheless, the close compositional and property-level agreement between the predicted candidates and recently reported high-performance alloys (Table 1, entries 1-12), together with the physics-guided plausibility analysis (Fig.5), provide preliminary evidence supporting the reliability of these predictions.

### 4.3. Elemental design principles for high-performance Fe-based amorphous alloys

Based on the GMTDL framework, we utilized the Shapley Additive Explanations (SHAP) method [53] to provide quantitative insights into the contribution of individual elements to the predicted outcomes. We first focused on the experimental datasets, and applied the SHAP method to scrutinize the influence of individual elements on $B_s$, $H_c$, and $D_c$. Fig. 8(a-c) presents the SHAP plots for the 12 most prevalent elements: Fe, B,

Si, P, C, Co, Nb, Ni, Mo, Zr, Ga, and Al, where positive/negative values indicate a positive/negative impact on the predicted properties. The analysis reveals a clear hierarchy of elemental influences on properties. Apparently, magnetic elements (Fe, Co) predominantly control $B_s$, showing strong positive SHAP values, validating that increase of the magnetic element content is the most straightforward way to enhance the saturation magnetization [54]. However, excessive Fe content may deteriorate the GFA, as the SHAP values are negative for $D_c$. B exhibits not only significantly positive impact on $D_c$ (enhancing GFA), but also largely negative SHAP values for $H_c$ (reducing coercivity), making it crucial for both structural stability and magnetic softness [10]. Diffusion-inhibiting elements (Zr, Nb, Mo) with large atomic radii effectively enhance the GFA, but exhibit strong detrimental effects on $B_s$ at high concentrations, necessitating precise compositional tuning [11].

To translate these global trends into practical alloy design strategies, we systematically analyzed the SHAP-derived dependence on the content of some key elements (see Supplementary Figs. S9–S11). For Fe, the SHAP values of $B_s$ increase almost linearly with increasing Fe content, clearly illustrating the contribution of Fe content to $B_s$ (see Fig. S9a). The SHAP values remain positive for Fe>72 at.%, showing positive contribution of Fe content to $B_s$, whereas the values are all negative for Fe<65 at.%, indicating negative contribution to $B_s$. On the contrary, the SHAP values of $D_c$ continuously decrease with increasing Fe content, and change from positive to negative at around 70 at.% (see Fig. S9c). This confirms the trade-off between $B_s$ and $D_c$. Moreover, the SHAP values of $B_s$ and $D_c$ as a function of Fe content suggest an optimal window of Fe content, i.e., 70~72 at.% to balance $B_s$ and $D_c$. As Fe content goes beyond 72 at.%, its effect on $D_c$ is always negative, indicating that it is necessary to introduce some other elements to improve the GFA of Fe-based alloys with Fe>72 at.%. For $H_c$, the SHAP values fluctuate between positive and negative in the whole range of Fe content (see Fig. S9b), indicating a random effect on the $H_c$. In contrast to Fe, Co provides positive contributions to $B_s$ as the content goes beyond 5 at.% (see Fig. S9d), while exerting a random contribution to $D_c$ (see Fig. S9f). In addition, Co may concurrently decrease $H_c$ (negative SHAP values, Fig. S9e). This demonstrates that Co addition (>5 at.%) may improve both $B_s$ and $H_c$ without weakening the GFA, and even has chances to enhance the GFA. For magnetic element Ni, it exhibits unique dual functionalities with a critical transition at 8 at.%, above which Ni simultaneously enhances $B_s$ (see Fig. S9g, positive SHAP values up to 0.4 T) and reduces $H_c$ (see Fig. S9h), negative SHAP values reaching -1.0 A/m). However, Ni has negative impact to the GFA, as the content is larger than 14 at.% (see Fig. S9i), which suggests an optimal regime of 8~14 at.% for enhancing the magnetic properties with acceptable GFA.

Metalloid elements show similar impact on these properties, but with distinct optimal content (see Fig. S10). For B, as the content is below 15 at.%, the SHAP values for $B_s$ are mostly positive, favoring the $B_s$ (see Fig. S10a). For $H_c$, the SHAP values are mostly positive below 15 at.%, but become mostly negative above 15 at.% (see Fig. S10b) which benefits a reduction of $H_c$. For $D_c$, the benefit emerges beyond ~6 at.% and strengthens with increasing content, as the SHAP values are mostly positive (see Fig. S10c). Thus, an optimal window of B content is around 15~25 at.%, favoring the $H_c$/GFA-prioritized designs, or 10~15 at.% when $B_s$ preservation is critical. The effect of P on $B_s$ is quite similar to B, but the critical content is around 6 at.%, beyond which the SHAP values for $B_s$ becomes negative, leading to worse $B_s$ (see Fig. S10d). However, the SHAP values for $D_c$ remain near zero below 4 at.%, but become strongly positive above 4 at.% (see Fig. S10f), strongly favoring the GFA. This confirms the role of P as a good GFA enhancer. In contrast to B, P exhibits almost random impact on $H_c$ (see Fig. S10e). Thus, for P, the content of 4~6 at.% may be the optimal range for tuning magnetic performance. For Si, its SHAP values for $B_s$ are positive only as the content is very low (Fig. S10g). However, SHAP values for $D_c$ are mostly positive as the content is above 2 at.% (Fig. S10i). Similar to P, Si addition shows nearly random effect on $H_c$ (Fig. S10h). There is still a narrow content range below 4 at.% for Si to tune both magnetic properties and GFA (see Fig. S10g&i). As C content is above 2 at.%, $B_s$ penalty becomes notable (see Fig. S10j), whereas $H_c$ benefits appear below 4 at.% (see Fig. S10k). In addition, $D_c$ gains emerge as the content ≥4 at.% (see Fig. S10l). This suggests 2~4 at.% of C content for balanced optimization or 4~6 at.% for GFA-prioritized compositions.

Transition metal modifiers impose more stringent trade-offs. Nb displays severe magnetic dilution—even small additions weaken the $B_s$ strongly, as SHAP values reach -2.0 (see Fig. S11a). On the other hand, Nb can effectively reduce $H_c$ as the content is below ~4 at.% (see Fig. S11b), and enhances $D_c$ above 2 at.% (see Fig. S11c). This creates distinct optimization strategies, that is, the content smaller than 1 at.% for improving $B_s$, and 2~4 at.% for $H_c$/GFA-focused designs with acceptable $B_S$, respectively.

Thus, the convergence of the optimal content window for these elements—Fe (70~72 at.%), Co (5~10 at.%), Ni (8~14 at.%), B (15~25 at.%, or 10~15 at.% for $B_s$-priority), P (4~6 at.%), Si (2~4 at.%), C (2~4 at.% for balanced optimization or 4~6 at.% for GFA-priority), and Nb (≤1 at.% for $B_s$-priority or 2~4 at.% for $H_c$/GFA-priority)—defines a highly constrained but achievable compositional space where high $B_s$, low $H_c$, and adequate $D_c$ for glass formation can be simultaneously realized.

Although such analyses of experimental data provide valuable insights, they are inherently limited by small sample sizes and high experimental costs. Our GMTDL framework circumvents this bottleneck by rapidly generating vast virtual libraries of alloy compositions, enabling systematic and robust statistical exploration at a scale which is unattainable previously. To elucidate the underlying elemental "design rules" that govern the simultaneous optimization of high $B_s$, low $H_c$, and superior GFA, we utilized a virtual library containing 500,000 alloy compositions with their predicted $B_s$, $H_c$, and $D_c$ values, generated by ten MTWAE models trained independently. From this virtual composition library, we extracted 7,500 candidates meeting stringent performance criteria ($B_s$>1.70 T, $D_c$>1 mm, and ln($H_c$)<1.5 A/m) and computed the occurrence frequencies and average contents for 28 elements. To validate the robustness of these findings, we generated 10,000 compositions using the MTWAE model and conducted the SHAP analysis, which reveals elemental trends, fundamentally consistent with those in the experimental datasets (see Supplementary Fig. S12).

Fig. 8(d) shows the occurrence frequency of the elements and their average contents among these generated alloys. It can be clearly seen that a fundamental Fe–B–Co–Si quaternary alloy system is essential for designing high-performance Fe-based alloys. The statistical results show that Fe and B are present in 100% of the retained alloys, demonstrating their indispensable roles in developing high-performance Fe-based alloys. The mean content of Fe is about 72.5 at.%, demonstrating its primary magnetic contributor and fundamentally enabling high $B_s$. This optimal content of Fe precisely aligns with that identified in the SHAP analysis (see Fig. S9a). B, with a mean content of ~12.2 at.%, is confirmed as critical for GFA. As the lightest metalloid element, B effectively improves GFA while having the weakest dilution effect on magnetic properties compared to other metalloids. The prevalence of B at this concentration corresponds to the spot identified in our SHAP analysis where B maintains low $H_\mathrm{c}$ while allowing sufficient magnetic element content for high $B_\mathrm{s}$.

Co has the third-highest frequency of occurrence (76.8%), with the mean content of ~9.5 at.%, consistent with the above SHAP analysis. According to the Slater-Pauling curve, partial substitution of Fe with Co enhances magnetic exchange interaction, leading to increased average magnetic moment per atom [55]. Completing this core quaternary system, Si exhibits the occurrence frequency as high as 74.9%, with mean content ~3.0 at.%, located in the optimal content window obtained by the SHAP analysis. While B is essential, excessive amounts lead to precipitation of Fe–B compounds during nanocrystallization, deteriorating soft magnetic properties [56-57]. Our results quantitatively validate that partial substitution of B with Si is highly effective, with Si improving the combination of GFA and magnetic softness through its intersolubility with bcc-Fe to form bcc-(Fe,Si) phase [58].

Apart from the primary Fe-B-Co-Si matrix, our analysis also highlights key secondary alloying elements which are critical for fine-tuning of material properties. P shows the fifth-highest frequency (64.5%), demonstrating its important role in tuning properties of Fe-based MGs. Its atomic radius (0.109 nm) is relatively larger, compared to B (0.087 nm). Therefore, partially substitution of B with P may enhance the exchange-coupling interactions, improving $B_\mathrm{s}$ [59]. Nb also shows significant roles with the frequency of 43.8%, confirming its roles in refining nanocrystalline grains, reducing $H_\mathrm{c}$ and improving GFA, despite its magnetic dilution effect [60]. C with the frequency 37.6% may effectively stabilize the amorphous structures [61]. Ni (24.0% frequency, ~3.4 at.% content) serves an essential role through partial substitution for Co to effectively reduce $H_\mathrm{c}$ while maintaining reasonable $B_\mathrm{s}$ [62].

The remarkable consistency between the SHAP analysis of experimental data and the statistical patterns in the virtual library validates the physical plausibility of the MTWAE-generated compositions and establishes quantitative, data-driven design rules for multi-objective optimization. These findings transform alloy design from empirical trial-and-error to a principled, predictive science, providing a roadmap for developing next-generation soft magnetic materials with tailored properties.

### 4.4. Online toolkit

To facilitate the application of the GMTDL framework, we developed an online interactive tool based on the Streamlit platform. By simply configuring parameters such as population size and the number of generations, the online toolkit offers users an efficient and convenient way to customize Fe-based alloys with desired properties for further experimental preparation and synthesis. The web application for property prediction and the GMTDL framework can be accessed at Fe-based amorphous alloys Property (https://multi-task-generative-model-predict-app.streamlit.app/) and GMTDL Generative Fe-based amorphous alloy Design (https://nsga-multi-task-generative-model-app.streamlit.app/).

**5. Conclusion**

In summary, we developed a generative multi-task deep learning (GMTDL) framework that enables efficient inverse design of Fe-based amorphous alloys with optimized soft-magnetic properties. The framework effectively navigates the high-dimensional composition space, generating a diverse range of novel alloy candidates that push performance beyond the limits of known alloys. Moreover, the GMTDL framework overcomes the long-standing challenge of trade-offs among $B_{\mathrm{s}}$, $H_{\mathrm{c}}$, and GFA, and identifies high-performance alloys with all three objectives optimized simultaneously. The GMTDL framework not only outperforms conventional ML-assisted alloy design approaches in success rate and accuracy, but also provides mechanistic insights into alloy design for guiding the development of next-generation soft magnetic alloys. Nevertheless, because $H_{\mathrm{c}}$ is highly sensitive to processing factors, such as annealing temperature, annealing time, and ribbon thickness, future improvement of the framework requires incorporating processing-related descriptors to establish a more comprehensive composition-processing-property model. In addition, the generative multi-task deep learning design strategy can be generalized to other material systems, providing a robust and intelligent tool to accelerate materials discovery for advanced technology.

**Supplementary Material**

The supplementary material contains additional details on dataset statistics, MTWAE architecture/training, and multi-objective optimization (NSGA-III), along with supporting figures (loss curves, hyperparameter selection, benchmarking, SHAP analyses) and tables (datasets, baseline hyperparameters, optimized alloy compositions).

**Acknowledgements**

This work was supported by National Natural Science Foundation of China (Nos. 12574220 and 52031016). Computational resources were provided by the Physical Laboratory of High-Performance Computing at Renmin University of China.

**Conflict of Interest**

The authors declare no conflict of interest.

**Data availability**

All data used in this work were provided in the manuscript and Supplementary Materials.

**Code availability**

The source code for the GMTDL framework is available at GitHub (https://github.com/kangyuanli/Generative_multi_task_deep_learning_framework).

The archived version with a persistent DOI is: https://doi.org/10.5281/zenodo.19272020.

Compd. **1005** 176172

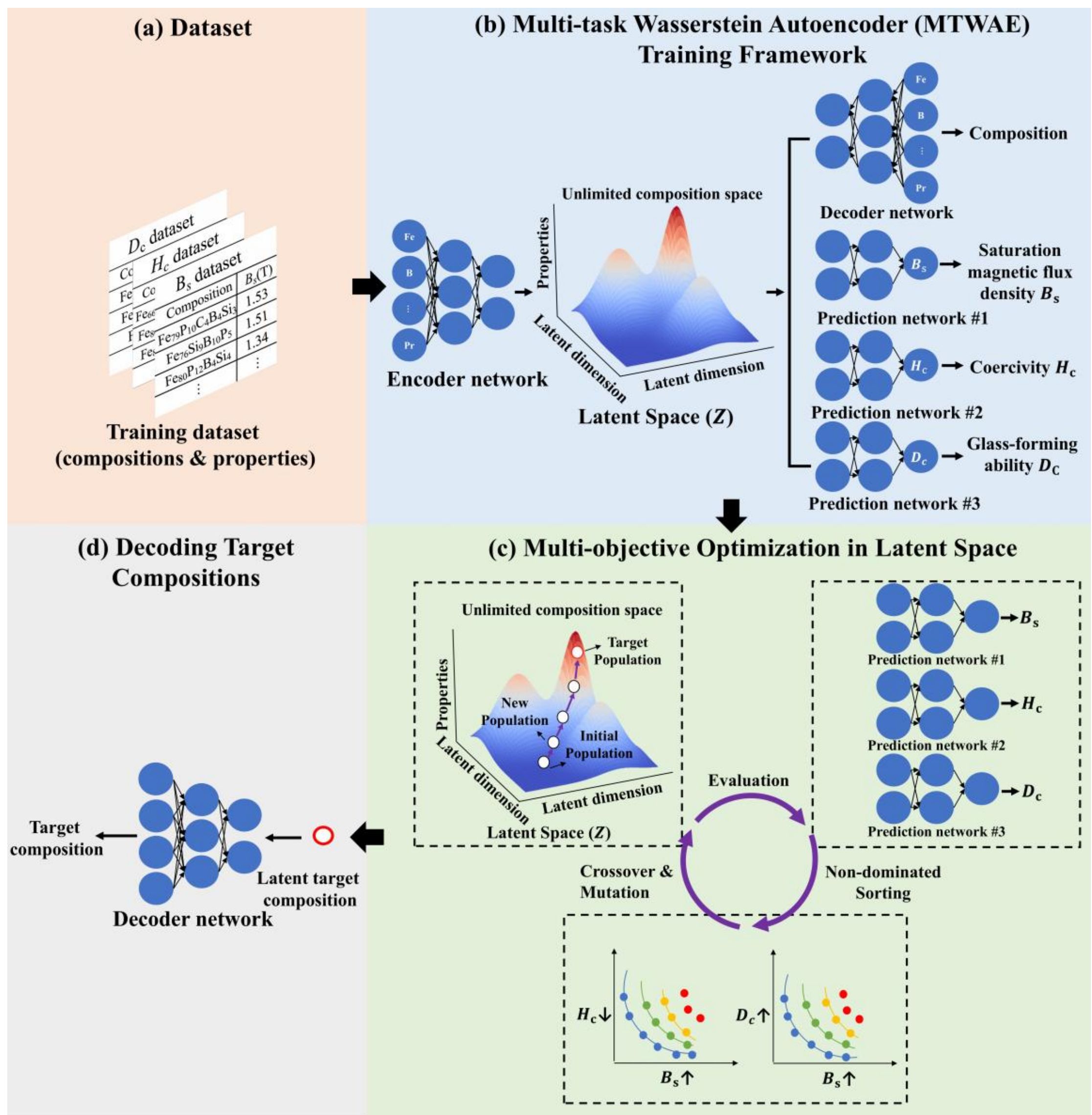

**Figure 1 Schematic of the generative multi-task deep learning framework. (a)** Dataset acquisition of alloy compositions and properties of $B_s$, $H_c$, and $D_c$. **(b)** The MTWAE model which maps compositions into a latent space and predicts properties simultaneously. **(c)** NSGA-III which performs the multi-objective optimization in latent space using the MTWAE predictors for evaluation, generating Pareto-optimal solutions. **(d)** The final optimal latent solutions are decoded back into candidate alloy compositions.

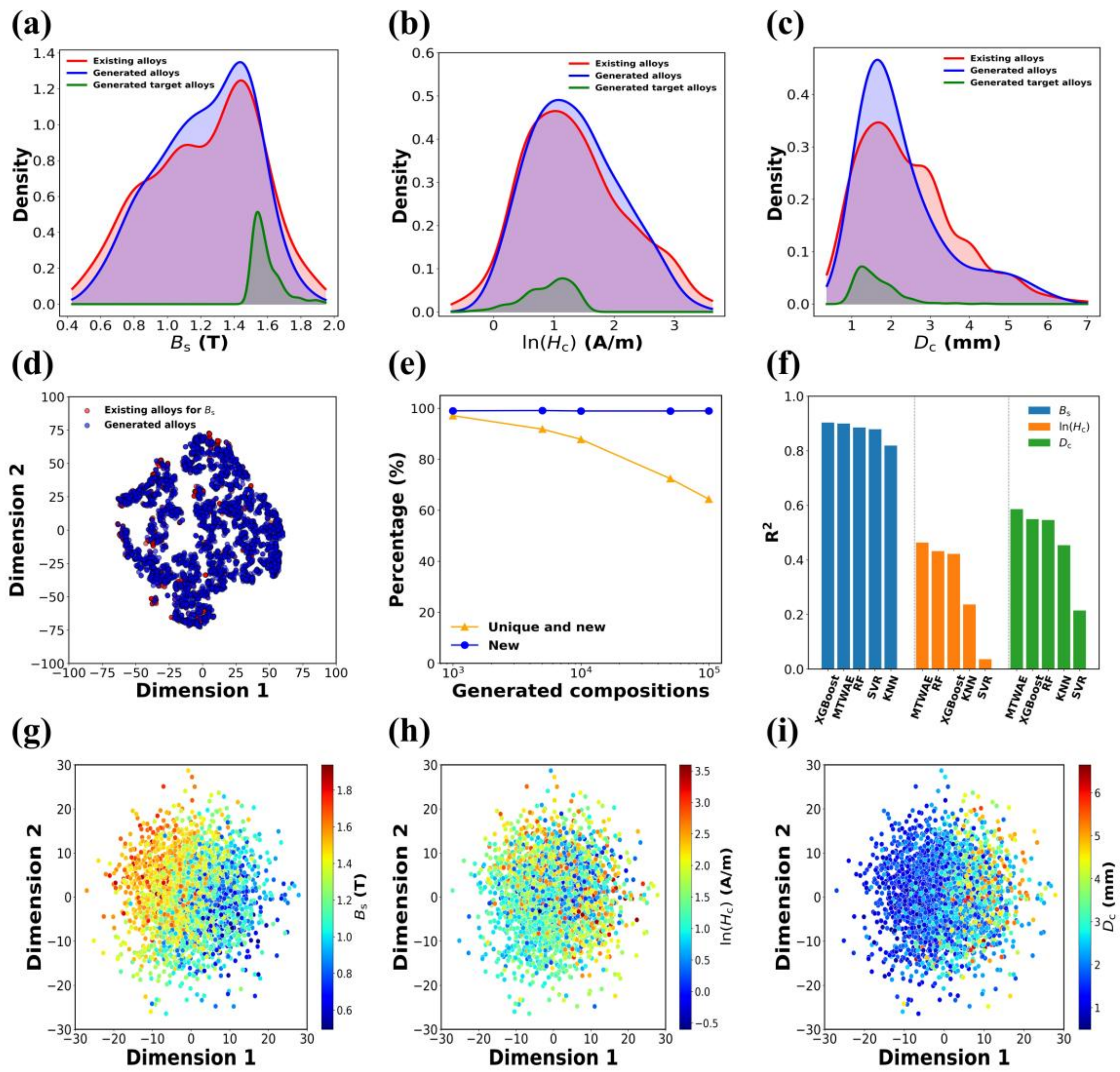

**Figure 2 Generation of diverse alloy compositions with multiple desired properties. (a-c)** Comparison of the property distribution of existing alloys (red) and generated alloys (blue) for saturation magnetic flux density ($B_\mathrm{s}$), coercivity (ln($H_\mathrm{c}$)), and critical casting diameter ($D_\mathrm{c}$), respectively. The green region denotes the target alloys that simultaneously satisfy $B_\mathrm{s}$ >1.5T, ln( $H_\mathrm{c}$ )<1.5A/m, and $D_\mathrm{c}$ >1mm. **(d)** t-SNE visualization of the compositional distribution of existing and generated alloys. **(e)** Analysis of the uniqueness of the generated alloys (ranging from 1000 to 100000). **(f)** Comparison of the property prediction accuracy between the MTWAE model and traditional ML models (SVR, KNN, RF, XGBoost). **(g-i)** Distribution of the 3000 randomly generated compositions in latent space colored by $B_\mathrm{s}$, ln($H_c$) and $D_\mathrm{c}$ values, respectively. Here principal component analysis (PCA) was used to compress the 3000 generated ones in the latent space from 8 dimensions to 2 dimensions.

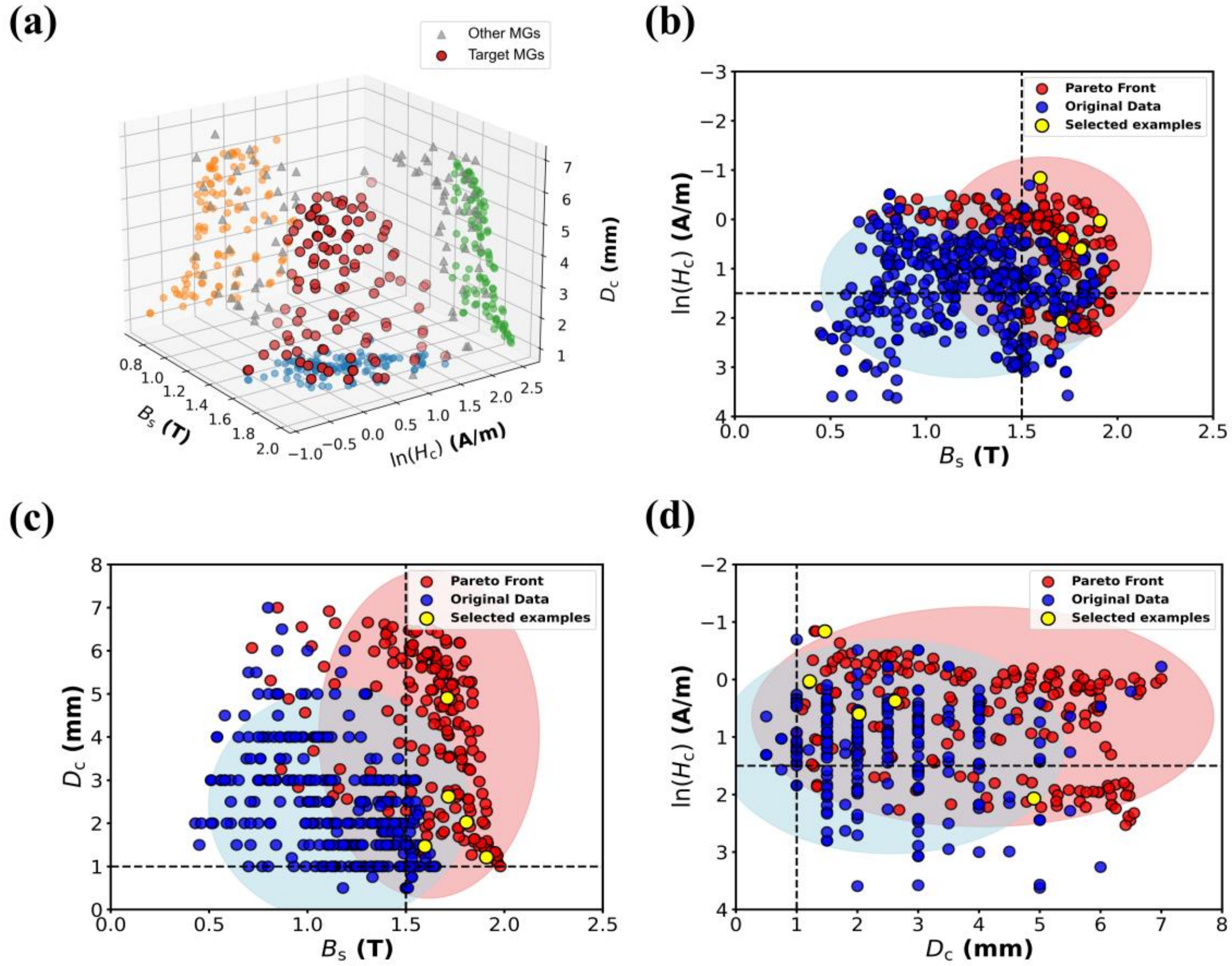

**Figure 3 Multi-objective optimization of alloy properties. (a)** Visualization of Pareto-optimal solutions obtained by the GMTDL framework in the three-dimensional property space of magnetic saturation flux density ($B_s$), coercivity ($\ln(H_c)$), and critical casting diameter ($D_c$). Red circles denote solutions simultaneously meeting the target cirteria ($B_s$>1.5T, $\ln(H_c)$<1.5A/m, $D_c$>1mm), while grey triangles represent non-ideal solutions. Green, blue, and orange circles indicate the projections of target solutions onto the $B_s$~$D_C$, $B_s$~$\ln(H_c)$, and $D_c$~$\ln(H_c)$ performance planes, respectively. **(b-d)** Comparison between the Pareto front obtained by the GMTDL optimization and experimental data on pairwise property planes ($B_s$~$\ln(H_c)$, $B_s$~$D_c$, $D_c$~$\ln(H_c)$), demonstrating the framework's capability to effectively expand the experimental performance boundaries.

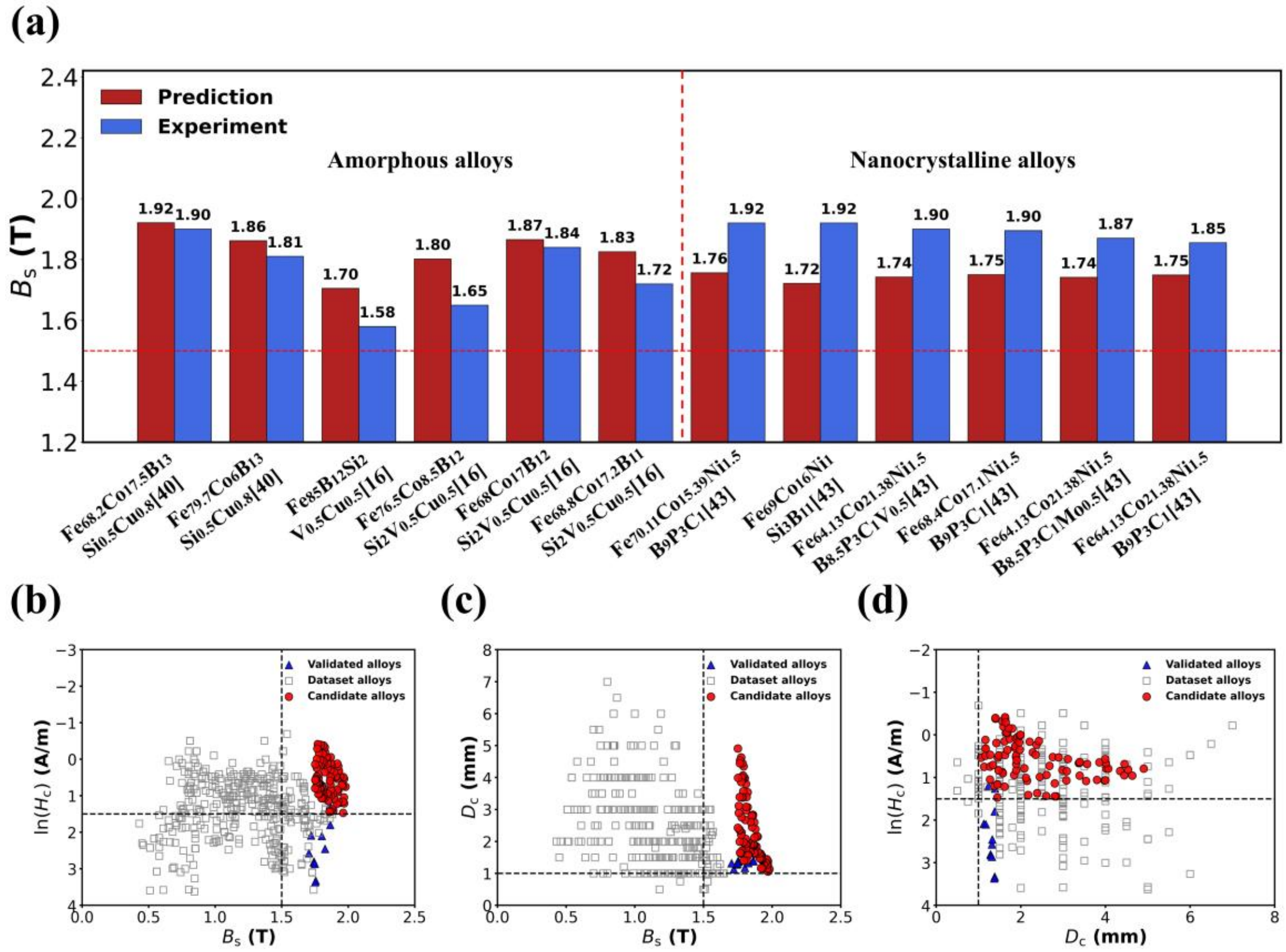

**Figure 4 Validation of the MTWAE model and discovery of high-performance Fe-based amorphous alloys. (a)** Comparison between MTWAE-predicted and experimentally measured saturation magnetic flux density ($B_s$) for twelve recently reported Fe-based amorphous or amorphous/nanocrystalline alloys, which were not included in the training set. Two-dimensional projections of the Pareto front obtained by performing NSGA-III optimization for 1000 generations on a population of 500 latent vectors in the MTWAE latent space: **(b)** $B_s$~ln($H_c$); **(c)** $B_s$~$D_c$; **(d)** $D_c$~ln($H_c$). Red circles indicate 95 newly discovered alloy compositions simultaneously satisfying $B_s$>1.75T, ln($H_c$)<1.5A/m, and $D_c$>1.0mm; squares represent experimental samples from the training dataset; triangles denote the literature alloys shown in (a).

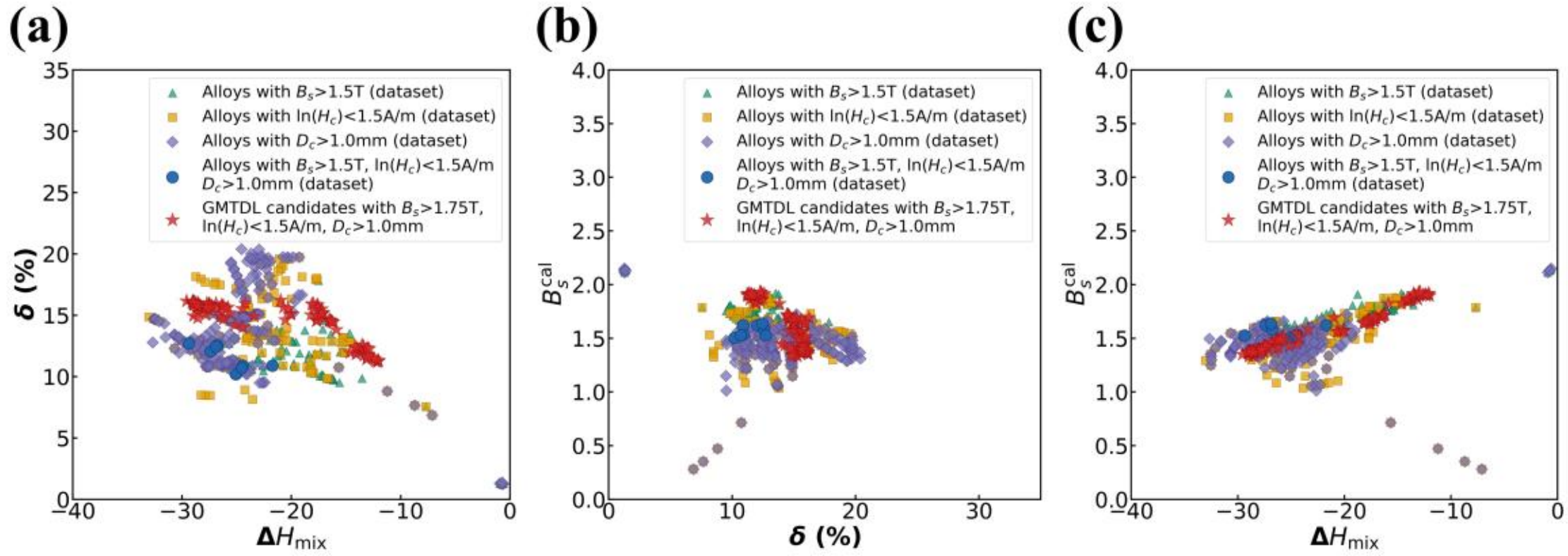

**Figure 5 Physics-guided descriptor mapping of GMTDL-discovered Fe-based amorphous-alloy candidates.** Projections of alloys in the datasets and GMTDL candidates on the physically motivated descriptor planes of $\delta$~$\Delta H_{\text{mix}}$(a), $B_s^{\text{cal}}$~$\delta$(b), and $B_s^{\text{cal}}$ ~ $\Delta H_{\text{mix}}$ (c), respectively. Here $\delta$ is the atomic-size mismatch, $\Delta H_{\text{mix}}$ mixing enthalpy, and $B_s^{\text{cal}}$ composition-only theoretical saturation-moment proxy, respectively. Green triangles denote alloys with $B_{\text{s}} > 1.5\,\text{T}$ in the dataset, orange squares denote alloys with $\ln(H_c) < 1.5\,\text{A/m}$ in the dataset, and purple diamonds denote alloys with $D_c > 1.0\text{mm}$ in the dataset. Blue circles indicate alloys that satisfy all three criteria simultaneously ($B_{\text{s}} > 1.5\text{T}$, $\ln(H_c) < 1.5\text{A/m}$, and $D_c > 1.0\text{mm}$) in the dataset. Red stars represent GMTDL candidates meeting more stringent design targets based on GMTDL predictions, i.e., $B_{\text{s}} > 1.75\text{T}$, $\ln(H_c) < 1.5\text{A/m}$, and $D_c >$ 1.0 mm.

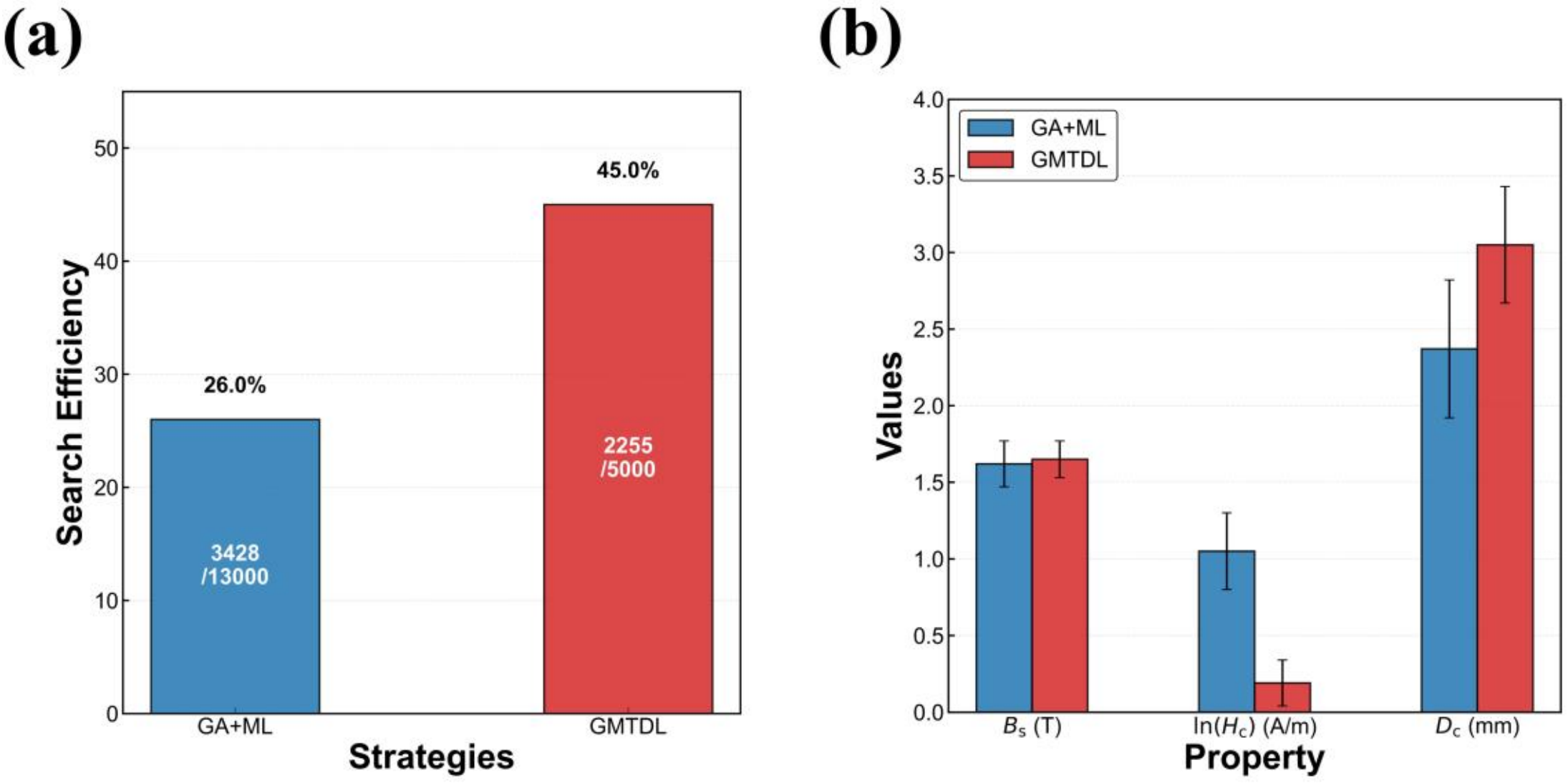

**Figure 6 Comparison of optimization strategies for alloy design. (a)** Search efficiency of the GA+ML strategies and the GMTDL framework, showing the fraction of generated samples meeting target criteria ($B_s$>1.5 T, ln($H_c$) < 1.5 A/m, $D_c$>1 mm). **(b)** Average property values of samples satisfying target criteria generated by GA+ML strategy and the GMTDL framework, respectively, demonstrating superior performance of the GMTDL-designed alloys across all three metrics. Error bars represent standard deviation.

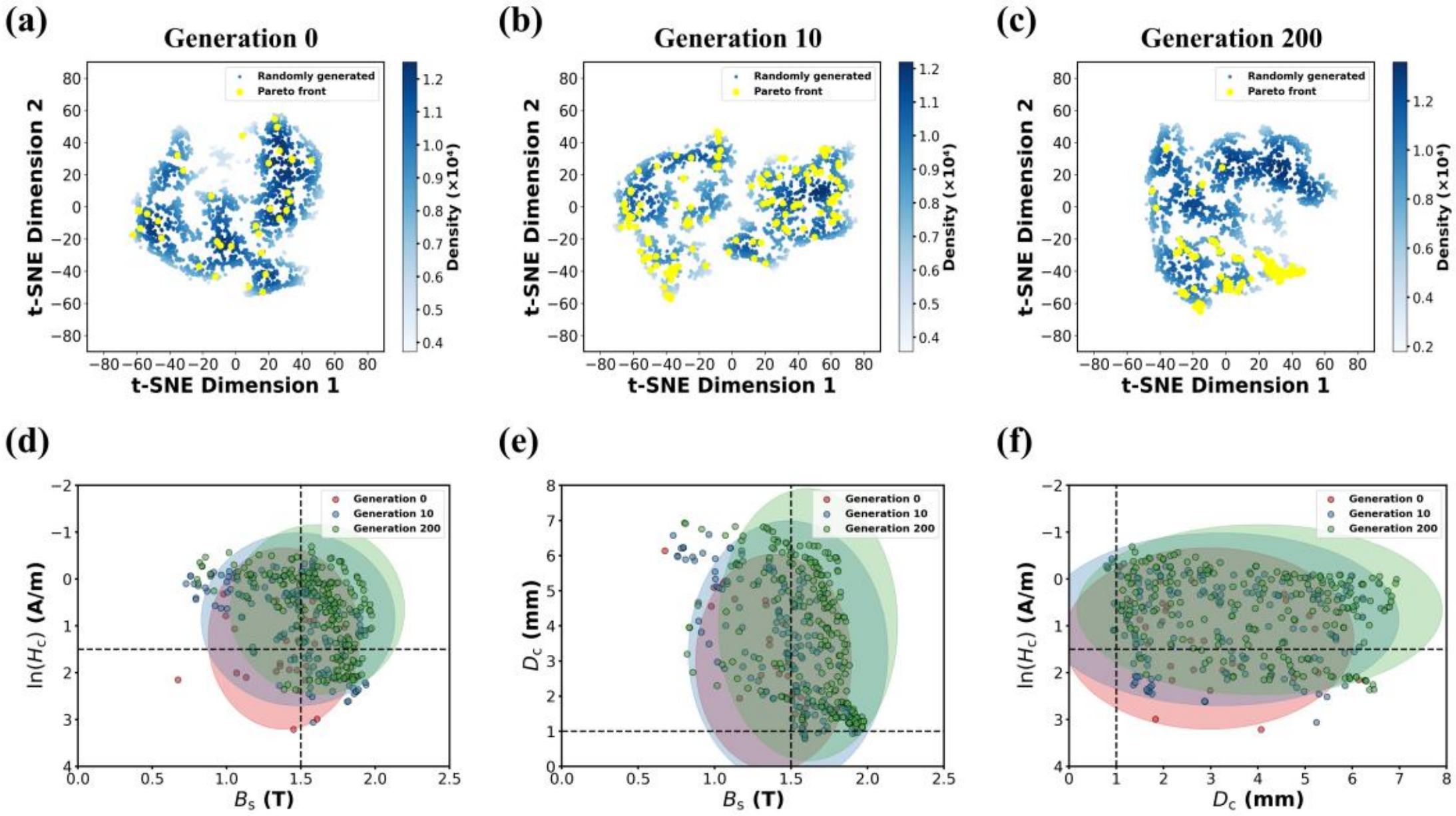

**Figure 7 Evolution of the Pareto front in the GMTDL framework**. **(a)** t-SNE representation of the 0th-generation (initial) Pareto front, highlighting the random distribution of the initial population. **(b)** t-SNE representation of the 10th-generation Pareto front, showing early progress in optimization. **(c)** t-SNE representation of the 200th-generation Pareto front, demonstrating a refined exploration of the design space. **(d–f)** Pairwise comparisons of $B_\mathrm{s}$~$\ln(H_\mathrm{c})$ **(d)**, $B_\mathrm{s}$~$D_\mathrm{c}$ **(e)**, and $D_\mathrm{c}$~$\ln(H_\mathrm{c})$ **(f)**, revealing how the solutions are progressively improved over iterations.

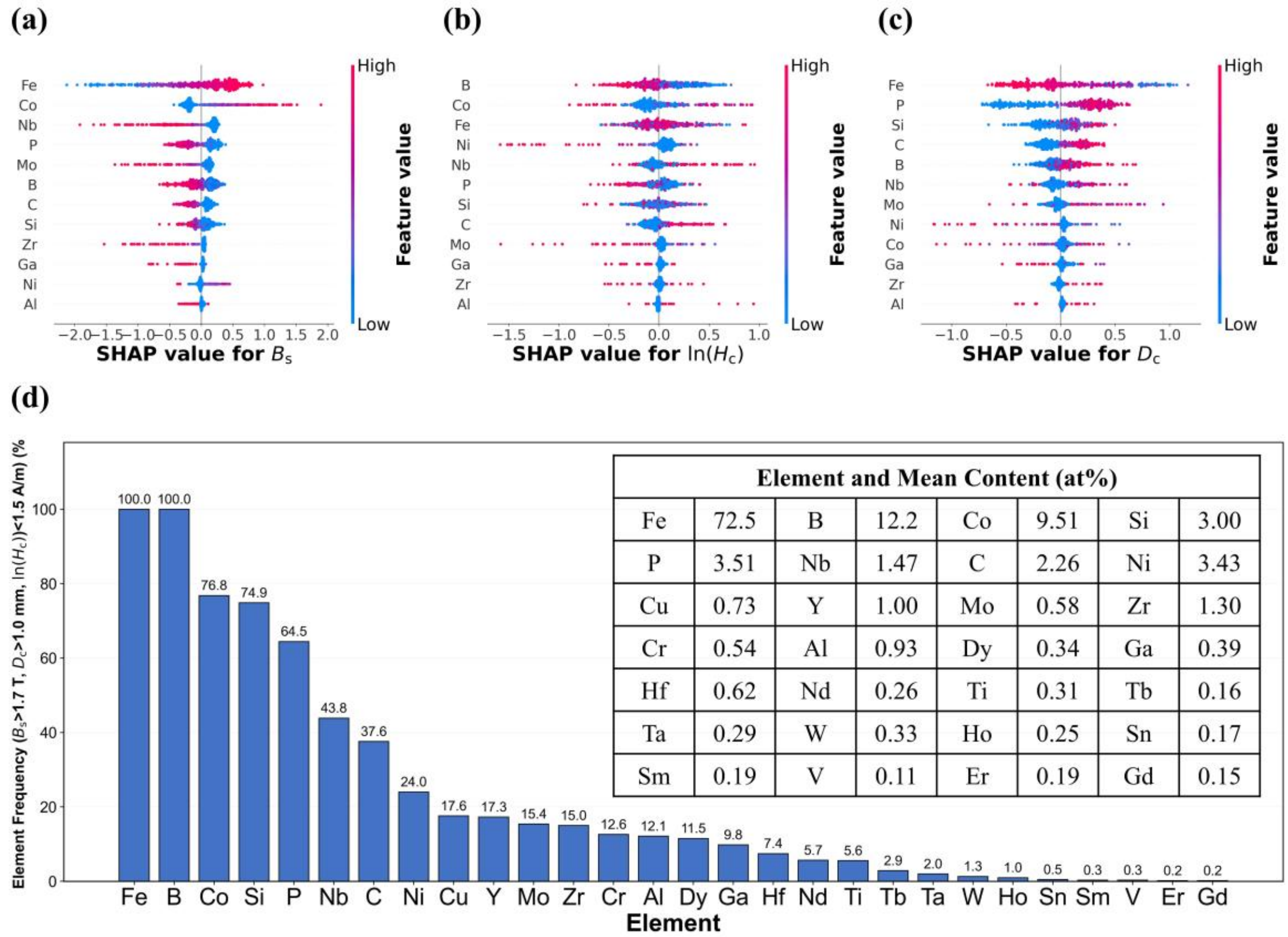

| Element and Mean Content (at%) | | | | | | | |
|---|---|---|---|---|---|---|---|
| Fe | 72.5 | B | 12.2 | Co | 9.51 | Si | 3.00 |
| P | 3.51 | Nb | 1.47 | C | 2.26 | Ni | 3.43 |
| Cu | 0.73 | Y | 1.00 | Mo | 0.58 | Zr | 1.30 |
| Cr | 0.54 | Al | 0.93 | Dy | 0.34 | Ga | 0.39 |
| Hf | 0.62 | Nd | 0.26 | Ti | 0.31 | Tb | 0.16 |
| Ta | 0.29 | W | 0.33 | Ho | 0.25 | Sn | 0.17 |
| Sm | 0.19 | V | 0.11 | Er | 0.19 | Gd | 0.15 |

**Figure 8 Quantitative insights into element-specific contributions to Fe-based amorphous alloy properties via SHAP analysis and frequency statistics.** SHAP values for $B_{\mathrm{s}}$ **(a)**, $H_{\mathrm{c}}$ **(b)**, and $D_{\mathrm{c}}$ **(c)**. The x-axis denotes the SHAP value for each element (positive values indicate a positive contribution, negative values indicate a negative contribution), and the y-axis ranks input features by their importance. Each point represents an individual sample from the dataset, with color indicating element concentration (red for higher concentration). **(d)** Statistical frequency analysis of elements within a subset of 7,500 MTWAE-generated compositions with $B_{\mathrm{s}}$>1.7T. The frequencies represent the fraction of compositions that additionally satisfy $D_{\mathrm{c}}$>1mm and ln($H_{\mathrm{c}}$)<1.5A/m. Bar height represents element frequency (%). The inset table shows the mean atomic concentrations (at.%) of each element.

**Table 1** Fe-based alloy compositions with the experimentally measured and predicted saturation magnetic flux density ($B_s$), predicted coercivity ($H_c$), and critical casting diameter ($D_c$) values. Compositions of 1-6 are Fe-based amorphous alloys, and compositions of 7-12 are Fe-based amorphous/nanocrystalline alloys. All the compositions of 1-12 are experimentally fabricated very recently. Compositions of 13-20 are predicted Fe-based amorphous alloys with ultra-high $B_s$ and ultra-low $H_c$ by our model.

| | Fe-based alloy compositions | experimental $B_s$ (T) | predicted $B_s$ (T) | predicted $\ln(H_c)/H_c$ (A/m) | predicted $D_c$ (mm) |
|---|---|---|---|---|---|
| 1 | $Fe_{68.2}Co_{17.5}B_{13}Si_{0.5}Cu_{0.8}$ [40] | 1.90 | 1.92 | 1.19/3.29 | 1.23 |
| 2 | $Fe_{79.7}Co_{6}B_{13}Si_{0.5}Cu_{0.8}$ [40] | 1.81 | 1.86 | 1.25/3.51 | 1.39 |
| 3 | $Fe_{85}B_{12}Si_{2}V_{0.5}Cu_{0.5}$ [16] | 1.58 | 1.70 | 2.57/13.07 | 1.31 |
| 4 | $Fe_{76.5}Co_{8.5}B_{12}Si_{2}V_{0.5}Cu_{0.5}$ [16] | 1.65 | 1.80 | 2.09/8.08 | 1.17 |
| 5 | $Fe_{68}Co_{17}B_{12}Si_{2}V_{0.5}Cu_{0.5}$ [16] | 1.84 | 1.87 | 1.79/6.03 | 1.38 |
| 6 | $Fe_{68.8}Co_{17.2}B_{11}Si_{2}V_{0.5}Cu_{0.5}$ [16] | 1.72 | 1.83 | 2.45/11.59 | 1.32 |
| 7 | $Fe_{70.11}Co_{15.39}Ni_{1.5}B_{9}P_{3}C_{1}$ [43] | 1.92 | 1.76 | 3.32/27.66 | 1.38 |
| 8 | $Fe_{69}Co_{16}Ni_{1}Si_{3}B_{11}$ [43] | 1.92 | 1.72 | 2.08/8.00 | 1.12 |
| 9 | $Fe_{64.13}Co_{21.38}Ni_{1.5}B_{8.5}P_{3}C_{1}V_{0.5}$ [43] | 1.90 | 1.74 | 2.80/16.44 | 1.28 |
| 10 | $Fe_{68.4}Co_{17.1}Ni_{1.5}B_{9}P_{3}C_{1}$ [43] | 1.90 | 1.75 | 3.36/28.79 | 1.37 |
| 11 | $Fe_{64.13}Co_{21.38}Ni_{1.5}B_{8.5}P_{3}C_{1}Mo_{0.5}$ [43] | 1.87 | 1.74 | 2.81/16.77 | 1.29 |
| 12 | $Fe_{64.13}Co_{21.38}Ni_{1.5}B_{9}P_{3}C_{1}$ [43] | 1.85 | 1.75 | 2.86/17.51 | 1.31 |
| 13 | $Fe_{78.57}Co_{7.49}B_{12.15}Si_{1.55}Cu_{0.11}$ | / | 1.91 | 0.48 | 1.48 |
| 14 | $Fe_{73.24}Co_{14.13}B_{11.29}Si_{1.18}$ | / | 1.96 | 0.74 | 1.27 |
| 15 | $Fe_{71.26}Co_{15.98}B_{11.35}Si_{1.22}Nb_{0.10}$ | / | 1.95 | 0.70 | 1.34 |
| 16 | $Fe_{65.76}Co_{19.43}B_{13.43}Si_{0.95}Nb_{0.33}$ | / | 1.98 | 0.75 | 1.13 |
| 17 | $Fe_{40.20}Co_{38.42}B_{19.75}Si_{0.19}Nb_{0.95}Y_{0.27}$ | / | 1.93 | 0.19 | 1.42 |
| 18 | $Fe_{67.99}Co_{0.29}B_{22.41}Si_{5.71}Nb_{1.35}Ni_{1.81}Cu_{0.13}Hf_{0.21}$ | / | 1.89 | 1.08 | 2.17 |
| 19 | $Fe_{83.61}B_{14.26}Si_{0.17}P_{0.15}Nb_{0.31}Ni_{1.19}Hf_{0.14}$ | / | 1.87 | 1.43 | 2.60 |
| 20 | $Fe_{63.27}Co_{2.23}B_{21.01}Si_{8.17}Nb_{4.73}Ni_{0.32}Cu_{0.14}$ | / | 1.85 | 0.98 | 2.45 |

**Supplementary Materials for**

# Multi-task deep-learning optimization of trade-off properties for superior-performance Fe-based soft magnetic alloys

Kang-Yuan Li(李康源)[1], Mao-Zhi Li(李茂枝)[1,*], Wei-Hua Wang(汪卫华)[2,3,4]

[1]*School of Physics and Key Laboratory of Quantum State Construction and Manipulation (Ministry of Education), Renmin University of China; Beijing, 100872, China*

[2]*Songshan Lake Materials Laboratory, Dongguan, Guangdong 523808, China*

[3]*Institute of Physics, Chinese Academy of Sciences, 100190 Beijing, China*

[4]*Center of Materials Science and Optoelectronics Engineering, University of Chinese Academy of Sciences, Beijing 100049, China*

*maozhili@ruc.edu.cn

## I. Data distribution analysis

We analyzed the distribution of $B_\mathrm{s}$, $H_\mathrm{c}$ and $D_\mathrm{c}$ values (see Supplementary Fig. S2). The $B_\mathrm{s}$ values range from 0.5T to 1.8T, with most values concentrated around 1.5T. Most $H_\mathrm{c}$ values are less than 10A/m, indicating a higher frequency in the lower value range, and the distribution decays rapidly with increasing $H_\mathrm{c}$. Such a skewed distribution may reduce the predictive accuracy of machine learning models. To improve the performance of our generative model, a log transformation was applied to the skewed data of $H_\mathrm{c}$ and a normal distribution is roughly conformed for $\ln(H_\mathrm{c})$. For $D_\mathrm{c}$, the values range from 1 mm to 7mm, with a main peak around 1mm. Each figure reflects a broad range of values, demonstrating the extensive variability and diversity of these datasets.

## II. The architecture and training of the multi-task Wasserstein autoencoder (MTWAE) model

The encoder network $q_\phi(Z|X)$ is designed to transform the input composition descriptor $X$ into a latent space representation $Z$. The latent space dimension k was scanned over $\{2, 4, 8, 16\}$. The encoding process involves mapping the composition data into a lower-dimensional latent space. This transformation is achieved through a multi-layer perceptron (MLP) comprising four linear layers with dimension of 90, 48, 30 and the latent space dimension, respectively. Each linear layer, except for the final latent space output layer, is followed by layer Normalization and a leaky rectified linear unit (Leaky ReLU) activation function with a negative slope of 0.01. Mathematically, the encoder can be expressed as:

$$Z = q_\phi(Z|X) = \mathrm{MLP}_\phi(X)$$

where $\phi$ denotes the parameters of the encoder network.

The decoder network $p_\theta(\hat{X}|Z)$ reconstructs the original composition descriptor $\hat{X}$ from the latent space representation $Z$. Essentially, it learns how to reverse the encoding process, ensuring that the model can regenerate the input from the learned latent space. Structurally, the decoder block has the same architecture as the encoder but is arranged in reverse order, consisting of four linear layers with dimension of 30, 48, 90, and input feature dimension. Similar to the encoder, except the output reconstruction layer, each linear layer is followed by layer normalization and Leaky ReLU activation. The output layer uses a Softmax activation function to ensure the validity of the reconstructed composition. The decoder is mathematically represented as:

$$\hat{X} = p_\theta(\hat{X}|Z) = \mathrm{Softmax}(\mathrm{MLP}_\theta(Z))$$

where $\theta$ denotes the parameters of the decoder network.

These predictor networks $f_{\omega_p}(y_p|Z)$ map the latent representation $Z$ to predicted values of the target properties $y_p$, where $p \in \{B_s, H_c, D_c\}$. Each property has a separate predictor that is tailored to predict the specific property based on the latent space. The predictors share the same latent space representation $Z$, allowing them to benefit from shared features while still being specialized to their respective property predictions. These predictors have the same architecture, consisting of three linear layers with 90 nodes in each one, followed by layer normalization and Leaky ReLU activation. The final layer outputs a single value corresponding to the predicted property. Formally, each predictor is defined as:

$$y_p = f_{\omega_p}(y_p|Z) = \mathrm{MLP}_{\omega_p}(Z)$$

where $\omega_p$ denotes the parameters of the predictor networks associated with property $y_p$. The MTWAE model integrates the encoder, decoder, and predictor networks within a unified framework. The architecture is intentionally streamlined to mitigate overfitting, focusing hyperparameter optimization on the number of training epochs (800), batch size (4), and learning rate ($10^{-3}$). These hyperparameters were selected through 30 iterations of cross-validation with random search, ensuring robust model performance.

The core objective of the MTWAE is to jointly optimize the encoder, decoder, and property predictors across multiple tasks, leveraging shared latent representations to facilitate mutual improvement. This optimization utilizes a comprehensive total loss function $L_{total}$, which aggregates the weighted task-specific losses:

$$L_{total} = \sum\nolimits_{task \in \{B_s, H_c, D_c\}} w_{task} L_{task}$$

where $w_{task}$ denotes the weight for each task to balance its contribution to the total loss. The loss $L_{task}$ for each task consists of three parts: a reconstruction loss $L_{recon}$, a property prediction loss $L_{pro}$, and a maximum mean discrepancy loss $L_{MMD}$:

$$L_{task} = L_{recon} + L_{pro} + L_{MMD}$$

where $L_{recon}$ is the cross-entropy loss, which makes the generated compositions as similar to the original ones as possible. $L_{pro}$ is defined as mean square error (MSE), which quantitatively measures the deviation between the predicted properties and the experimentally measured ones of the alloys. $L_{MMD}$ represents the maximum mean discrepancy (MMD) [1] between the latent space distribution $P_z$ and a Gaussian prior distribution $Q$, where MMD is a kernel-based statistical measure used to determine whether two probability distributions are identical by comparing their moments in a reproducing kernel Hilbert space (RKHS), $H$. Minimizing $L_{MMD}$ enable the latent space distribution to align more closely with the Gaussian prior, thereby increasing the diversity of the generated samples. Mathematically, $L_{MMD}$ is defined as:

$$L_{MMD}(P_z, Q) = \left\| E_{Z \sim p_z}[\varphi(Z)] - E_{\tilde{Z} \sim Q}\left[\varphi(\tilde{Z})\right] \right\|_H$$

where $\varphi$ is the feature mapping that projects the random variables $Z$ and $\tilde{Z}$ from distributions $P_z$ and $Q$ into the $H$, respectively, and $E$ denotes the expectation operator. Such a loss function can effectively regularize the encoder variance [2]. We used an inverse multiquadric kernel to measure the MMD, which is a common choice for such a task [3]. Here $\lambda_{MMD} = 10^{-4}$ was set to balance the effect of MMD relative to the reconstruction and prediction losses.

Given the significantly different dataset sizes across the $B_s$ (574 samples), $H_c$ (383 samples), and $D_c$ (311 samples) tasks, the larger dataset or the relatively easier task ($B_s$) may potentially dominate the model training. We mitigate this issue through a dual strategy comprising resampling and weight adjustment. Firstly, a resampling procedure was implemented, where smaller-sized datasets ($H_c$ and $D_c$) were repeated within each epoch to match the mini-batch number of the largest task ($B_s$). This approach equalizes the training frequency across tasks, mitigating the dominance of the largest-sample dataset. To further address task imbalance, three distinct weighting strategies were systematically explored: equal weighting, inverse sample size weighting, and uncertainty weighting. These strategies manifest distinct behaviors in convergence (see Supplementary Fig. S3). Equal weighting assigns the same weight ($w_{task} = 1$) across tasks, treating them equivalently irrespective of their dataset size or intrinsic complexity. As a consequence, the $B_s$ task—owing to its larger size—tends to disproportionately influence parameter updates, leading to suboptimal learning for the smaller-sized tasks ($H_c$ and $D_c$). Inverse sample size weighting was thus proposed to inversely scale weights based on dataset size, effectively boosting the impact of smaller tasks. The raw weights are computed via:

$$w_{task}^{raw} = \frac{T}{N_{task} \times K}$$

where $T$ is the total number of samples across all tasks, $N_{task}$ the number of samples for a specific task, and $K$ the number of tasks. These raw weights are then normalized so that the sum of weights for all tasks equals 1:

$$w_{task} = \frac{w_{task}^{raw}}{\sum_{task} w_{task}^{raw}}$$

This weighting method notably reduced dominance by the largest dataset ($B_s$) and improved the learning of smaller tasks ($H_c$ and $D_c$). The third strategy, uncertainty weighting, introduces learnable parameters $\sigma$, dynamically adjusting each task's weight based on its intrinsic difficulty (uncertainty) [4]:

$$w_{task}L_{task} = \left(1/(2\sigma^2)\right) \cdot \left(L_{recon} + L_{pro} + L_{MMD}\right) + \log(\sigma)$$

This adaptive mechanism down-weights tasks with higher predictive uncertainty, preventing noisy or difficult tasks from destabilizing training while allowing the model to allocate more capacity to better-behaved tasks.

Subsequent quantitative evaluations via five-fold cross-validation with the latent space dimension fixed at k=8 (see Supplementary Fig. S4) further validated these insights. Specifically, the inverse sample size weighting strategy consistently delivered superior predictive performance across tasks, significantly enhancing the prediction accuracy for the $D_{\mathrm{c}}$ task, moderately improving $H_{\mathrm{c}}$ performance, and slightly reducing $B_{\mathrm{s}}$ performance. In contrast, equal weighting exhibited strong performance on $B_{\mathrm{s}}$ but insufficient accuracy on $D_{\mathrm{c}}$, whereas uncertainty weighting performed comparably to equal weighting for the $B_{\mathrm{s}}$ task but underperformed relative to inverse sample size weighting on the $H_{\mathrm{c}}$ and $D_{\mathrm{c}}$ tasks. The combined mean $R^2$ across all tasks further highlights that inverse sample size weighting provides the most effective compromise, adeptly addressing both the imbalance in dataset sizes and variations in task complexity (see Supplementary Fig. S5).

Moreover, we systematically examined the impact of latent space dimensionality (k) on MTWAE performance (see supplementary Fig. S6). An increase of the latent dimensionality generally improves the composition reconstruction fidelity (lower CV-MAE) and property prediction accuracy (higher CV- R2) up to a certain threshold (k=8). As the dimensionality goes beyond k=16, performance deteriorated, particularly for property predictions of $D_{\mathrm{c}}$ and $H_{\mathrm{c}}$, indicative of overfitting due to limited data. Thus, an appropriate k value (k=8) can balance the model complexity with the risk of overfitting, achieving optimal reconstruction fidelity and prediction accuracy. Here we chose k=8 as the optimal latent dimensionality and trained the MTWAE model on the entire training dataset containing $B_{\mathrm{s}}$, $H_{\mathrm{c}}$ and $D_{\mathrm{c}}$ data. Supplementary Fig. S7 confirms this optimal choice, showcasing the robust predictive capabilities of the MTWAE model (k=8) across multiple critical properties of Fe-based metallic glasses. The predictive accuracy is demonstrated by high correlations between predicted and measured values for $B_{\mathrm{s}}$ ($R^2 = 0.88$), $\ln(H_{\mathrm{c}})$ ($R^2 = 0.49$), and $D_{\mathrm{c}}$ ($R^2 = 0.63$) on test datasets.

The PyTorch implementation of our model is available via GitHub (https://github.com/kangyuanli/Generative_multi_task_deep_learning_framework).

**III. Multi-objective evolutionary optimization.**

To realize efficient multi-objective evolutionary optimization within our inverse design framework, we utilized the Genetic and Evolutionary Algorithm Toolbox for Python with High Performance (GEATPY) [5]. In this work, we employed the Non-dominated Sorting Genetic Algorithm III (NSGA-III), an advanced variant of genetic algorithm for handing optimization problems involving multiple objectives. NSGA-III is well-suited for addressing the challenges posed by many-objective optimization, effectively maintaining diversity in solutions and uniformly approximating complex Pareto fronts by leveraging a reference-point-based approach.

The fundamental optimization procedure of NSGA-III operates as follows: A population comprising a predefined number of candidate solutions (individuals) undergoes iterative processes of selection, crossover, and mutation to progressively refine the solutions. Initially, latent alloy representations are randomly generated from a multivariate Gaussian distribution informed by the latent space distribution learned by the MTWAE encoder. Each individual within the population is assigned a fitness score based on its performance across the targeted objectives. Specifically, these fitness values are derived from predictions obtained through the pre-trained MTWAE predictor networks for $B_s$, $H_c$, and $D_c$.

The NSGA-III algorithm then employs a non-dominated sorting mechanism to rank the individuals based on Pareto dominance criteria, systematically identifying Pareto-optimal solutions-those for which no other solution can achieve superior performance across all objectives simultaneously. This sorting facilitates the preservation of high-quality solutions exhibiting diverse trade-offs among the targeted properties. Subsequently, solutions from superior Pareto fronts are preferentially selected to undergo genetic operators-namely, simulated crossover and mutation-to generate new candidate solutions for the next generation. The iterative nature of this procedure ensures that the population progressively converges towards optimal regions of the multi-dimensional objective space.

The evolutionary optimization continues iteratively until meeting predefined termination criteria, such as reaching a maximum number of generations, or convergence to a specified performance threshold. This structured optimization process is essential for effectively balancing competing property requirements, specifically high $B_s$, low $H_c$, and high $D_c$, ultimately generating latent representations that correspond to superior Fe-based metallic glass compositions. Upon convergence, the optimal latent representations identified via NSGA-III are decoded into explicit alloy compositions using the trained MTWAE decoder network.

**IV. Physical descriptors.**

For each alloy composition expressed by atomic fractions $c_i (\sum_i c_i = 1)$, we compute physically motivated descriptors used in Fig. 5, including the atomic-size mismatch $\delta$, mixing enthalpy $\Delta H_{mix}$, and a theoretical saturation magnetic flux density $B_s^{cal}$. These descriptors are used to assess whether the discovered candidates remain within empirically plausible thermodynamic–structural regimes.

**Atomic-size mismatch.** The average metallic radius is $\bar{r} = \sum_i c_i r_i$, where $r_i$ is the metallic radius of element $i$. The atomic-size mismatch is defined as

$$\delta(\%) = 100 \times \sqrt{\sum_{i=1}^{n} c_i \left(1 - \frac{r_i}{\bar{r}}\right)^2}.$$

**Mixing enthalpy.** The mixing enthalpy is computed by a Miedema-type pairwise approximation:

$$\Delta H_{mix} = \sum_{i<j} 4\, c_i c_j \Delta H_{ij}^{mix},$$

where $\Delta H_{ij}^{mix}$ is the equiatomic binary mixing enthalpy for element pair $(i, j)$ taken from a Miedema-type compilation [6].

**Theoretical saturation magnetic flux density $B_s^{cal}$.** According to theories in magnetism, $B_s$ is proportional to the mean magnetic moment of all the atom in an alloy [8]. Therefore, a theoretical saturation magnetic flux density can be estimated by

$$B_s^{cal} = \frac{N_A\, \bar{\mu}\, \mu_B}{V_m},$$

where $N_A$ is the Avogadro constant, $\mu_B$ is the Bohr magneton, and $V_m$ is the theoretical molar volume estimated by a rule-of-mixtures:

$$V_m = \sum_i c_i \frac{M_i}{\rho_i},$$

with $M_i$ and $\rho_i$ being the atomic weight and elemental density, respectively. $\bar{\mu}$ is the mean atomic magnetic moment and can be estimated by

$$\bar{\mu} = \sum_i c_i\, m_i,$$

where $m_i$ is the elemental saturation magnetic moment in $\mu_B/\text{atom}$. We set $m_{\text{Fe}} = 2.22$, $m_{\text{Co}} = 1.72$, and $m_{\text{Ni}} = 0.60\ \mu_B/\text{atom}$, respectively, which are consistent with commonly reported saturation moments for $3d$ ferromagnets and the Slater–Pauling context [7]. For non-ferromagnetic elements, we use $m_i = 0$. It can be seen that $B_s^{cal}$ depends only on the compositions.

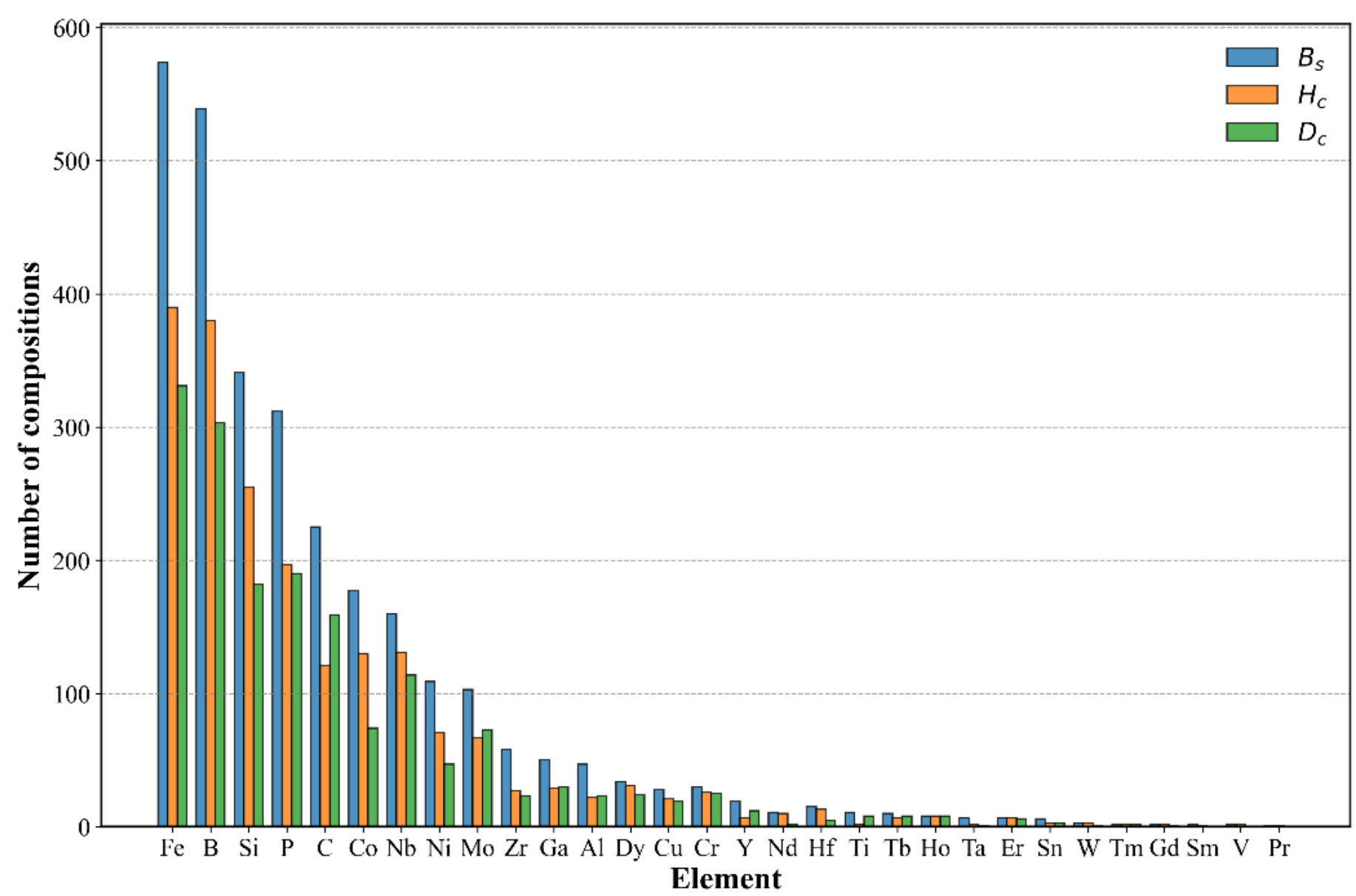

**Figure S1** Number of elements' occurrence in the datasets for predicting $B_\mathrm{s}$, $H_\mathrm{c}$ and $D_\mathrm{c}$, respectively.

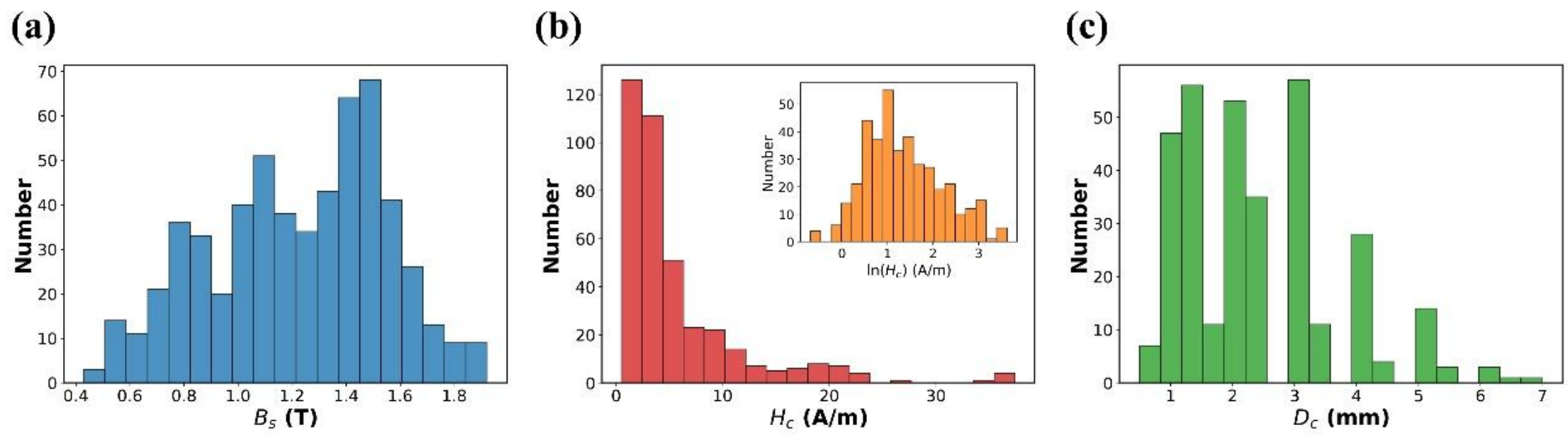

**Figure S2** Distribution of magnetic properties of $B_\mathrm{s}$ (a) and $H_\mathrm{c}$ (b), and critical diameter ($D_\mathrm{c}$) (c) in the dataset.

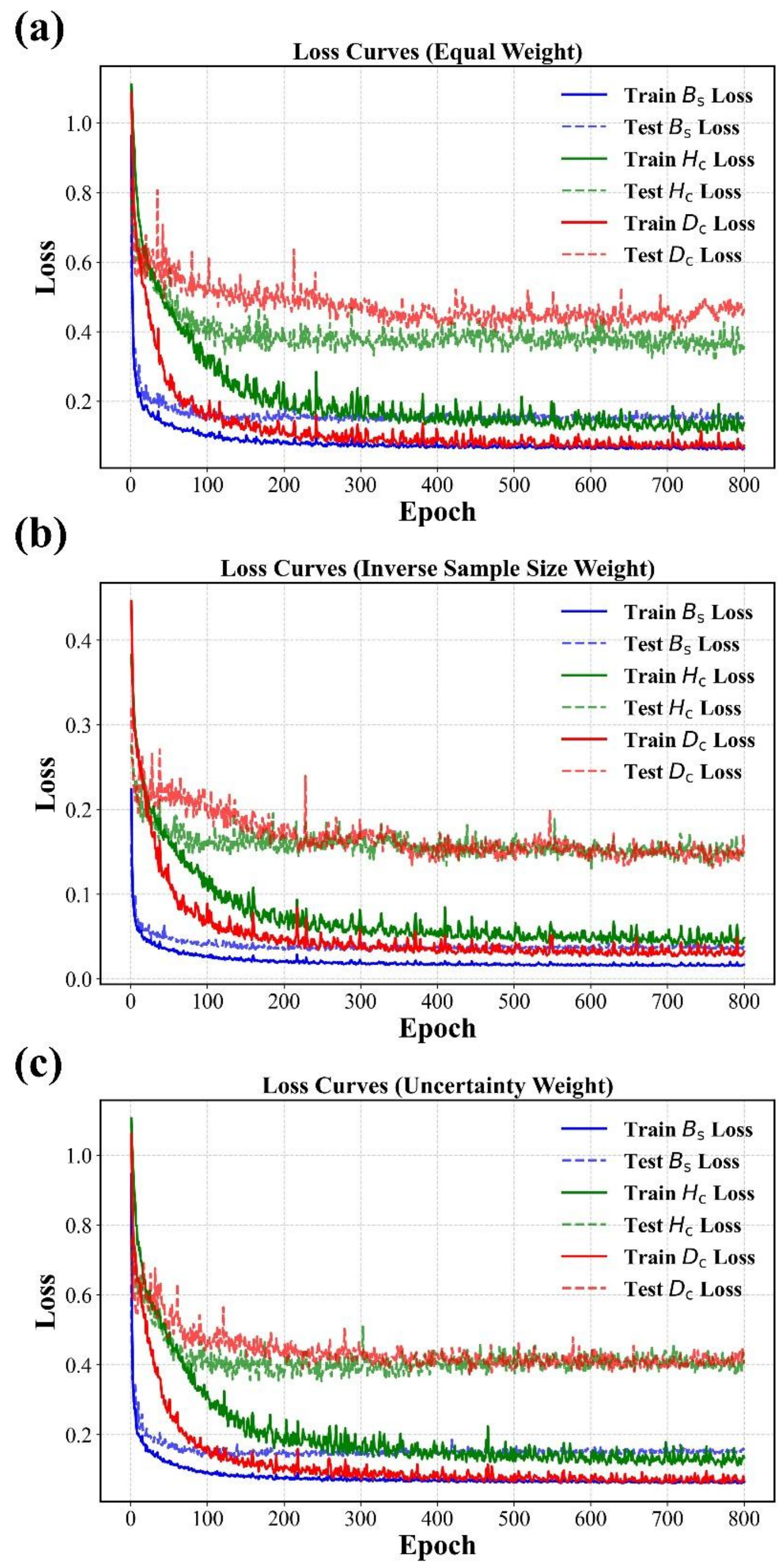

**Figure S3** The training curves of MTWAE under three different weight adjustment strategies (equal weight, inverse sample size weight, uncertainty weight). Each subplot shows the evolution of the training and testing losses for $B_\mathrm{s}$, $H_\mathrm{c}$, and $D_\mathrm{c}$ over 800 epochs. These curves illustrate how each weighting method affects the learning dynamics and the final model performance in the three tasks.

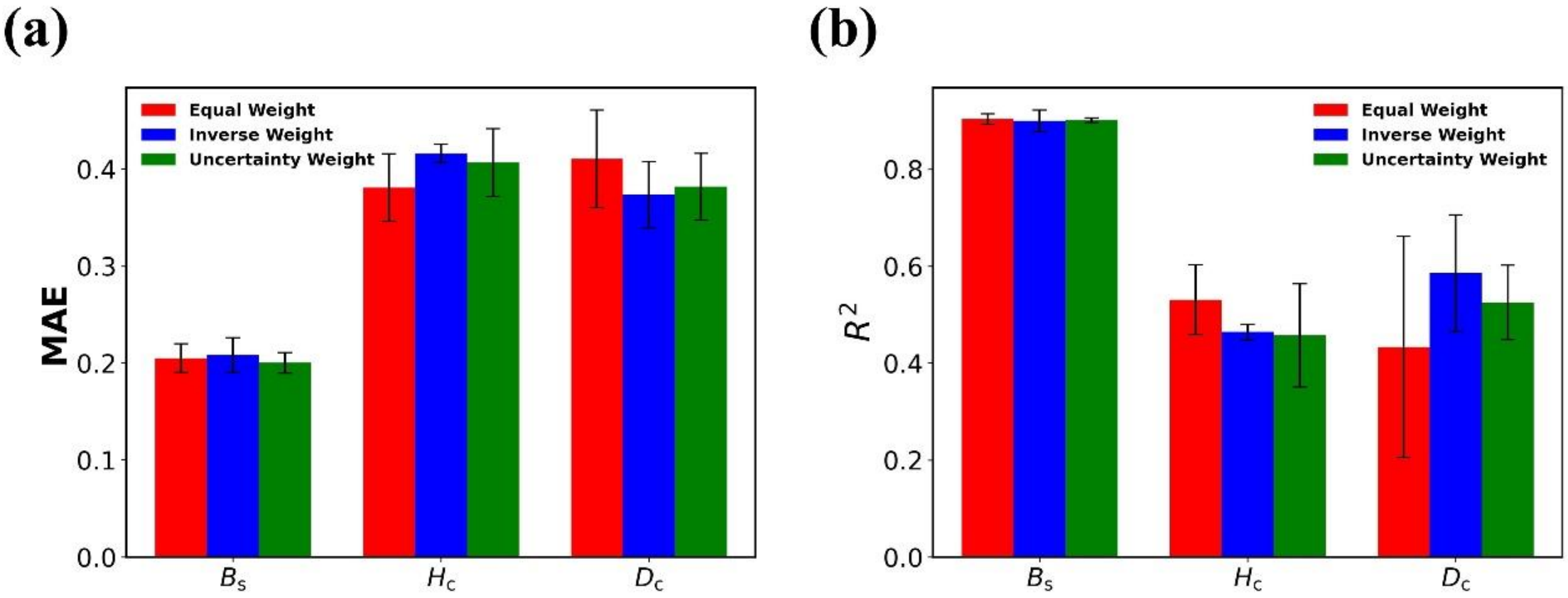

**Figure S4** Performance of the three weight adjustment strategies on the $B_{\mathrm{s}}$, $H_{\mathrm{c}}$, and $D_{\mathrm{c}}$ tasks under 5-fold cross-validation at the latent space dimension k=8 (error bars denote the standard deviation across folds). (a) MAE: a smaller mean absolute error indicates less prediction deviation. (b) $R^2$: a coefficient of determination closer to 1 indicates better model fit.

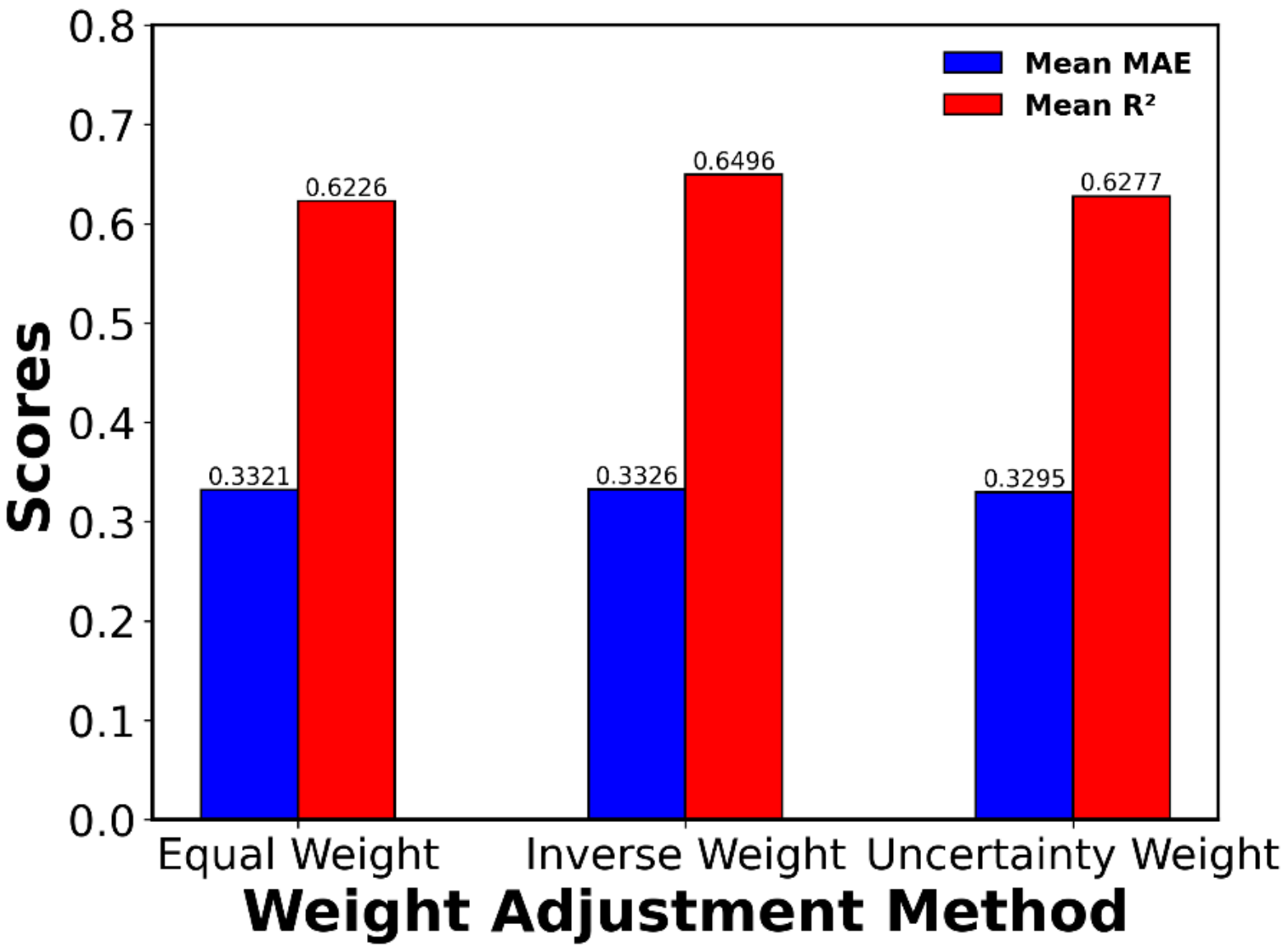

**Figure S5** Overall performance comparison of the three weight adjustment strategies—no weighting, inverse weighting, and uncertainty weighting—in terms of Mean-MAE and Mean-$R^2$. Specifically, the 5-fold cross-validation results for each of the three tasks ($B_s$, $H_c$, and $D_c$) presented in Figure S4 were averaged within each task, and then calculated the average of all tasks to evaluate the overall performance.

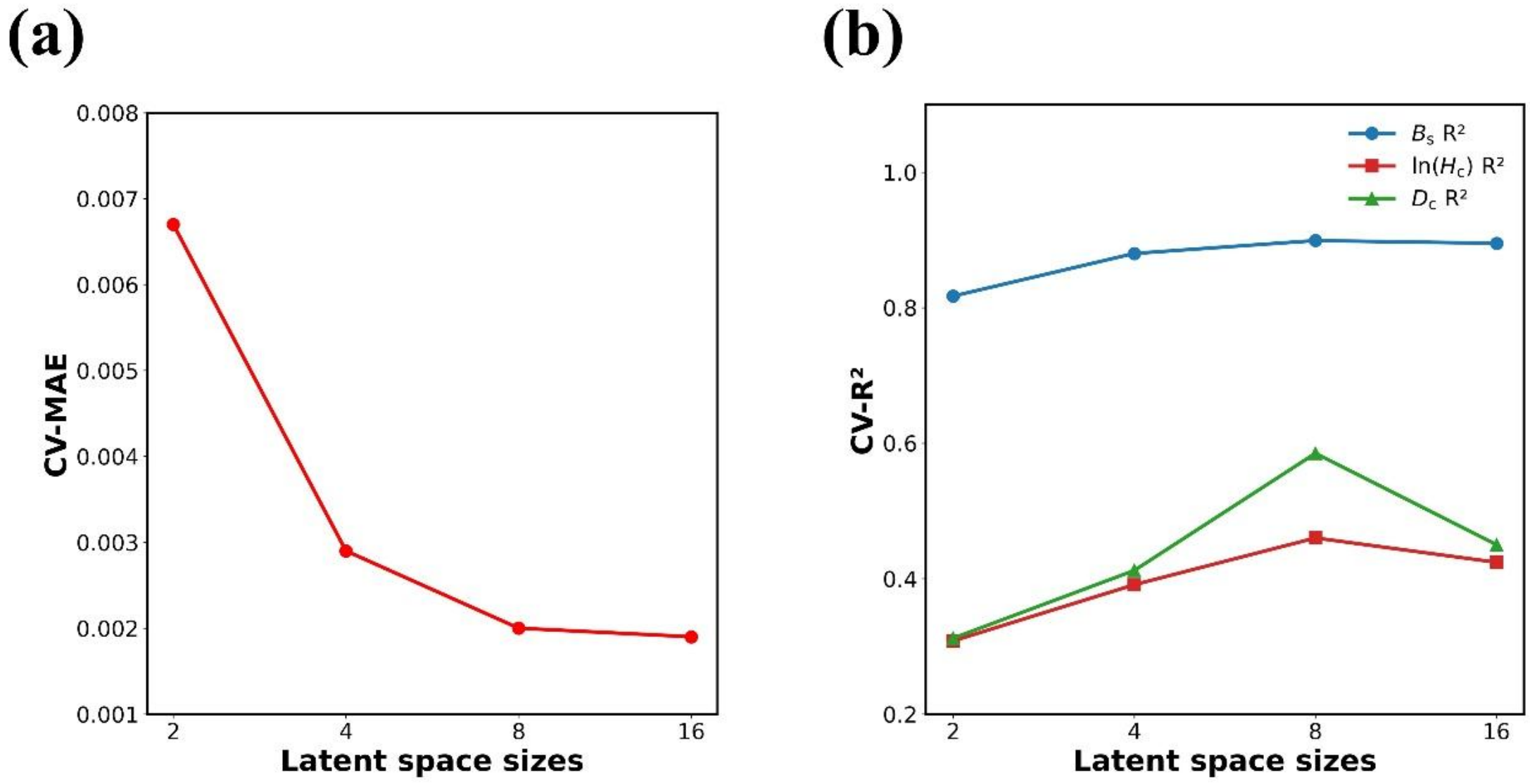

**Figure S6 Effect of latent space dimension ($k$) on the performance of the MTWAE model (five-fold cross-validation).** (a) Relationship between the latent space dimension ($k$) and the mean absolute error (MAE), illustrating the improvement in reconstruction accuracy as $k$ increase. (b) Relationship between the latent space dimension ($k$) and the coefficient of determination ($R^2$), demonstrating that larger $k$ enhances property prediction accuracy; overfitting occurs when $k$ becomes excessively large, leading to a decline in prediction performance.

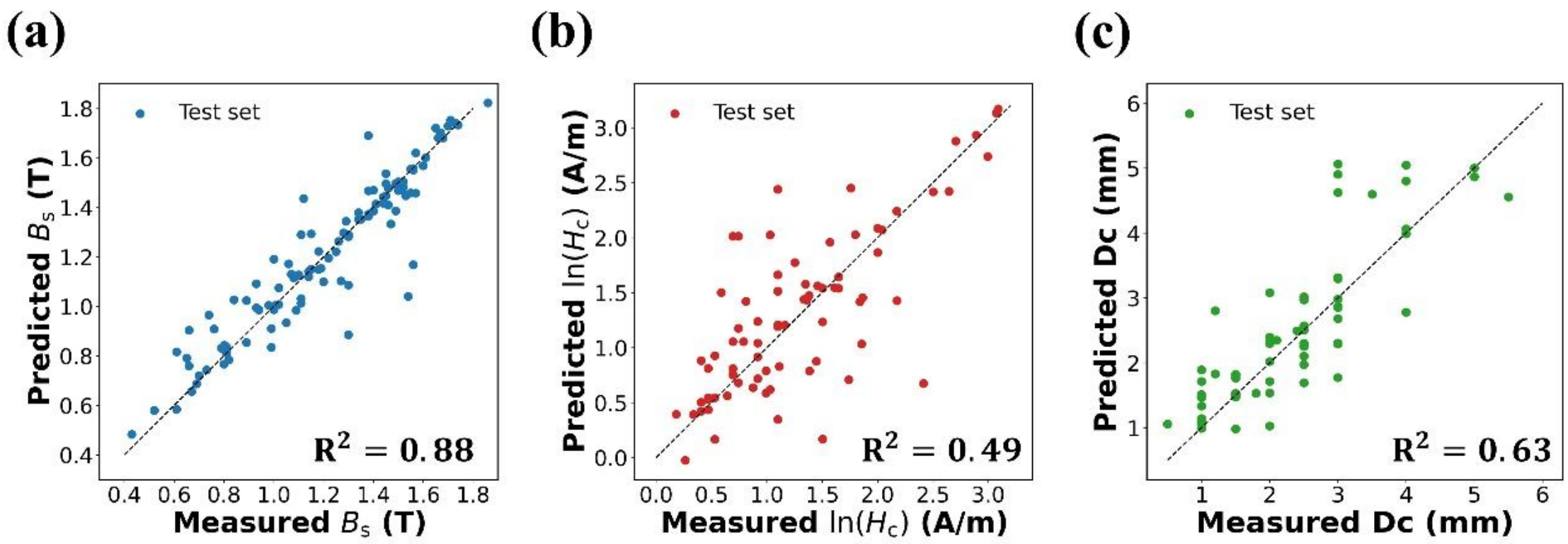

**Figure S7 Performance of the MTWAE model in predicting key properties of Fe-based metallic glasses.** (a) Predicted versus measured saturation magnetic flux density ($B_{\mathrm{s}}$), (b) Predicted versus measured coercivity ($\ln(H_{\mathrm{c}})$), and (c) predicted versus measured critical casting diameter ($D_{\mathrm{c}}$) for testing datasets. The dashed line indicates perfect correlation (y=x). The coefficient of determination ($R^2$) is shown for each property on the testing set ($R^2 = 0.88$ for $B_{\mathrm{s}}$, $R^2 = 0.49$ for $\ln(H_{\mathrm{c}})$ and $R^2 = 0.63$), highlighting the model's ability to accurately predict multiple target properties.

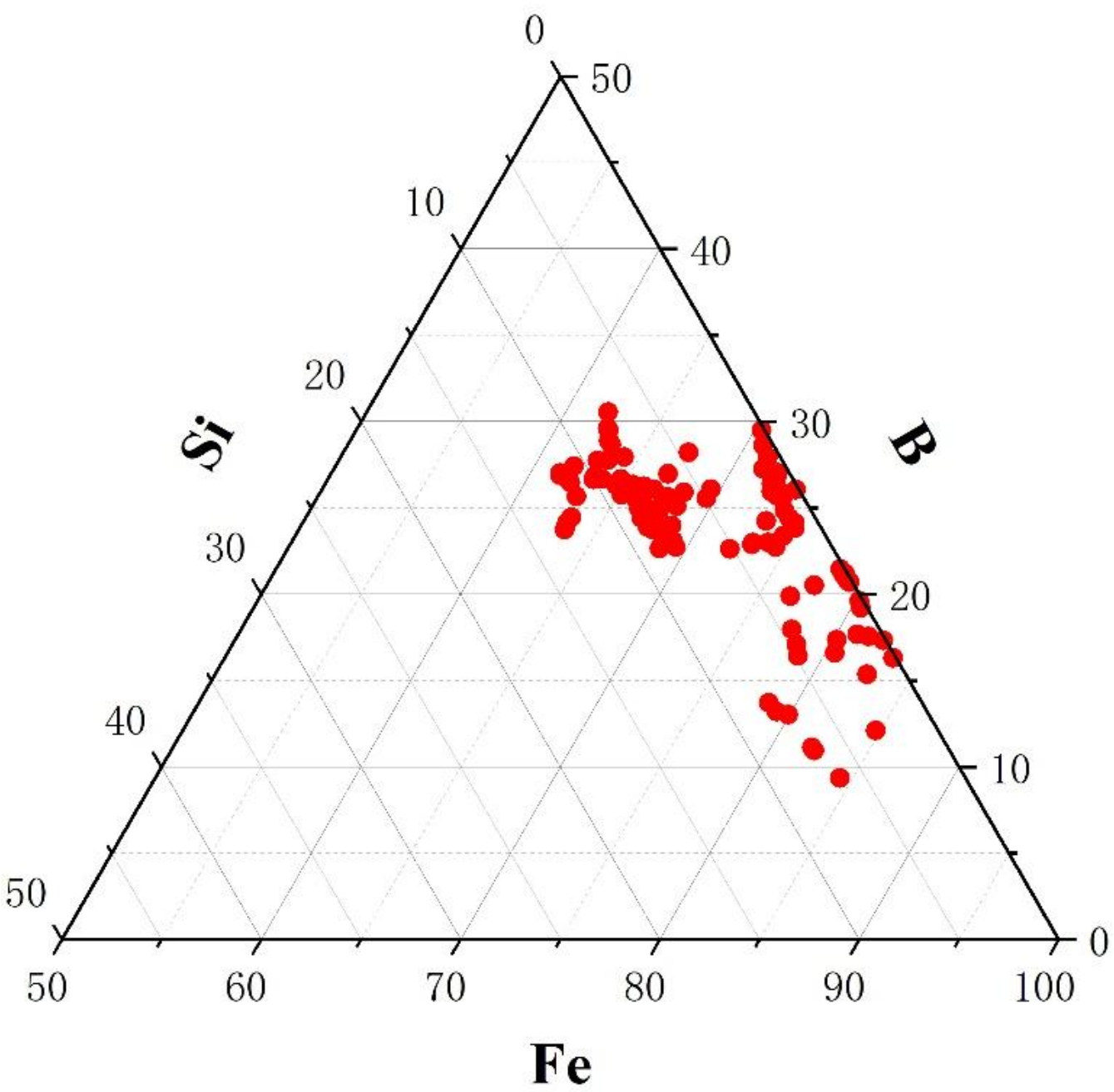

**Figure S8 Compositional diversity of Fe-based candidate alloys.** Ternary Fe–B–Si plot of selected candidate alloys derived from our multi-objective generative design framework. Each point represents a candidate alloy's relative content of Fe, B, and Si, clearly illustrating the substantial compositional diversity among alloys with similarly optimized target properties. Although these alloys include additional elements beyond Fe, B, and Si, this ternary representation effectively highlights their compositional variations and distribution across the design space.

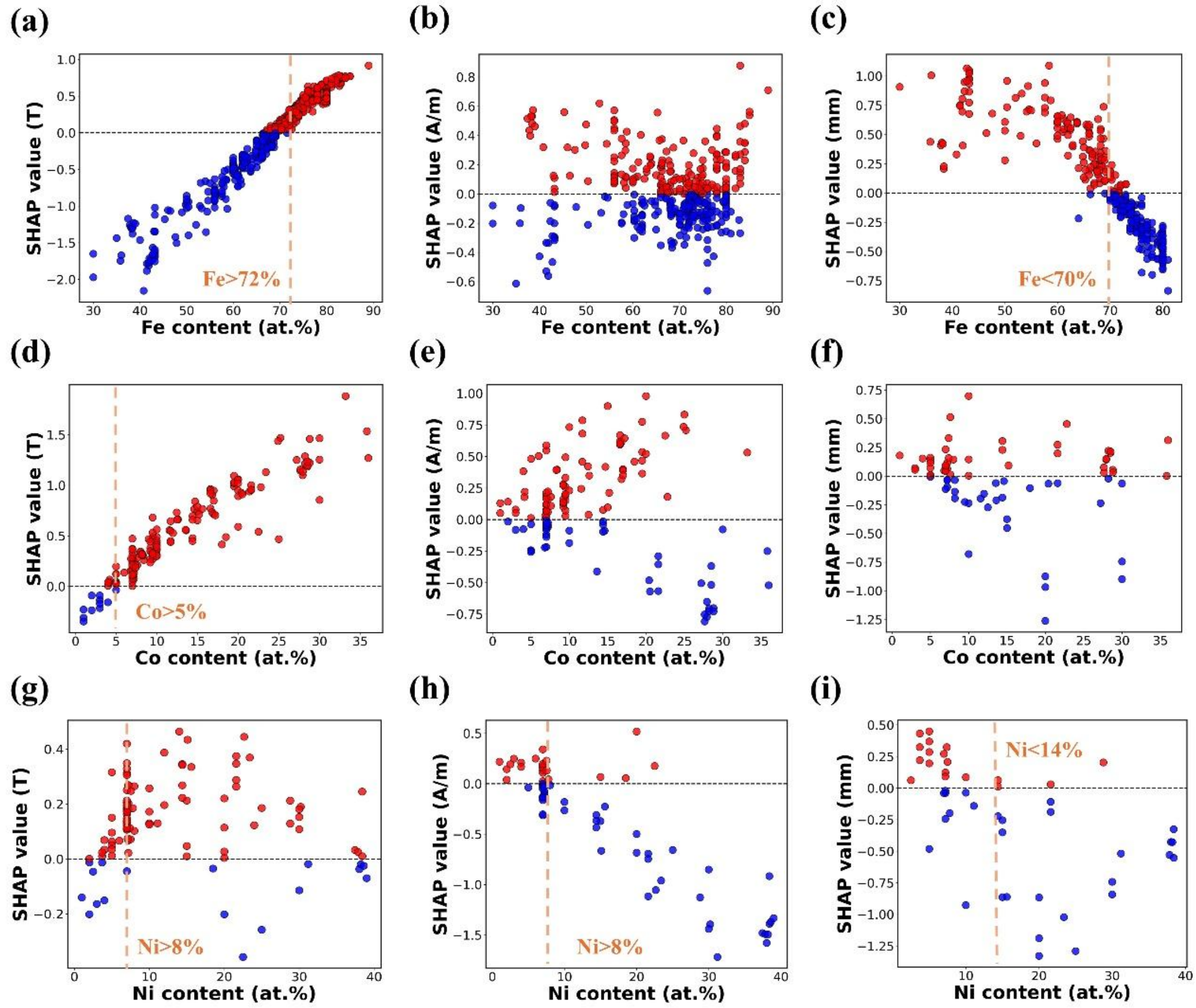

**Figure S9** SHAP value as a function of content of Fe (a-c), Co (d-f), and Ni (g-i) for saturation magnetic flux density $B_\mathrm{s}$, coercivity $\ln(H_\mathrm{c})$, and critical casting diameter $D_\mathrm{c}$, respectively. Positive SHAP values (red) indicate that increasing the elemental content enhances the corresponding property, whereas negative values (blue) indicate a detrimental effect. The vertical dashed lines mark key compositional thresholds discussed in the main text, including Fe > 72 at.% ($B_\mathrm{s}$ gain) and < 70 at.% ($D_\mathrm{c}$ gain), Co > 5 at.% ($B_\mathrm{s}$ gain), and Ni > 8 at.% (simultaneous $B_\mathrm{s}$ enhancement and $H_\mathrm{c}$ reduction).

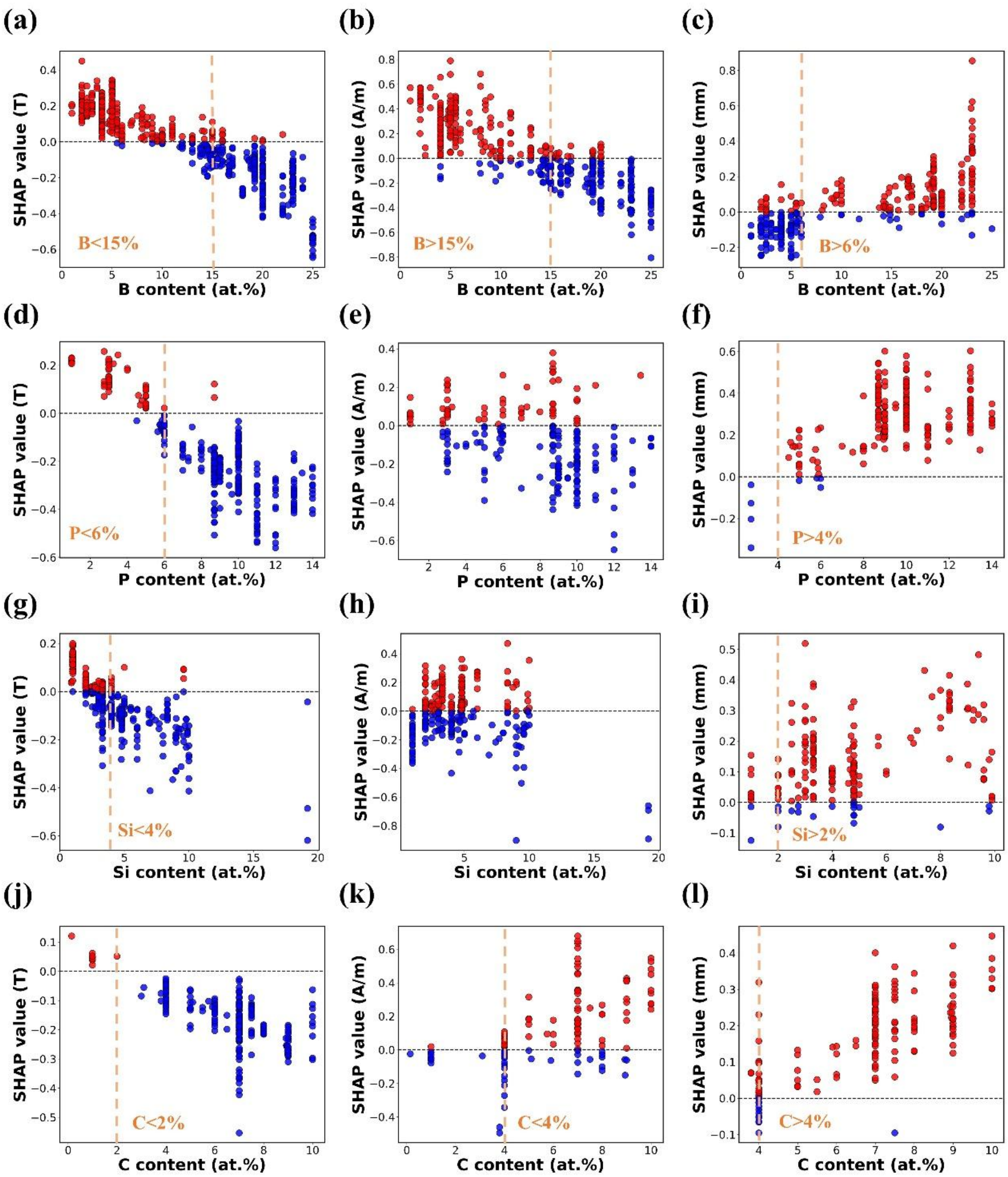

**Figure S10** SHAP value as a function of content of B (a-c), P (d-f), Si (g-i), and C (j-l) for saturation magnetic flux density $B_\mathrm{s}$, coercivity $\ln(H_\mathrm{c})$, and critical casting diameter $D_\mathrm{c}$, respectively. Red (blue) points represent positive (negative) SHAP contributions. The vertical dashed lines indicate the critical compositional thresholds where the SHAP sign or trend changes: B ≈ 15 at.% for $B_\mathrm{s}$ and $\ln(H_\mathrm{c})$ and B≈ 6 at.% for $D_\mathrm{c}$, P ≈ 6 at.% for $B_\mathrm{s}$ and ≈ 4 at.% for $D_\mathrm{c}$, Si ≈ 4 at.% for $B_\mathrm{s}$ and ≈ 2 at.% for $D_\mathrm{c}$, and C ≈ 2 and 4 at.% marking the onset of $B_\mathrm{s}$ penalty and $H_\mathrm{c}/D_\mathrm{c}$ benefits, respectively.

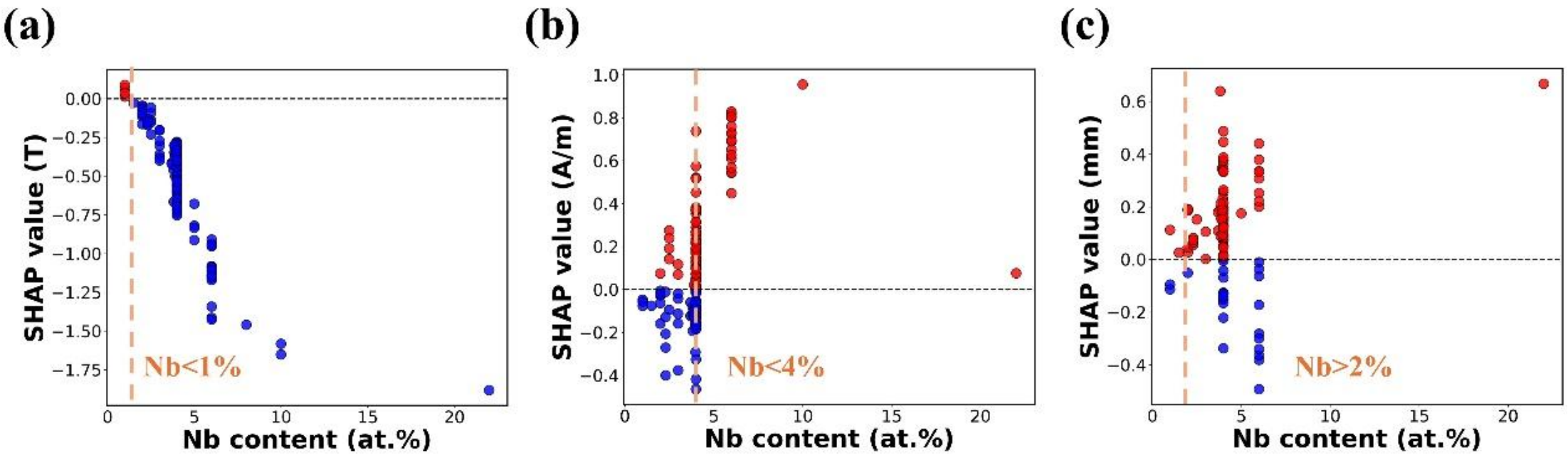

**Figure S11** SHAP value as a function of Nb content for saturation magnetic flux density $B_s$ (a), coercivity $\ln(H_c)$ (b), and critical casting diameter $D_c$ (c), respectively. Vertical dashed lines highlight the practical design windows inferred from these trends: Nb<1 at.% for $B_s$-prioritized designs and Nb≈2–4 at.% for $H_c$/GFA-prioritized compositions with acceptable $B_s$. Even small Nb additions strongly reduce $B_s$, whereas moderate Nb content decreases $\ln(H_c)$ and enhances $D_c$.

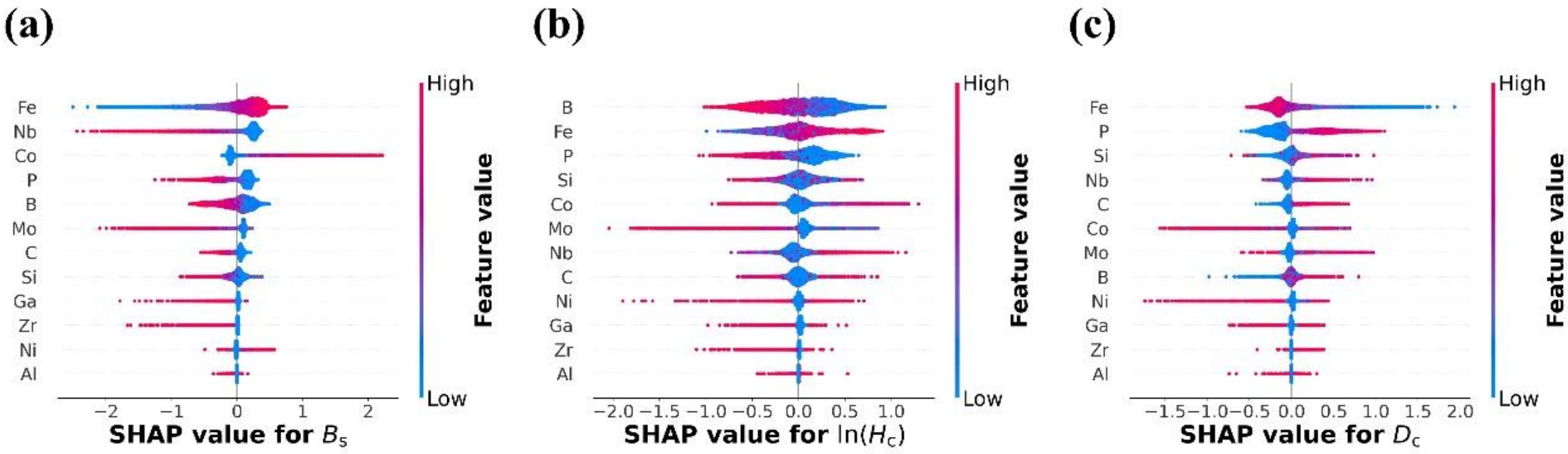

**Figure S12. SHAP value analysis of 10,000 virtual alloy compositions generated by MTWAE.** (a) SHAP values for $B_s$, (b) SHAP values for $H_c$, and (c) SHAP values for $D_c$. The x-axis denotes the SHAP value for each element (positive values indicate a positive contribution, negative values indicate a negative contribution), and the y-axis ranks input features by their importance. Each point represents an individual sample from the expanded virtual dataset, with color indicating element concentration (red for higher concentration, blue for lower concentration). This analysis validates the robustness of elemental trends identified in the original experimental dataset.

**Table S1** 574 Fe-based MG compositions and their $B_s$ values collected from experimental measurements in literature.

| | Amorphous Alloy | $B_s$ (T) | Reference |
|---|---|---|---|
| 1. | Fe79P10C4B4Si3 | 1.53 | [1] |
| 2. | Fe78Mo1P10C4B4Si3 | 1.44 | |
| 3. | Fe77Mo2P10C4B4Si3 | 1.39 | |
| 4. | Fe76Mo3P10C4B4Si3 | 1.32 | |
| 5. | Fe75Mo4P10C4B4Si3 | 1.27 | |
| 6. | Fe74Mo5P10C4B4Si3 | 1.14 | |
| 7. | Fe76C7.0Si3.3B5P8.7 | 1.52 | [2] |
| 8. | Fe75C7.0Si3.3B5P8.7Mo1 | 1.41 | |
| 9. | Fe73C7.0Si3.3B5P8.7Mo3 | 1.3 | |
| 10. | Fe71C7.0Si3.3B5P8.7Mo5 | 1.1 | |
| 11. | (Fe0.76Si0.096B0.084P0.06)100Mo0 | 1.51 | [3] |
| 12. | (Fe0.76Si0.096B0.084P0.06)98Mo2 | 1.35 | |
| 13. | (Fe0.76Si0.096B0.084P0.06)96Mo4 | 1.1 | |
| 14. | (Fe0.76Si0.096B0.084P0.06)94Mo6 | 0.98 | |
| 15. | Fe75B15Si10 | 1.56 | [4] |
| 16. | (Fe0.75B0.15Si0.10)99Zr1 | 1.53 | |
| 17. | (Fe75B15Si10)99Nb1 | 1.5 | |
| 18. | (Fe75B15Si10)98Nb2 | 1.47 | |
| 19. | (Fe75B15Si10)96Nb4 | 1.4 | |
| 20. | Fe77Ga3P9.5C4B4Si2.5 | 1.36 | [5] |
| 21. | Fe78Ga2P9.5C4B4Si2.5 | 1.4 | |
| 22. | (Fe0.74Tb0.01B0.2Si0.05)96Nb4 | 1.14 | [6] |
| 23. | (Fe0.73Tb0.02B0.2Si0.05)96Nb4 | 1.01 | |
| 24. | (Fe0.72Tb0.03B0.2Si0.05)96Nb4 | 0.92 | |
| 25. | (Fe0.71Tb0.04B0.2Si0.05)96Nb4 | 0.76 | |
| 26. | (Fe0.70Tb0.05B0.2Si0.05)96Nb4 | 0.69 | |
| 27. | (Fe0.69Tb0.06B0.2Si0.05)96Nb4 | 0.61 | |
| 28. | (Fe0.68Tb0.07B0.2Si0.05)96Nb4 | 0.52 | |
| 29. | [(Fe0.9Co0.1)0.75B0.2Si0.05]96Nb4 | 1.13 | [7] |
| 30. | [(Fe0.8Co0.2)0.75B0.2Si0.05]96Nb4 | 1.05 | |
| 31. | [(Fe0.7Co0.3)0.75B0.2Si0.05]96Nb4 | 0.98 | |
| 32. | [(Fe0.6Co0.4)0.75B0.2Si0.05]96Nb4 | 0.93 | |
| 33. | [(Fe0.5Co0.5)0.75B0.2Si0.05]96Nb4 | 0.84 | |
| 34. | Fe76Mo3.5P10C4B4Si2.5 | 1.21 | [8] |
| 35. | Fe71Ni5Mo3.5P10C4B4Si2.5 | 1.17 | |
| 36. | Fe66Ni10Mo3.5P10C4B4Si2.5 | 1.12 | |
| 37. | Fe61Ni15Mo3.5P10C4B4Si2.5 | 1.03 | |
| 38. | Fe56Ni20Mo3.5P10C4B4Si2.5 | 0.93 | |

| 39. | Fe46Ni30Mo3.5P10C4B4Si2.5 | 0.75 | |
|---|---|---|---|
| 40. | (Fe0.75B0.15Si0.1)96Nb4 | 1.2 | [9] |
| 41. | [(Fe0.8Co0.1Ni0.1)0.75B0.2Si0.05]96Nb4 | 1.1 | |
| 42. | [(Fe0.6Co0.1Ni0.3)0.75B0.2Si0.05]96Nb4 | 0.8 | |
| 43. | [(Fe0.6Co0.2Ni0.2)0.75B0.2Si0.05]96Nb4 | 0.86 | |
| 44. | [(Fe0.6Co0.3Ni0.1)0.75B0.2Si0.05]96Nb4 | 0.9 | |
| 45. | Fe76Mo2Ga2P10C4B4Si2 | 1.32 | [10] |
| 46. | Fe74Mo4Ga2P10C4B4Si2 | 1.16 | |
| 47. | Fe75Mo2Ga3P10C4B4Si2 | 1.27 | |
| 48. | Fe73Mo4Ga3P10C4B4Si2 | 1.11 | |
| 49. | (Fe0.75B0.15Si0.10)99Nb1 | 1.5 | [11] |
| 50. | (Fe0.75B0.15Si0.10)98Nb2 | 1.49 | |
| 51. | (Fe0.75B0.15Si0.10)96Nb4 | 1.47 | |
| 52. | (Fe0.775B0.125Si0.10)98Nb2 | 1.51 | |
| 53. | Fe76Si9B10P5 | 1.51 | [12] |
| 54. | Fe73Al5Ga2P11C5Si4 | 1.29 | |
| 55. | Fe72Al5Ga2P10C6B4Si1 | 1.14 | |
| 56. | Fe30Co30Ni15Si8B17 | 0.92 | |
| 57. | Fe74Nb6Y3B17 | 0.81 | |
| 58. | Fe74Cr2Mo2Ga2P10C4B4Si2 | 1.11 | [13] |
| 59. | Fe72Cr4Mo2Ga2P10C4B4Si2 | 0.96 | |
| 60. | Fe70Cr6Mo2Ga2P10C4B4Si2 | 0.84 | |
| 61. | Fe73Al5Ga2P11C5B4 | 1.07 | [14] |
| 62. | Fe72Al5Ga2P11C6B4 | 1.04 | |
| 63. | Fe72Al5Ga2P11C5B4Si1 | 1.06 | |
| 64. | Fe71Al5Ga2P11C6B4Si1 | 1.02 | |
| 65. | Fe70Al5Ga2P11C6B4Si2 | 0.93 | |
| 66. | Fe69Al5Ga2P11C6B4Si3 | 0.95 | |
| 67. | Fe68Al5Ga2P11C6B4Si4 | 0.89 | |
| 68. | Fe67Al5Ga2P11C6B4Si5 | 0.81 | |
| 69. | Fe66Al5Ga2P11C6B4Si6 | 0.82 | |
| 70. | Fe65Al5Ga2P11C6B4Si7 | 0.67 | |
| 71. | Fe72Al5Ga2P9C6B4Si2 | 1.1 | |
| 72. | Fe72Al5Ga2P8C6B4Si3 | 1.1 | |
| 73. | Fe72Al5Ga2P7C6B4Si4 | 1.13 | |
| 74. | Fe72Al4Ga2P11C6B4Si1 | 1.06 | |
| 75. | Fe72Al3Ga2P11C6B4Si2 | 1.09 | |
| 76. | Fe72Al5Ga2P11C4B4Si2 | 1.07 | |
| 77. | Fe80P12B4Si4 | 1.34 | [15] |
| 78. | Fe76Al4P12B4Si4 | 1.24 | |
| 79. | Fe74Al4Ga2P12B4Si4 | 1.14 | |
| 80. | Fe56Co7Ni7Zr10B20 | 0.93 | [16] |
| 81. | Fe56Co7Ni7Zr8Nb2B20 | 0.74 | |

| 82. | Fe56Co7Ni7Zr6Nb4B20 | 0.71 | |
|---|---|---|---|
| 83. | Fe56Co7Ni7Zr4Nb6B20 | 0.69 | |
| 84. | Fe56Co7Ni7Zr2Nb8B20 | 0.67 | |
| 85. | Fe56Co7Ni7Nb10B20 | 0.55 | |
| 86. | Fe66Co7Ni7B20 | 1.55 | [17] |
| 87. | Fe62Co7Ni7Hf4B20 | 1.2 | |
| 88. | Fe58Co7Ni7Hf8B20 | 1 | |
| 89. | Fe56Co7Ni7Hf10B20 | 0.82 | |
| 90. | Fe54Co7Ni7Hf12B20 | 0.61 | |
| 91. | Fe76Co7Ni7Hf10 | 1.27 | |
| 92. | Fe72Co7Ni7Hf10B4 | 1.18 | |
| 93. | Fe68Co7Ni7Hf10B8 | 1.09 | |
| 94. | Fe60Co7Ni7Hf10B16 | 0.88 | |
| 95. | Fe67Co10Sm3B20 | 1.5 | [18] |
| 96. | Fe67Co10Nd3B20 | 1.48 | |
| 97. | Fe67Co10Dy3B20 | 1.36 | |
| 98. | Fe67Co10Tb3B20 | 1.31 | |
| 99. | Fe75Mo5P10C7.5B2.5 | 1.1 | [19] |
| 100. | Fe61Co8Zr3Y3Ni5Nb5B15 | 0.88 | [20] |
| 101. | Fe61Co8Zr4Y2Ni5Nb5B15 | 1.05 | |
| 102. | Fe70Ni2B24Nb4 | 1.12 | [21] |
| 103. | Fe68Ni4B24Nb4 | 1.08 | |
| 104. | Fe64Ni8B24Nb4 | 1.03 | |
| 105. | Fe60Ni12B24Nb4 | 1.01 | |
| 106. | Fe58Ni14B24Nb4 | 0.98 | |
| 107. | Fe74Al4Sn2P12Si4B4 | 1.32 | [22] |
| 108. | Fe74Al4Sn2P10Si4B4C2 | 1.34 | |
| 109. | Fe74Al4Sn2P6Si4B4C6 | 1.42 | |
| 110. | Fe80P9C8B2Si1 | 1.55 | [23] |
| 111. | Fe79Sn1P9C8B2Si1 | 1.51 | |
| 112. | Fe78Sn2P9C8B2Si1 | 1.46 | |
| 113. | Fe77Sn3P9C8B2Si1 | 1.42 | |
| 114. | Fe62Co9.5Pr3.5B25 | 1.36 | [24] |
| 115. | Fe62Co9.5Nd3Dy0.5B25 | 1.41 | |
| 116. | Fe62Co9.5Sm3.5B25 | 1.35 | |
| 117. | Fe62Co9.5Gd3.5B25 | 1.41 | |
| 118. | Fe62Co9.5Dy3.5B25 | 1.43 | |
| 119. | Fe62Co9.5Tb3.5B25 | 1.36 | |
| 120. | Fe62Co9.5Er3.5B25 | 1.38 | |
| 121. | Fe76P5(Si0.3B0.5C0.2)19 | 1.44 | [25] |
| 122. | Fe80P11C9 | 1.37 | [26] |
| 123. | Fe77.3C5.9Si3.3B4.8P8.7 | 1.52 | [27] |
| 124. | Fe77Si8B10P5 | 1.54 | [28] |
| 125. | Fe78Si7B10P5 | 1.57 | |
| 126. | Fe79Si6B10P5 | 1.6 | |
| 127. | Fe80Si5B10P5 | 1.62 | |
| 128. | Fe81Si4B10P5 | 1.64 | |
| 129. | Fe81B10Si5.5P3.5 | 1.65 | [29] |
| 130. | Fe82B10Si5P3 | 1.68 | |
| 131. | Fe83B9Si5P3 | 1.68 | |
| 132. | Fe84B8.5Si4.5P3 | 1.7 | |
| 133. | Fe85B8Si4P3 | 1.65 | |
| 134. | Fe76Si8P9C7 | 1.33 | [30] |
| 135. | Fe80P13C7 | 1.48 | [31] |
| 136. | (Fe95Co5)80P13C7 | 1.54 | |
| 137. | (Fe90Co10)80P13C7 | 1.57 | |
| 138. | (Fe85Co15)80P13C7 | 1.56 | |
| 139. | (Fe90Co10)82P11C7 | 1.6 | |
| 140. | (Fe90Co10)82P8C7B3 | 1.62 | |
| 141. | (Fe90Co10)82P6C7B3Si2 | 1.65 | |
| 142. | Fe83B9Si4.5P3.5 | 1.69 | [32] |
| 143. | Fe83B9Si4.5P3.1C0.4 | 1.71 | |
| 144. | Fe83B9Si4.35P3.45C0.2 | 1.70 | |
| 145. | Fe84B8.5Si4.25P3.25 | 1.70 | |
| 146. | Fe84B8.5Si4.25P3C0.25 | 1.66 | |
| 147. | Fe84B8.5Si4.1P3.25C0.15 | 1.72 | |
| 148. | Fe85B8Si4.25P2.75 | 1.68 | |
| 149. | Fe85B8Si3.75P3.25 | 1.67 | |
| 150. | Fe85B8Si3.5P3.5 | 1.70 | |
| 151. | Fe85B8Si4P2.85C0.15 | 1.65 | |
| 152. | Fe85B8Si3.85P3.05C0.15 | 1.65 | |
| 153. | Fe81P8.5C5.5B2Si3 | 1.56 | [33] |
| 154. | Fe82Mo1P6.5C5.5B2Si3 | 1.59 | |
| 155. | Fe82Si4B11P3 | 1.66 | [34] |
| 156. | Fe83C1B11Si2P3 | 1.67 | [35] |
| 157. | (Fe0.95Co0.05)83Si1B16 | 1.79 | [36] |
| 158. | (Fe0.9Co0.1)83Si1B16 | 1.82 | |
| 159. | (Fe0.85Co0.15)83Si1B16 | 1.83 | |
| 160. | (Fe0.8Co0.2)83Si1B16 | 1.86 | |
| 161. | (Fe0.7Co0.3)83Si1B16 | 1.82 | |
| 162. | (Fe0.6Co0.4)83Si1B16 | 1.78 | |
| 163. | (Fe0.8Co0.2)85B14Si1 | 1.92 | [37] |
| 164. | (Fe0.95Co0.05)84B15Si1 | 1.82 | |
| 165. | (Fe0.9Co0.1)84B15Si1 | 1.86 | |
| 166. | (Fe0.85Co0.15)84B15Si1 | 1.87 | |
| 167. | (Fe0.825Co0.175)84B15Si1 | 1.88 | |
| 168. | (Fe0.8Co0.2)84B15Si1 | 1.88 | |
| 169. | (Fe0.7Co0.3)84B15Si1 | 1.83 | |
| 170. | (Fe0.8Co0.2)83B16Si1 | 1.86 | |
| 171. | Fe72B20Si4Nb4 | 1.19 | [38] |
| 172. | Fe71.8Cu0.2B20Si4Nb4 | 1.22 | |
| 173. | Fe71.6Cu0.4B20Si4Nb4 | 1.24 | |
| 174. | Fe71.4Cu0.6B20Si4Nb4 | 1.25 | |
| 175. | Fe71.2Cu0.8B20Si4Nb4 | 1.21 | |
| 176. | Fe71Cu1B20Si4Nb4 | 1.17 | |
| 177. | Fe72B20Si4Nb3.9Cu0.1 | 1.2 | [39] |
| 178. | Fe72B20Si4Nb3.8Cu0.2 | 1.23 | |
| 179. | Fe72B20Si4Nb3.7Cu0.3 | 1.27 | |
| 180. | Fe68Nb5Al4Si3B20 | 1.19 | [40] |
| 181. | Fe63Nb10Al4Si3B20 | 0.58 | |
| 182. | Fe67.7C7.0Si3.3B5.5P8.7Cr2.3Mo2.5Al2.0Co1 | 0.846 | [41] |
| 183. | Fe68.7C7.0Si3.3B5.5P8.7Cr2.3Mo2.5Al2.0 | 0.844 | |
| 184. | Fe65.7C7.0Si3.3B5.5P8.7Cr2.3Mo2.5Al2.0Co3 | 0.802 | |
| 185. | Fe63.7C7.0Si3.3B5.5P8.7Cr2.3Mo2.5Al2.0Co5 | 0.852 | |
| 186. | Fe61.7C7.0Si3.3B5.5P8.7Cr2.3Mo2.5Al2.0Co7 | 0.775 | |
| 187. | Fe58.7C7.0Si3.3B5.5P8.7Cr2.3Mo2.5Al2.0Co10 | 0.837 | |
| 188. | Fe68.2C7Si3.3B5.5P8.7Cr2.3Al2Co3 | 1.03 | [42] |
| 189. | Fe71.2C7Si3.3B5.5P8.7Cr2.3Al2 | 0.9 | |
| 190. | Fe73.5C7Si3.3B5.5P8.7Nb2.0 | 1.1 | [43] |
| 191. | Fe75.5C7.0Si3.3B5.5P8.7 | 1.36 | |
| 192. | Fe75Cr5P9B4C7 | 1.03 | [44] |
| 193. | (Fe0.8P0.09C0.09B0.02)99.3Cu0.7 | 1.44 | [45] |
| 194. | (Fe36Co36B19.2Si4.8Nb4)99.5Cu0.5 | 1.1 | |
| 195. | (Fe40Ni40P14B6)96Ga4 | 0.75 | |
| 196. | Fe77.3C5.1B6.9Si2.7P7.3Cu0.7 | 1.45 | |
| 197. | Fe79.3C3.1B6.9Si2.7P7.3Cu0.7 | 1.51 | |
| 198. | Fe75.3C7.0Si3.3B5.0P8.7Cu0.7 | 1.61 | |
| 199. | Fe75.7C7.0Si3.3B5.0P8.7Cu0.3 | 1.56 | |
| 200. | (Fe83C13Si3Mn0.6P0.3S0.1)90Ga2B4Al4 | 1.21 | |
| 201. | (Fe83C13Si3Mn0.6P0.3S0.1)92Al4B4 | 1.11 | |
| 202. | Fe72Y6B22 | 1.47 | [46] |
| 203. | Fe76B20Y4 | 1.56 | |
| 204. | Fe62.8Co10B13.5Si10Nb3Cu0.7 | 1.35 | [47] |
| 205. | (Fe0.39Ni0.39B0.16P0.06)96Nb4 | 0.66 | [48] |
| 206. | (Fe0.39Ni0.39B0.16P0.06)97Nb3 | 0.71 | |
| 207. | (Fe0.39Ni0.39B0.16P0.06)97.5Nb2.5 | 0.73 | |
| 208. | (Fe0.39Ni0.39B0.16P0.06)98Nb2 | 0.75 | |
| 209. | (Fe0.39Ni0.39B0.16P0.06)98.5Nb1.5 | 0.77 | |
| 210. | (Fe0.39Ni0.39B0.16P0.06)99Nb1 | 0.8 | |

| 211. | (Fe0.39Ni0.39B0.16P0.06)100 | 0.85 | |
|---|---|---|---|
| 212. | Fe82.75Si4B8P4Cu1.25 | 1.83 | [49] |
| 213. | Fe83Si4B8P4Cu1 | 1.82 | |
| 214. | Fe77B18Zr5 | 1.33 | [50] |
| 215. | Fe77B18Ti1Zr4 | 1.35 | |
| 216. | Fe77B18Ti1.5Zr3.5 | 1.36 | |
| 217. | Fe77B18Ti2Zr3 | 1.38 | |
| 218. | Fe77B18Ti2.5Zr2.5 | 1.38 | |
| 219. | Fe77B18Ti3Zr2 | 1.39 | |
| 220. | Fe77B18Ti3.5Zr1.5 | 1.39 | |
| 221. | Fe77B18Ti4Zr1 | 1.41 | |
| 222. | Fe77B18Ti5 | 1.38 | |
| 223. | Fe73Ga4P11C5B4Si3 | 1.3 | [51] |
| 224. | (Fe0.8Co0.2)73Ga4P11C5B4Si3 | 1.3 | |
| 225. | (Fe0.85Co0.15)77Ga2P10C5B3.5Si2.5 | 1.4 | |
| 226. | Fe76Si9.6B9.6P4.8 | 1.51 | [52] |
| 227. | (Fe0.76Si0.096B0.096P0.048)98Cr2 | 1.38 | |
| 228. | (Fe0.76Si0.096B0.096P0.048)96Cr4 | 1.29 | |
| 229. | (Fe0.76Si0.096B0.096P0.048)94Cr6 | 1.11 | |
| 230. | (Fe0.72B0.24Nb0.04)95.5Y4.5 | 0.8 | [53] |
| 231. | Fe80B11Si9 | 1.59 | |
| 232. | Fe77.7B15.5Si5Zr1.8 | 1.56 | [54] |
| 233. | Fe77B16Si5.5Zr1.5 | 1.57 | |
| 234. | Fe77B14Si7.7Zr1.3 | 1.54 | |
| 235. | Fe77.5B15.5Si5.5Zr1.5 | 1.6 | |
| 236. | Fe73Cr5Si10B12 | 1.12 | [55] |
| 237. | Fe61Ni12Cr5Si10B12 | 0.93 | |
| 238. | Fe49Ni24Cr5Si10B12 | 0.8 | |
| 239. | Fe71Nb6B23 | 1.09 | [56] |
| 240. | Fe70Ho1Nb6B23 | 0.99 | |
| 241. | Fe69Ho2Nb6B23 | 0.84 | |
| 242. | Fe68Ho3Nb6B23 | 0.68 | |
| 243. | Fe67Ho4Nb6B23 | 0.61 | |
| 244. | Fe66Ho5Nb6B23 | 0.51 | |
| 245. | Fe65Ho6Nb6B23 | 0.45 | |
| 246. | Fe70Er1Nb6B23 | 1.01 | [57] |
| 247. | Fe68Er3Nb6B23 | 0.71 | |
| 248. | Fe66Er5Nb6B23 | 0.54 | |
| 249. | Fe64Er7Nb6B23 | 0.43 | |
| 250. | Fe78P13C9 | 1.4 | [58] |
| 251. | Fe79.9Cu0.1P13C7 | 1.43 | [59] |
| 252. | Fe79.7Cu0.3P13C7 | 1.43 | |
| 253. | Fe79.4Cu0.6P13C7 | 1.42 | |

| 254. | (Fe0.8Co0.2)83B9Si5P3 | 1.72 | [60] |
|---|---|---|---|
| 255. | (Fe0.8Co0.2)84B8.5Si4.5P3 | 1.74 | |
| 256. | (Fe0.8Co0.2)85B8Si4P3 | 1.76 | |
| 257. | (Fe0.8Co0.2)86B7.5Si3.5P3 | 1.69 | |
| 258. | Fe83B17 | 1.65 | [61] |
| 259. | Fe83B15Si2 | 1.68 | |
| 260. | Fe83B14Si2C1 | 1.67 | |
| 261. | Fe83B12Si2P3 | 1.66 | |
| 262. | Fe83B11Si2P3C1 | 1.67 | |
| 263. | Fe76Si3.3P8.7C7B5 | 1.52 | [62] |
| 264. | Fe75Si3.3P8.7C7B5Co1 | 1.56 | |
| 265. | Fe73Si3.3P8.7C7B5Co3 | 1.54 | |
| 266. | Fe71Si3.3P8.7C7B5Co5 | 1.52 | |
| 267. | Fe69Si3.3P8.7C7B5Co7 | 1.5 | |
| 268. | Fe67Si3.3P8.7C7B5Co9 | 1.48 | |
| 269. | Fe75Si3.3P8.7C7B5Mo1 | 1.41 | |
| 270. | Fe73Si3.3P8.7C7B5Mo3 | 1.29 | |
| 271. | Fe71Si3.3P8.7C7B5Mo5 | 1.11 | |
| 272. | Fe75Si3.3P8.7C7B5Ga1 | 1.55 | |
| 273. | Fe74Si3.3P8.7C7B5Ga2 | 1.49 | |
| 274. | Fe72Si3.3P8.7C7B5Ga4 | 1.44 | |
| 275. | Fe75.7Cu0.3Co7Si3.3P8.7C7B5 | 1.61 | |
| 276. | Fe75.5C7Si3.3B5.5P8.7 | 1.3 | [63] |
| 277. | Fe74.5C7Si3.3B5.5P8.7Co1 | 1.1 | |
| 278. | Fe73.5C7Si3.3B5.5P8.7Co2 | 1.07 | |
| 279. | Fe72.5C7Si3.3B5.5P8.7Co3 | 1.04 | |
| 280. | Fe71.5C7Si3.3B5.5P8.7Co4 | 1.01 | |
| 281. | Fe74.5C7Si3.3B5.5P8.7Ni1 | 1.11 | |
| 282. | Fe73.5C7Si3.3B5.5P8.7Ni2 | 0.99 | |
| 283. | Fe72.5C7Si3.3B5.5P8.7Ni3 | 0.85 | |
| 284. | Fe71.5C7Si3.3B5.5P8.7Ni4 | 0.66 | |
| 285. | Fe78Mo1B15Si6 | 1.52 | [64] |
| 286. | Fe78Mo1B5P10Si6 | 1.42 | |
| 287. | Fe78Mo1B13P6Si2 | 1.48 | |
| 288. | Fe78Mo1B5P6Si10 | 1.46 | |
| 289. | Fe78Mo1B9P9Si3 | 1.43 | |
| 290. | Fe78Mo1B9P6Si6 | 1.46 | |
| 291. | Fe78Mo1B12P3Si6 | 1.51 | |
| 292. | Fe78Mo1B9P3Si9 | 1.45 | |
| 293. | (Fe0.85Co0.15)78Mo1B15Si6 | 1.56 | |
| 294. | (Fe0.75Co0.25)78Mo1B15Si6 | 1.51 | |
| 295. | (Fe0.85Co0.15)78Mo1B13P6Si2 | 1.52 | |
| 296. | (Fe0.75Co0.25)78Mo1B13P6Si2 | 1.49 | |

| 297. | (Fe0.85Co0.15)78Mo1B5P6Si10 | 1.47 | |
|---|---|---|---|
| 298. | (Fe0.75Co0.25)78Mo1B5P6Si10 | 1.44 | |
| 299. | (Fe0.75Co0.25)78Mo1B9P6Si6 | 1.48 | |
| 300. | Fe75Co5P13C7 | 1.55 | [65] |
| 301. | Fe70Co10P13C7 | 1.55 | |
| 302. | Fe65Co15P13C7 | 1.52 | |
| 303. | Fe60Co20P13C7 | 1.52 | |
| 304. | Fe75.5B14.5P7Nb3 | 1.2 | [66] |
| 305. | (Fe0.9Ni0.1)75.5B14.5P7Nb3 | 1.13 | |
| 306. | (Fe0.8Ni0.2)75.5B14.5P7Nb3 | 1.07 | |
| 307. | (Fe0.7Ni0.3)75.5B14.5P7Nb3 | 0.96 | |
| 308. | (Fe0.6Ni0.4)75.5B14.5P7Nb3 | 0.79 | |
| 309. | (Fe0.9Co0.1)76Si9B10P5 | 1.49 | [67] |
| 310. | (Fe0.75Si0.1B0.15)96Nb4 | 1.4 | |
| 311. | Fe74Al14Ga2P12B4Si4 | 1.14 | |
| 312. | Fe73Al15Ga2P11C5B4 | 1.29 | |
| 313. | Fe72Al15Ga2P10C6B4Si1 | 1.14 | |
| 314. | Fe75Ga5P12C4B4 | 1.28 | [68] |
| 315. | Fe70Al5Ga2P9.65C5.75B4.6Si3 | 1.2 | [69] |
| 316. | Fe65Co10Ga5P12C4B4 | 1.3 | [70] |
| 317. | Fe60Co15Ga5P12C4B4 | 1.27 | |
| 318. | Fe62.5Co12.5Ga5P12C4B4 | 1.28 | |
| 319. | Fe74Mo4Ga2P12C4B4 | 1.2 | |
| 320. | Fe78Ga2P12C4B4 | 1.27 | |
| 321. | Fe74Mo6P10C7.5B2.5 | 1.02 | [71] |
| 322. | Fe62.9Ni11.1Mo6P10C7.5B2.5 | 0.9 | |
| 323. | Fe56Co7Ni7Zr7.5Nb2.5B20 | 1.01 | [72] |
| 324. | Fe56Co7Ni7Zr6Nb2.5Ta1.5B20 | 0.89 | |
| 325. | Fe56Co7Ni7Zr6Nb2.5Ti1.5B20 | 1.06 | |
| 326. | Fe56Co7Ni7Zr6Nb2.5Mo1.5B20 | 1.07 | |
| 327. | Fe61Co10Zr5W4B20 | 0.8 | [73] |
| 328. | (Fe0.75B0.2Si0.05)96Nb4 | 1.2 | [74] |
| 329. | [(Fe0.9Ni0.1)0.75B0.2Si0.05]96Nb4 | 1.1 | |
| 330. | [(Fe0.8Ni0.2)0.75B0.2Si0.05]96Nb4 | 1 | |
| 331. | [(Fe0.7Ni0.3)0.75B0.2Si0.05]96Nb4 | 0.9 | |
| 332. | [(Fe0.6Ni0.4)0.75B0.2Si0.05]96Nb4 | 0.8 | |
| 333. | [(Fe0.6Co0.4)0.75B0.2Si0.05]0.96Nb0.04 | 1.04 | [75] |
| 334. | {[(Fe0.6Co0.4)0.75B0.2Si0.05]0.96Nb0.04}99Cr1 | 0.98 | |
| 335. | {[(Fe0.6Co0.4)0.75B0.2Si0.05]0.96Nb0.04}98Cr2 | 0.91 | |
| 336. | {[(Fe0.6Co0.4)0.75B0.2Si0.05]0.96Nb0.04}97Cr3 | 0.86 | |
| 337. | {[(Fe0.6Co0.4)0.75B0.2Si0.05]0.96Nb0.04}96Cr4 | 0.81 | |
| 338. | (Fe0.75Si0.1B0.15)99Zr1 | 1.53 | [76] |
| 339. | Fe76.0C7.0Si3.3B5.0P8.7 | 1.52 | |
| 340. | Fe75.0C7.0Si3.3B5.0P8.7Ga1.0 | 1.55 | |
| 341. | Fe74.0C7.0Si3.3B5.0P8.7Ga2.0 | 1.49 | |
| 342. | Fe72.0C7.0Si3.3B5.0P8.7Ga4.0 | 1.44 | |
| 343. | Fe60.3Co9.2Cr2Nd3Dy0.5B25 | 1.16 | [77] |
| 344. | Fe60.3Co9.2Mo2Nd3Dy0.5B25 | 1.15 | |
| 345. | Fe60.3Co9.2W2Nd3Dy0.5B25 | 1.14 | |
| 346. | Fe60.3Co9.2V2Nd3Dy0.5B25 | 1.13 | |
| 347. | Fe60.3Co9.2Nb2Nd3Dy0.5B25 | 1.15 | |
| 348. | Fe60.3Co9.2Ta2Nd3Dy0.5B25 | 1.19 | |
| 349. | Fe77Nb6B17 | 0.74 | [78] |
| 350. | Fe78Si9B13 | 1.56 | [79] |
| 351. | Fe83Si2B11P3C1 | 1.67 | [80] |
| 352. | Fe78Co5Si2B11P3C1 | 1.71 | |
| 353. | Fe73Co10Si2B11P3C1 | 1.73 | |
| 354. | Fe68Co15Si2B11P3C1 | 1.74 | |
| 355. | Fe63Co20Si2B11P3C1 | 1.73 | |
| 356. | Fe60Co20P11C9 | 1.61 | |
| 357. | Fe72Si19.2B4.8Nb4 | 1.4 | |
| 358. | (Fe0.9Co0.1)72Si19.2B4.8Nb4 | 1.13 | |
| 359. | (Fe0.8Co0.2)72Si19.2B4.8Nb4 | 1.05 | |
| 360. | Fe75B16.67Si8.33 | 1.57 | [81] |
| 361. | Fe73.33B16.67Si8.33Hf1.67 | 1.47 | |
| 362. | Fe72.5B16.67Si8.33Hf2.5 | 1.45 | |
| 363. | Fe71.67B16.67Si8.33Hf3.33 | 1.28 | |
| 364. | Fe70.83B16.67Si8.33Hf4.17 | 1.18 | |
| 365. | Fe70B16.67Si8.33Hf5 | 1.14 | |
| 366. | Fe82.55B13.79Si0.9Zr2.76 | 1.61 | [82] |
| 367. | Fe81.55B14.79Si0.9Zr2.76 | 1.59 | |
| 368. | Fe78.8B13.79Si3.9Ta0.75Zr2.76 | 1.5 | |
| 369. | Fe77.05B14.79Si3.9Ta1.5Zr2.76 | 1.42 | |
| 370. | Fe75.55B14.79Si6.9Zr2.76 | 1.49 | |
| 371. | Fe75.05B13.79Si6.9Ta1.5Zr2.76 | 1.36 | |
| 372. | Fe73.8B15.79Si6.9Ta0.75Zr2.76 | 1.34 | |
| 373. | Fe70Zr10B20 | 0.72 | [83] |
| 374. | Fe70Zr6Nb4B20 | 0.7 | |
| 375. | Fe70Zr4Nb4Ti2B20 | 0.64 | |
| 376. | Fe43Co18.5Ni18.5P14B6 | 1 | [84] |
| 377. | Fe40Co20Ni20P14B6 | 0.95 | |
| 378. | Fe35Co22.5Ni22.5P14B6 | 0.86 | |
| 379. | Fe30Co25Ni25P14B6 | 0.78 | |
| 380. | Fe53Co18Mo9P14B6 | 0.76 | [85] |
| 381. | Fe71Mo9P14B6 | 0.71 | |
| 382. | Fe78Co6Si2B13P1 | 1.7 | [86] |

| 383. | Fe80Co4Si2B13P1 | 1.69 | |
|---|---|---|---|
| 384. | Fe82Co2Si2B13P1 | 1.67 | |
| 385. | Fe84Si2B13P1 | 1.67 | |
| 386. | Fe82Ni2Si2B13P1 | 1.65 | |
| 387. | Fe80Ni4Si2B13P1 | 1.61 | |
| 388. | Fe78Ni6Si2B13P1 | 1.57 | |
| 389. | Fe76B10P5Si9 | 1.5 | [87] |
| 390. | Fe75.5B10P5Si9Cu0.5 | 1.52 | |
| 391. | Fe75B10P5Si9Cu1 | 1.52 | |
| 392. | Fe74.5B10P5Si9Cu1.5 | 1.45 | |
| 393. | Fe72Ni8Nb4Si2B14 | 1.09 | [88] |
| 394. | Fe73Mo4B5Si5P8C5 | 1.15 | [89] |
| 395. | Fe80Al2B5P9C4 | 1.56 | |
| 396. | Fe72B17Si5Mo6 | 1.08 | [90] |
| 397. | Fe72B17Si5(Mo0.5Nb0.5)6 | 1.06 | |
| 398. | Fe66Dy5Nb6B23 | 0.46 | [91] |
| 399. | Fe66Tm5Nb6B23 | 0.58 | |
| 400. | Fe84Si3B13 | 1.66 | [92] |
| 401. | Fe78B14.2Si2.75P2.75Nb2.3 | 1.39 | [93] |
| 402. | Fe70.2Ni7.8B14.2Si2.75P2.75Nb2.3 | 1.35 | |
| 403. | Fe62.4Ni15.6B14.2Si2.75P2.75Nb2.3 | 1.24 | |
| 404. | Fe54.6Ni23.4B14.2Si2.75P2.75Nb2.3 | 1.13 | |
| 405. | Fe46.5Ni31.2B14.2Si2.75P2.75Nb2.3 | 0.99 | |
| 406. | Fe80P10C7B3 | 1.5 | [94] |
| 407. | Fe80P8C7B5 | 1.53 | |
| 408. | Fe80P13C4B3 | 1.49 | |
| 409. | Fe80P8C9B3 | 1.52 | |
| 410. | Fe80P12C8 | 1.35 | [95] |
| 411. | Fe80P11C8B1 | 1.42 | |
| 412. | Fe80P10C8B2 | 1.44 | |
| 413. | Fe80P9C8B3 | 1.51 | |
| 414. | Fe80P8C8B4 | 1.57 | |
| 415. | (Fe0.75B0.15Si0.1)99Zr1 | 1.44 | [96] |
| 416. | (Fe0.75B0.15Si0.1)97Zr3 | 1.49 | |
| 417. | Fe80P9C9B2 | 1.38 | [97] |
| 418. | Fe76Ga1P8.7C7Si3.3B5 | 1.55 | |
| 419. | Fe66Co10Mo4P9C4B4Si3 | 1 | |
| 420. | Fe76Mo2P10C7.5B2.5Si2 | 1.34 | |
| 421. | Fe78Mo1P9C6.5B3.5Si2 | 1.45 | |
| 422. | Fe77Al3P9C9B2 | 1.36 | |
| 423. | Fe67.5Mo7.5P10C10B5 | 0.81 | [98] |
| 424. | Fe62.5Co5Mo7.5P10C10B5 | 0.83 | |
| 425. | Fe60Ni7.5Mo7.5P10C10B5 | 0.77 | |

| 426. | Fe60Co5Ni2.5Mo7.5P10C10B5 | 0.8 | |
|---|---|---|---|
| 427. | Fe68Dy6B22Nb4 | 0.58 | [99] |
| 428. | (Fe0.9Co0.1)68Dy6B22Nb4 | 0.65 | |
| 429. | (Fe0.8Co0.2)68Dy6B22Nb4 | 0.7 | |
| 430. | (Fe0.7Co0.3)68Dy6B22Nb4 | 0.75 | |
| 431. | (Fe0.6Co0.4)68Dy6B22Nb4 | 0.61 | |
| 432. | (Fe0.72Mo0.04B0.24)94Dy6 | 0.59 | |
| 433. | [(Fe0.8Co0.2)72Mo4B24]94Dy6 | 0.51 | |
| 434. | Fe80Si8.75B10Cu1.25 | 1.46 | [100] |
| 435. | Fe72.8B16Si8Zr3.2 | 1.34 | [101] |
| 436. | Fe73.85B15.38Si7.69Zr3.08 | 1.37 | |
| 437. | Fe74.81B14.8Si7.41Zr2.96 | 1.43 | |
| 438. | Fe75.71B14.29Si7.14Zr2.86 | 1.5 | |
| 439. | Fe76.55B13.79Si6.9Zr2.76 | 1.54 | |
| 440. | Fe70Ni10P13C7 | 1.47 | [102] |
| 441. | Fe60Ni20P13C7 | 1.28 | |
| 442. | Fe50Ni30P13C7 | 1.25 | |
| 443. | Fe73.33B16.67Si8.33Zr1.67 | 1.48 | [103] |
| 444. | Fe72.50B16.67Si8.33Zr2.50 | 1.4 | |
| 445. | Fe71.67B16.67Si8.33Zr3.33 | 1.41 | |
| 446. | Fe70.83B16.67Si8.33Zr4.17 | 1.26 | |
| 447. | Fe77Al3P9B2C9 | 1.37 | [104] |
| 448. | Fe89Hf7Al3Zr1 | 1.6 | [105] |
| 449. | Fe75Ni5P13C7 | 1.4 | [106] |
| 450. | Fe65Ni15P13C7 | 1.3 | |
| 451. | Fe55Ni25P13C7 | 1.21 | |
| 452. | Fe60Co20P14B6 | 1.53 | [107] |
| 453. | Fe50Co30P14B6 | 1.46 | |
| 454. | (Fe0.76B0.24)96Nb4 | 1.25 | [108] |
| 455. | (Fe0.75Dy0.01B0.24)96Nb4 | 1.18 | |
| 456. | (Fe0.74Dy0.02B0.24)96Nb4 | 1.01 | |
| 457. | (Fe0.73Dy0.03B0.24)96Nb4 | 0.89 | |
| 458. | (Fe0.72Dy0.04B0.24)96Nb4 | 0.85 | |
| 459. | (Fe0.71Dy0.05B0.24)96Nb4 | 0.7 | |
| 460. | (Fe0.70Dy0.06B0.24)96Nb4 | 0.68 | |
| 461. | (Fe0.69Dy0.07B0.24)96Nb4 | 0.56 | |
| 462. | (Fe0.9Ni0.1)72B20Si4Nb4 | 1.08 | [109] |
| 463. | (Fe0.8Ni0.2)72B20Si4Nb4 | 0.99 | |
| 464. | (Fe0.7Ni0.3)72B20Si4Nb4 | 0.91 | |
| 465. | (Fe0.6Ni0.4)72B20Si4Nb4 | 0.78 | |
| 466. | (Fe0.71Gd0.05B0.24)96Nb4 | 0.76 | [110] |
| 467. | (Fe0.71Tb0.05B0.24)96Nb4 | 0.71 | |
| 468. | (Fe0.71Ho0.05B0.24)96Nb4 | 0.75 | |

| 469. | (Fe0.71Er0.05B0.24)96Nb4 | 0.74 | |
|---|---|---|---|
| 470. | (Fe0.71Tm0.05B0.24)96Nb4 | 0.87 | |
| 471. | Fe60Ni20P14B6 | 1.13 | [111] |
| 472. | Fe50Ni30P14B6 | 1.02 | |
| 473. | [(Fe60Co40)0.75Si0.05B0.20]95Nb4Zr1 | 1.04 | [112] |
| 474. | [(Fe60Co30Ni10)0.75Si0.05B0.20]95Nb4Zr1 | 0.92 | |
| 475. | Fe75P10C10B5 | 1.38 | [113] |
| 476. | Fe72.5Mo2.5P10C10B5 | 1.11 | |
| 477. | Fe70Mo5P10C10B5 | 0.93 | |
| 478. | Fe65Mo10P10C10B5 | 0.65 | |
| 479. | Fe62Ni10Y6B22 | 1.25 | [114] |
| 480. | Fe67Ni5Y6B22 | 1.35 | |
| 481. | Fe42Co30Y6B22 | 1.15 | |
| 482. | Fe52Co20Y6B22 | 1.22 | |
| 483. | Fe62Co10Y6B22 | 1.33 | |
| 484. | ((Fe0.7Co0.3)71.2B24Y4.8)96Nb4 | 0.85 | [115] |
| 485. | ((Fe0.9Co0.1)71.2B24Y4.8)96Nb4 | 0.9 | |
| 486. | (Fe71.2B24Y4.8)96Nb4 | 0.84 | |
| 487. | Fe72Y2Ta4B22 | 1.32 | [116] |
| 488. | Fe72Y1Nb5B22 | 1.25 | |
| 489. | Fe64Co7Zr6Nd3B20 | 1.38 | [117] |
| 490. | Fe56Co7Ni7Zr8W2B20 | 0.7 | [118] |
| 491. | Fe56Co7Ni7Zr8Mo2B20 | 0.73 | |
| 492. | Fe56Co7Ni7Zr8Cr2B20 | 0.75 | |
| 493. | Fe56Co7Ni7Zr8V2B20 | 0.83 | |
| 494. | Fe56Co7Ni7Zr8Hf2B20 | 0.8 | |
| 495. | Fe56Co7Ni7Zr8Ti2B20 | 0.82 | |
| 496. | Fe61Co7Ni7Zr8Nb2B15 | 0.85 | [119] |
| 497. | Fe56Co7Ni7Zr8Ta2B20 | 0.74 | |
| 498. | (Fe0.68Dy0.07B0.2Si0.05)96Nb4 | 0.51 | [120] |
| 499. | (Fe0.69Dy0.06B0.2Si0.05)96Nb4 | 0.58 | |
| 500. | (Fe0.7Dy0.05B0.2Si0.05)96Nb4 | 0.65 | |
| 501. | (Fe0.71Dy0.04B0.2Si0.05)96Nb4 | 0.75 | |
| 502. | (Fe0.72Dy0.03B0.2Si0.05)96Nb4 | 0.86 | |
| 503. | (Fe0.73Dy0.02B0.2Si0.05)96Nb4 | 0.99 | |
| 504. | (Fe0.74Dy0.01B0.2Si0.05)96Nb4 | 1.12 | |
| 505. | Fe69Nb6B17Y3Co5 | 1 | [121] |
| 506. | Fe81Mo1Si3P7.5C5.5B2 | 1.64 | [122] |
| 507. | Fe78Mo1Si2P9C6.5B3.5 | 1.45 | |
| 508. | (Fe0.8P0.09C0.09B0.02)99Cu1 | 1.38 | [123] |
| 509. | (Fe0.8P0.09C0.09B0.02)99.5Cu0.5 | 1.4 | |
| 510. | (Fe0.8P0.09C0.09B0.02)99.7Cu0.3 | 1.4 | |
| 511. | (Fe0.8P0.09C0.09B0.02)99.9Cu0.1 | 1.39 | |
| 512. | Fe83B9C3Si4P1 | 1.64 | [124] |
| 513. | Fe83B10C2Si4P1 | 1.64 | |
| 514. | Fe76Mo4(P0.45C0.2B0.2Si0.15)20 | 1.21 | [125] |
| 515. | (Fe0.9Co0.1)76Mo4(P0.45C0.2B0.2Si0.15)20 | 1.19 | |
| 516. | (Fe0.7Co0.3)76Mo4(P0.45C0.2B0.2Si0.15)20 | 1.08 | |
| 517. | (Fe0.8Co0.2)76Mo4(P0.45C0.2B0.2Si0.15)20 | 1.14 | |
| 518. | Fe67Co7Mo4Si2P10C7.5B2.5 | 1.24 | [126] |
| 519. | Fe70.3Ni3.7Mo6P10C7.5B2.5 | 0.99 | |
| 520. | Fe69Ni5Mo6P10C7.5B2.5 | 0.97 | |
| 521. | Fe66.6Ni7.4Mo6P10C7.5B2.5 | 0.94 | |
| 522. | Fe70.3Ni3.7Mo5Si1P10C7.5B2.5 | 1.08 | |
| 523. | Fe69Ni5Mo5Si1P10C7.5B2.5 | 1.06 | |
| 524. | Fe62.9Co7.4Ni3.7Mo6P10C7.5B2.5 | 1.05 | |
| 525. | Fe60Co7Ni7Mo6P10C7.5B2.5 | 0.96 | |
| 526. | Fe62Co7Ni7Mo4P10C7.5B2.5 | 1.01 | |
| 527. | Fe74Mo5Nb1P10C7.5B2.5 | 1.11 | |
| 528. | Fe74Mo5Cr1P10C7.5B2.5 | 1.04 | |
| 529. | Fe66.6Co7.4Mo6P10C7.5B2.5 | 1.05 | |
| 530. | Fe77Al3P6B5C9 | 1.41 | |
| 531. | Fe77Al3P7B4C9 | 1.38 | |
| 532. | Fe77Al3P8B3C9 | 1.36 | |
| 533. | Fe80P8B3C9 | 1.5 | |
| 534. | Fe80P9B2C8Si1 | 1.44 | |
| 535. | Fe80P9B2C7Si2 | 1.43 | |
| 536. | Fe80P9B2C6Si3 | 1.41 | |
| 537. | Fe80P9B2C9 | 1.46 | |
| 538. | Fe80P10B1C9 | 1.4 | |
| 539. | Fe77Al3P10B1C9 | 1.32 | |
| 540. | Fe77Al3P11C9 | 1.29 | |
| 541. | Fe74Mo5P10C7.5B2.5Si1 | 1.1 | [127] |
| 542. | Fe78Mo1B5C5P10Si1 | 1.43 | [128] |
| 543. | Fe78Mo1B3C7P10Si1 | 1.42 | |
| 544. | (Fe0.85Co0.15)78Mo1B3C7P10Si1 | 1.43 | |
| 545. | (Fe0.8Co0.2)78Mo1B3C7P10Si1 | 1.44 | |
| 546. | (Fe0.7Co0.3)78Mo1B3C7P10Si1 | 1.38 | |
| 547. | Fe78Mo1B7C3P10Si1 | 1.44 | |
| 548. | Fe84B8P4Si2Mo2 | 1.47 | [129] |
| 549. | Fe84B8P3.75Si1.75Mo2Cu0.5 | 1.34 | |
| 550. | Fe85B8P3.5Si1.5Mo1Cu1 | 1.64 | |
| 551. | Fe84B8P3.5Si1.5Mo2Cu1 | 1.40 | |
| 552. | Fe83.2P9B1C6Cu0.8 | 1.54 | [130] |
| 553. | Fe83.2P8B2C6Cu0.8 | 1.59 | |
| 554. | Fe83.2P7B3C6Cu0.8 | 1.62 | |

| 555. | Fe83.2P6B4C6Cu0.8 | 1.62 | |
|---|---|---|---|
| 556. | Fe76P12C10.8Si1.2 | 1.25 | [131] |
| 557. | Fe78P11C9.9Si1.1 | 1.35 | |
| 558. | Fe80P10C9Si1 | 1.45 | |
| 559. | Fe82P9C8.1Si0.9 | 1.48 | |
| 560. | Fe76Mo4P12C4B4 | 1.11 | [132] |
| 561. | Fe66Ni10Mo4P12C4B4 | 1.07 | |
| 562. | Fe56Ni20Mo4P12C4B4 | 0.88 | |
| 563. | Fe56Ni20Mo4P11C4B4Si1 | 0.93 | |
| 564. | Fe56Ni20Mo4P10C4B4Si2 | 0.91 | |
| 565. | Fe80P16C3B1 | 1.49 | [133] |
| 566. | Fe77.9Mo2.05Si2.05P9C6.75B2.25 | 1.47 | [134] |
| 567. | Fe80Mo1Si2P8C6B3 | 1.60 | |
| 568. | Fe79.8Mo2.1Si2.1P8C6B2 | 1.55 | |
| 569. | Fe79Si5B13P3 | 1.66 | [135] |
| 570. | Fe82B13Si4P1 | 1.63 | |
| 571. | Fe73.5B9Si13.5Nb3Cu1 | 1.24 | |
| 572. | Fe83.5B10C6Cu0.5 | 1.74 | |
| 573. | Fe81B10P5Nb4 | 1.60 | [136] |
| 574. | Fe81Si4B10P4Cu1 | 1.71 | [137] |

**Table S2** 383 Fe-based MG compositions and their $H_c$ values collected from experimental measurements in literature.

| | Amorphous Alloy | $H_c$ (A/m) | Reference |
|---|---|---|---|
| 1. | Fe79P10C4B4Si3 | 3.2 | [1] |
| 2. | Fe78Mo1P10C4B4Si3 | 2.7 | |
| 3. | Fe77Mo2P10C4B4Si3 | 2.1 | |
| 4. | Fe76Mo3P10C4B4Si3 | 1.7 | |
| 5. | Fe75Mo4P10C4B4Si3 | 1.5 | |
| 6. | Fe74Mo5P10C4B4Si3 | 1.5 | |
| 7 | Fe76C7.0Si3.3B5P8.7 | 1.6 | [2] |
| 8. | Fe73C7.0Si3.3B5P8.7Mo3 | 4.2 | |
| 9. | (Fe0.76Si0.096B0.084P0.06)100Mo0 | 1.7 | [3] |
| 10. | (Fe0.76Si0.096B0.084P0.06)98Mo2 | 1.9 | |
| 11. | (Fe0.76Si0.096B0.084P0.06)96Mo4 | 1.9 | |
| 12. | (Fe0.76Si0.096B0.084P0.06)94Mo6 | 2.1 | |
| 13. | Fe75B15Si10 | 4.7 | [4] |
| 14. | (Fe0.75B0.15Si0.10)99Zr1 | 2.8 | |
| 15. | (Fe75B15Si10)99Nb1 | 3.7 | |
| 16. | (Fe75B15Si10)98Nb2 | 3.5 | |
| 17. | (Fe75B15Si10)96Nb4 | 2.9 | |
| 18. | Fe77Ga3P9.5C4B4Si2.5 | 4.25 | [5] |
| 19. | Fe78Ga2P9.5C4B4Si2.5 | 3.35 | |
| 20. | (Fe0.74Tb0.01B0.2Si0.05)96Nb4 | 2.5 | [6] |
| 21. | (Fe0.73Tb0.02B0.2Si0.05)96Nb4 | 3.4 | |

| 22. | (Fe0.72Tb0.03B0.2Si0.05)96Nb4 | 6.7 | |
|---|---|---|---|
| 23. | (Fe0.71Tb0.04B0.2Si0.05)96Nb4 | 12.2 | |
| 24. | (Fe0.70Tb0.05B0.2Si0.05)96Nb4 | 19.1 | |
| 25. | (Fe0.69Tb0.06B0.2Si0.05)96Nb4 | 35.8 | |
| 26. | [(Fe0.9Co0.1)0.75B0.2Si0.05]96Nb4 | 2.7 | [7] |
| 27. | [(Fe0.8Co0.2)0.75B0.2Si0.05]96Nb4 | 2.5 | |
| 28. | [(Fe0.7Co0.3)0.75B0.2Si0.05]96Nb4 | 2.1 | |
| 29. | [(Fe0.6Co0.4)0.75B0.2Si0.05]96Nb4 | 1.7 | |
| 30. | [(Fe0.5Co0.5)0.75B0.2Si0.05]96Nb4 | 1.5 | |
| 31. | Fe76Mo3.5P10C4B4Si2.5 | 2 | [8] |
| 32. | Fe71Ni5Mo3.5P10C4B4Si2.5 | 1.9 | |
| 33. | Fe66Ni10Mo3.5P10C4B4Si2.5 | 1.6 | |
| 34. | Fe61Ni15Mo3.5P10C4B4Si2.5 | 1.5 | |
| 35. | Fe56Ni20Mo3.5P10C4B4Si2.5 | 1.3 | |
| 36. | Fe46Ni30Mo3.5P10C4B4Si2.5 | 1.1 | |
| 37. | (Fe0.75B0.15Si0.1)96Nb4 | 3.7 | [9] |
| 38. | [(Fe0.8Co0.1Ni0.1)0.75B0.2Si0.05]96Nb4 | 3 | |
| 39. | [(Fe0.6Co0.1Ni0.3)0.75B0.2Si0.05]96Nb4 | 2.5 | |
| 40. | [(Fe0.6Co0.2Ni0.2)0.75B0.2Si0.05]96Nb4 | 2.5 | |
| 41. | [(Fe0.6Co0.3Ni0.1)0.75B0.2Si0.05]96Nb4 | 2 | |
| 42. | Fe76Mo2Ga2P10C4B4Si2 | 2.9 | [10] |
| 43. | Fe74Mo4Ga2P10C4B4Si2 | 3.3 | |
| 44. | Fe75Mo2Ga3P10C4B4Si2 | 2.4 | |
| 45. | Fe73Mo4Ga3P10C4B4Si2 | 3 | |
| 46. | (Fe0.75B0.15Si0.10)99Nb1 | 3.7 | [11] |
| 47. | (Fe0.75B0.15Si0.10)98Nb2 | 3.5 | |
| 48. | (Fe0.75B0.15Si0.10)96Nb4 | 2.9 | |
| 49. | (Fe0.775B0.125Si0.10)98Nb2 | 3.7 | |
| 50. | Fe76Si9B10P5 | 0.8 | [12] |
| 51. | Fe73Al5Ga2P11C5Si4 | 6.3 | |
| 52. | Fe72Al5Ga2P10C6B4Si1 | 2.8 | |
| 53. | Fe30Co30Ni15Si8B17 | 3.4 | |
| 54. | Fe74Nb6Y3B17 | 15 | |
| 55. | Fe74Cr2Mo2Ga2P10C4B4Si2 | 2.75 | [13] |
| 56. | Fe72Cr4Mo2Ga2P10C4B4Si2 | 2.5 | |
| 57. | Fe70Cr6Mo2Ga2P10C4B4Si2 | 2.25 | |
| 58. | Fe80P12B4Si4 | 1.1 | [15] |
| 59. | Fe76Al4P12B4Si4 | 2.6 | |
| 60. | Fe74Al4Ga2P12B4Si4 | 6.4 | |
| 61. | Fe56Co7Ni7Zr10B20 | 5 | [16] |
| 62. | Fe56Co7Ni7Zr8Nb2B20 | 5 | |
| 63. | Fe66Co7Ni7B20 | 12 | [17] |
| 64. | Fe62Co7Ni7Hf4B20 | 2 | |

| 65. | Fe58Co7Ni7Hf8B20 | 2 | |
|---|---|---|---|
| 66. | Fe56Co7Ni7Hf10B20 | 2.5 | |
| 67. | Fe54Co7Ni7Hf12B20 | 4 | |
| 68. | Fe72Co7Ni7Hf10B4 | 10 | |
| 69. | Fe68Co7Ni7Hf10B8 | 3 | |
| 70. | Fe60Co7Ni7Hf10B16 | 4 | |
| 71. | Fe75Mo5P10C7.5B2.5 | 1.8 | [19] |
| 72. | Fe80P9C8B2Si1 | 3 | [23] |
| 73. | Fe79Sn1P9C8B2Si1 | 3.8 | |
| 74. | Fe78Sn2P9C8B2Si1 | 5 | |
| 75. | Fe77Sn3P9C8B2Si1 | 6.2 | |
| 76. | Fe62Co9.5Pr3.5B25 | 3.7 | [24] |
| 77. | Fe62Co9.5Nd3Dy0.5B25 | 2.6 | |
| 78. | Fe62Co9.5Sm3.5B25 | 3.8 | |
| 79. | Fe62Co9.5Gd3.5B25 | 1.9 | |
| 80. | Fe62Co9.5Dy3.5B25 | 3 | |
| 81. | Fe62Co9.5Tb3.5B25 | 3.1 | |
| 82. | Fe62Co9.5Er3.5B25 | 1 | |
| 83. | Fe76P5(Si0.3B0.5C0.2)19 | 4.5 | [25] |
| 84. | Fe80P11C9 | 16.5 | [26] |
| 85. | Fe77.3C5.9Si3.3B4.8P8.7 | 11.2 | [27] |
| 86. | Fe79Si6B10P5 | 1.6 | |
| 87. | Fe80Si5B10P5 | 1.6 | |
| 88. | Fe81B10Si5.5P3.5 | 3.3 | [29] |
| 89. | Fe82B10Si5P3 | 3.9 | |
| 90. | Fe83B9Si5P3 | 5.4 | |
| 91. | Fe84B8.5Si4.5P3 | 6.2 | |
| 92. | Fe85B8Si4P3 | 17.4 | |
| 93. | Fe83B9Si4.5P3.5 | 4.4 | [32] |
| 94. | Fe83B9Si4.5P3.1C0.4 | 3.5 | |
| 95. | Fe83B9Si4.35P3.45C0.2 | 6.5 | |
| 96. | Fe84B8.5Si4.25P3.25 | 6.2 | |
| 97. | Fe84B8.5Si4.25P3C0.25 | 8 | |
| 98. | Fe84B8.5Si4.1P3.25C0.15 | 4.8 | |
| 99. | Fe85B8Si3.75P3.25 | 8.3 | |
| 100. | Fe85B8Si3.5P3.5 | 7.8 | |
| 101. | Fe85B8Si3.85P3.05C0.15 | 7.9 | |
| 102. | Fe82Si4B11P3 | 2.2 | [34] |
| 103. | Fe83C1B11Si2P3 | 2.4 | [35] |
| 104. | (Fe0.95Co0.05)83Si1B16 | 5.2 | [36] |
| 105. | (Fe0.9Co0.1)83Si1B16 | 4.5 | |
| 106. | (Fe0.85Co0.15)83Si1B16 | 3.8 | |
| 107. | (Fe0.8Co0.2)83Si1B16 | 3 | |
| 108. | (Fe0.7Co0.3)83Si1B16 | 5.4 | |
| 109. | (Fe0.6Co0.4)83Si1B16 | 7.5 | |
| 110. | (Fe0.8Co0.2)85B14Si1 | 2.2 | [37] |
| 111. | (Fe0.95Co0.05)84B15Si1 | 5 | |
| 112. | (Fe0.9Co0.1)84B15Si1 | 5 | |
| 113. | (Fe0.85Co0.15)84B15Si1 | 3.7 | |
| 114. | (Fe0.825Co0.175)84B15Si1 | 4.5 | |
| 115. | (Fe0.8Co0.2)84B15Si1 | 3.6 | |
| 116. | (Fe0.7Co0.3)84B15Si1 | 4 | |
| 117. | (Fe0.8Co0.2)83B16Si1 | 3 | |
| 118. | Fe72B20Si4Nb4 | 2.5 | [38] |
| 119. | Fe71.8Cu0.2B20Si4Nb4 | 2 | |
| 120. | Fe71.6Cu0.4B20Si4Nb4 | 1.6 | |
| 121. | Fe71.4Cu0.6B20Si4Nb4 | 1.4 | |
| 122. | Fe71.2Cu0.8B20Si4Nb4 | 1.5 | |
| 123. | Fe71Cu1B20Si4Nb4 | 1.6 | |
| 124. | Fe72B20Si4Nb3.9Cu0.1 | 2 | [39] |
| 125. | Fe72B20Si4Nb3.8Cu0.2 | 1.7 | |
| 126. | Fe72B20Si4Nb3.7Cu0.3 | 1.5 | |
| 127. | Fe63Nb10Al4Si3B20 | 11 | |
| 128. | Fe67.7C7.0Si3.3B5.5P8.7Cr2.3Mo2.5Al2.0Co1 | 26.066 | [41] |
| 129. | Fe68.7C7.0Si3.3B5.5P8.7Cr2.3Mo2.5Al2.0 | 37.434 | |
| 130. | Fe65.7C7.0Si3.3B5.5P8.7Cr2.3Mo2.5Al2.0Co3 | 35.424 | |
| 131. | Fe63.7C7.0Si3.3B5.5P8.7Cr2.3Mo2.5Al2.0Co5 | 11.53 | |
| 132. | Fe61.7C7.0Si3.3B5.5P8.7Cr2.3Mo2.5Al2.0Co7 | 20.122 | |
| 133. | Fe58.7C7.0Si3.3B5.5P8.7Cr2.3Mo2.5Al2.0Co10 | 17.977 | |
| 134. | Fe68.2C7Si3.3B5.5P8.7Cr2.3Al2Co3 | 9.39 | [42] |
| 135. | Fe71.2C7Si3.3B5.5P8.7Cr2.3Al2 | 12.64 | |
| 136. | Fe73.5C7Si3.3B5.5P8.7Nb2.0 | 11.9 | [43] |
| 137. | Fe75.5C7.0Si3.3B5.5P8.7 | 18.9 | |
| 138. | Fe75Cr5P9B4C7 | 2.3 | [44] |
| 139. | (Fe0.8P0.09C0.09B0.02)99.3Cu0.7 | 2 | [45] |
| 140. | (Fe36Co36B19.2Si4.8Nb4)99.5Cu0.5 | 2 | |
| 141. | (Fe40Ni40P14B6)96Ga4 | 5 | |
| 142. | Fe77.3C5.1B6.9Si2.7P7.3Cu0.7 | 9.8 | |
| 143. | Fe79.3C3.1B6.9Si2.7P7.3Cu0.7 | 11.1 | |
| 144. | Fe75.3C7.0Si3.3B5.0P8.7Cu0.7 | 6.7 | |
| 145. | Fe75.7C7.0Si3.3B5.0P8.7Cu0.3 | 3.6 | |
| 146. | (Fe83C13Si3Mn0.6P0.3S0.1)90Ga2B4Al4 | 18.3 | |
| 147. | Fe72Y6B22 | 4 | [46] |
| 148. | (Fe0.39Ni0.39B0.16P0.06)96Nb4 | 1.8 | [48] |
| 149. | (Fe0.39Ni0.39B0.16P0.06)97Nb3 | 1.2 | |
| 150. | (Fe0.39Ni0.39B0.16P0.06)97.5Nb2.5 | 1 | |
| 151. | (Fe0.39Ni0.39B0.16P0.06)98Nb2 | 1.1 | |
| 152. | (Fe0.39Ni0.39B0.16P0.06)98.5Nb1.5 | 1.4 | |
| 153. | (Fe0.39Ni0.39B0.16P0.06)99Nb1 | 1.5 | |
| 154. | (Fe0.39Ni0.39B0.16P0.06)100 | 1.6 | |
| 155. | Fe82.75Si4B8P4Cu1.25 | 2.1 | [49] |
| 156. | Fe83Si4B8P4Cu1 | 2.3 | |
| 157. | Fe73Ga4P11C5B4Si3 | 1.9 | [51] |
| 158. | (Fe0.8Co0.2)73Ga4P11C5B4Si3 | 3 | |
| 159. | (Fe0.85Co0.15)77Ga2P10C5B3.5Si2.5 | 4.8 | |
| 160. | Fe76Si9.6B9.6P4.8 | 0.8 | [52] |
| 161. | (Fe0.76Si0.096B0.096P0.048)98Cr2 | 3 | |
| 162. | (Fe0.76Si0.096B0.096P0.048)96Cr4 | 2.4 | |
| 163. | (Fe0.76Si0.096B0.096P0.048)94Cr6 | 1.4 | |
| 164. | (Fe0.72B0.24Nb0.04)95.5Y4.5 | 0.8 | [53] |
| 165. | Fe71Nb6B23 | 3.39 | [56] |
| 166. | Fe70Ho1Nb6B23 | 4.46 | |
| 167. | Fe69Ho2Nb6B23 | 6.55 | |
| 168. | Fe68Ho3Nb6B23 | 7.27 | |
| 169. | Fe67Ho4Nb6B23 | 9.56 | |
| 170. | Fe66Ho5Nb6B23 | 11.28 | |
| 171. | Fe65Ho6Nb6B23 | 9.97 | |
| 172. | Fe70Er1Nb6B23 | 5.34 | [57] |
| 173. | Fe68Er3Nb6B23 | 6.12 | |
| 174. | Fe66Er5Nb6B23 | 8.78 | |
| 175. | Fe64Er7Nb6B23 | 5.81 | |
| 176. | (Fe0.8Co0.2)83B9Si5P3 | 3.5 | [60] |
| 177. | (Fe0.8Co0.2)84B8.5Si4.5P3 | 4.9 | |
| 178. | (Fe0.8Co0.2)85B8Si4P3 | 5.8 | |
| 179. | (Fe0.8Co0.2)86B7.5Si3.5P3 | 15.2 | |
| 180. | Fe83B17 | 18 | [61] |
| 181. | Fe83B15Si2 | 3 | |
| 182. | Fe83B14Si2C1 | 4.2 | |
| 183. | Fe83B12Si2P3 | 1.9 | |
| 184. | Fe83B11Si2P3C1 | 2.1 | |
| 185. | Fe75.5C7Si3.3B5.5P8.7 | 9.54 | [63] |
| 186. | Fe74.5C7Si3.3B5.5P8.7Co1 | 8.26 | |
| 187. | Fe73.5C7Si3.3B5.5P8.7Co2 | 7.27 | |
| 188. | Fe72.5C7Si3.3B5.5P8.7Co3 | 4.81 | |
| 189. | Fe71.5C7Si3.3B5.5P8.7Co4 | 6.77 | |
| 190. | Fe74.5C7Si3.3B5.5P8.7Ni1 | 8.48 | |
| 191. | Fe73.5C7Si3.3B5.5P8.7Ni2 | 9.79 | |
| 192. | Fe72.5C7Si3.3B5.5P8.7Ni3 | 8.78 | |
| 193. | Fe71.5C7Si3.3B5.5P8.7Ni4 | 9.65 | |
| 194. | Fe78Mo1B15Si6 | 16 | [64] |
| 195. | Fe78Mo1B5P10Si6 | 12 | |
| 196. | Fe78Mo1B13P6Si2 | 18 | |
| 197. | Fe78Mo1B5P6Si10 | 20 | |
| 198. | Fe78Mo1B9P9Si3 | 18 | |
| 199. | Fe78Mo1B9P6Si6 | 18 | |
| 200. | Fe78Mo1B12P3Si6 | 19 | |
| 201. | Fe78Mo1B9P3Si9 | 20 | |
| 202. | (Fe0.85Co0.15)78Mo1B15Si6 | 20 | |
| 203. | (Fe0.75Co0.25)78Mo1B15Si6 | 21 | |
| 204. | (Fe0.85Co0.15)78Mo1B13P6Si2 | 22 | |
| 205. | (Fe0.75Co0.25)78Mo1B13P6Si2 | 14 | |
| 206. | (Fe0.85Co0.15)78Mo1B5P6Si10 | 19 | |
| 207. | (Fe0.75Co0.25)78Mo1B5P6Si10 | 10 | |
| 208. | (Fe0.75Co0.25)78Mo1B9P6Si6 | 13 | |
| 209. | Fe75.5B14.5P7Nb3 | 3.3 | [66] |
| 210. | (Fe0.9Ni0.1)75.5B14.5P7Nb3 | 2.8 | |
| 211. | (Fe0.8Ni0.2)75.5B14.5P7Nb3 | 2.7 | |
| 212. | (Fe0.7Ni0.3)75.5B14.5P7Nb3 | 2.5 | |
| 213. | (Fe0.6Ni0.4)75.5B14.5P7Nb3 | 2.3 | |
| 214. | (Fe0.9Co0.1)76Si9B10P5 | 1.2 | [67] |
| 215. | (Fe0.75Si0.1B0.15)96Nb4 | 2.9 | |
| 216. | Fe74Al14Ga2P12B4Si4 | 6.4 | |
| 217. | Fe73Al15Ga2P11C5B4 | 6.3 | |
| 218. | Fe72Al15Ga2P10C6B4Si1 | 2.8 | |
| 219. | Fe75Ga5P12C4B4 | 1.6 | [68] |
| 220. | Fe70Al5Ga2P9.65C5.75B4.6Si3 | 2.2 | [69] |
| 221. | Fe74Mo4Ga2P12C4B4 | 2.5 | |
| 222. | Fe78Ga2P12C4B4 | 1.6 | |
| 223. | Fe56Co7Ni7Zr7.5Nb2.5B20 | 9.5 | [72] |
| 224. | Fe56Co7Ni7Zr6Nb2.5Ta1.5B20 | 6.05 | |
| 225. | Fe56Co7Ni7Zr6Nb2.5Ti1.5B20 | 6.1 | |
| 226. | Fe56Co7Ni7Zr6Nb2.5Mo1.5B20 | 7.4 | |
| 227. | Fe61Co10Zr5W4B20 | 1.4 | [73] |
| 228. | (Fe0.75B0.2Si0.05)96Nb4 | 2.5 | [74] |
| 229. | [(Fe0.9Ni0.1)0.75B0.2Si0.05]96Nb4 | 2.25 | |
| 230. | [(Fe0.8Ni0.2)0.75B0.2Si0.05]96Nb4 | 1.8 | |
| 231. | [(Fe0.7Ni0.3)0.75B0.2Si0.05]96Nb4 | 1.6 | |
| 232. | [(Fe0.6Ni0.4)0.75B0.2Si0.05]96Nb4 | 1.2 | |
| 233. | [(Fe0.6Co0.4)0.75B0.2Si0.05]0.96Nb0.04 | 1.7 | [75] |
| 234. | {[(Fe0.6Co0.4)0.75B0.2Si0.05]0.96Nb0.04}99Cr1 | 1.6 | |
| 235. | {[(Fe0.6Co0.4)0.75B0.2Si0.05]0.96Nb0.04}98Cr2 | 1.4 | |
| 236. | {[(Fe0.6Co0.4)0.75B0.2Si0.05]0.96Nb0.04}97Cr3 | 0.8 | |

| 237. | {[(Fe0.6Co0.4)0.75B0.2Si0.05]0.96Nb0.04}96Cr4 | 0.6 | |
|---|---|---|---|
| 238. | (Fe0.75Si0.1B0.15)99Zr1 | 2.8 | [76] |
| 239. | Fe76.0C7.0Si3.3B5.0P8.7 | 1.6 | |
| 240. | Fe75.0C7.0Si3.3B5.0P8.7Ga1.0 | 4.9 | |
| 241. | Fe74.0C7.0Si3.3B5.0P8.7Ga2.0 | 6.8 | |
| 242. | Fe72.0C7.0Si3.3B5.0P8.7Ga4.0 | 5.2 | |
| 243. | Fe60.3Co9.2Cr2Nd3Dy0.5B25 | 3.85 | [77] |
| 244. | Fe60.3Co9.2Mo2Nd3Dy0.5B25 | 4.98 | |
| 245. | Fe60.3Co9.2W2Nd3Dy0.5B25 | 3.98 | |
| 246. | Fe60.3Co9.2V2Nd3Dy0.5B25 | 4.52 | |
| 247. | Fe60.3Co9.2Nb2Nd3Dy0.5B25 | 4.78 | |
| 248. | Fe60.3Co9.2Ta2Nd3Dy0.5B25 | 3.98 | |
| 249. | Fe77Nb6B17 | 22 | [78] |
| 250. | Fe78Si9B13 | 2.4 | [79] |
| 251. | Fe83Si2B11P3C1 | 5.2 | [80] |
| 252. | Fe78Co5Si2B11P3C1 | 7.2 | |
| 253. | Fe73Co10Si2B11P3C1 | 22.2 | |
| 254. | Fe68Co15Si2B11P3C1 | 35.6 | |
| 255. | Fe63Co20Si2B11P3C1 | 22 | |
| 256. | Fe72Si19.2B4.8Nb4 | 2.9 | |
| 257. | (Fe0.9Co0.1)72Si19.2B4.8Nb4 | 2.7 | |
| 258. | (Fe0.8Co0.2)72Si19.2B4.8Nb4 | 2.5 | |
| 259. | Fe75B16.67Si8.33 | 4.3 | [81] |
| 260. | Fe73.33B16.67Si8.33Hf1.67 | 1.6 | |
| 261. | Fe72.5B16.67Si8.33Hf2.5 | 1.5 | |
| 262. | Fe71.67B16.67Si8.33Hf3.33 | 1.7 | |
| 263. | Fe70.83B16.67Si8.33Hf4.17 | 8.9 | |
| 264. | Fe43Co18.5Ni18.5P14B6 | 8.87 | [84] |
| 265. | Fe40Co20Ni20P14B6 | 9.11 | |
| 266. | Fe35Co22.5Ni22.5P14B6 | 5.36 | |
| 267. | Fe30Co25Ni25P14B6 | 6.46 | |
| 268. | Fe78Co6Si2B13P1 | 10.9 | [86] |
| 269. | Fe80Co4Si2B13P1 | 7.7 | |
| 270. | Fe82Co2Si2B13P1 | 6.5 | |
| 271. | Fe84Si2B13P1 | 6.1 | |
| 272. | Fe82Ni2Si2B13P1 | 3.4 | |
| 273. | Fe80Ni4Si2B13P1 | 5.2 | |
| 274. | Fe78Ni6Si2B13P1 | 5.6 | |
| 275. | Fe76B10P5Si9 | 14.7 | [87] |
| 276. | Fe75.5B10P5Si9Cu0.5 | 16.6 | |
| 277. | Fe75B10P5Si9Cu1 | 12.7 | |
| 278. | Fe74.5B10P5Si9Cu1.5 | 12.5 | |
| 279. | Fe72Ni8Nb4Si2B14 | 3.95 | [88] |
| 280. | Fe73Mo4B5Si5P8C5 | 5.9 | [89] |
| 281. | Fe80Al2B5P9C4 | 5.3 | |
| 282. | Fe72B17Si5Mo6 | 3 | [90] |
| 283. | Fe72B17Si5(Mo0.5Nb0.5)6 | 2.5 | |
| 284. | Fe66Dy5Nb6B23 | 10.49 | [91] |
| 285. | Fe66Tm5Nb6B23 | 8.57 | |
| 286. | Fe84Si3B13 | 4.3 | [92] |
| 287. | Fe78B14.2Si2.75P2.75Nb2.3 | 1.7 | [93] |
| 288. | Fe70.2Ni7.8B14.2Si2.75P2.75Nb2.3 | 1.5 | |
| 289. | Fe62.4Ni15.6B14.2Si2.75P2.75Nb2.3 | 1.5 | |
| 290. | Fe54.6Ni23.4B14.2Si2.75P2.75Nb2.3 | 1.1 | |
| 291. | Fe46.5Ni31.2B14.2Si2.75P2.75Nb2.3 | 0.9 | |
| 292. | Fe80P12C8 | 7.7 | [95] |
| 293. | Fe80P11C8B1 | 3.6 | |
| 294. | Fe80P10C8B2 | 2.2 | |
| 295. | Fe80P9C8B3 | 3.9 | |
| 296. | Fe80P8C8B4 | 2.9 | |
| 297. | Fe76Ga1P8.7C7Si3.3B5 | 4.9 | |
| 298. | Fe66Co10Mo4P9C4B4Si3 | 1.6 | |
| 299. | Fe76Mo2P10C7.5B2.5Si2 | 1.7 | |
| 300. | Fe77Al3P9C9B2 | 2.1 | |
| 301. | Fe67.5Mo7.5P10C10B5 | 2.07 | [98] |
| 302. | Fe62.5Co5Mo7.5P10C10B5 | 3.04 | |
| 303. | Fe60Ni7.5Mo7.5P10C10B5 | 4.37 | |
| 304. | Fe60Co5Ni2.5Mo7.5P10C10B5 | 3.2 | |
| 305. | Fe68Dy6B22Nb4 | 4.61 | [99] |
| 306. | (Fe0.9Co0.1)68Dy6B22Nb4 | 6.49 | |
| 307. | (Fe0.8Co0.2)68Dy6B22Nb4 | 4.73 | |
| 308. | (Fe0.7Co0.3)68Dy6B22Nb4 | 6.53 | |
| 309. | (Fe0.6Co0.4)68Dy6B22Nb4 | 6.63 | |
| 310. | Fe80Si8.75B10Cu1.25 | 9.21 | [100] |
| 311. | Fe72.8B16Si8Zr3.2 | 1.7 | [101] |
| 312. | Fe73.85B15.38Si7.69Zr3.08 | 0.6 | |
| 313. | Fe74.81B14.8Si7.41Zr2.96 | 1.3 | |
| 314. | Fe75.71B14.29Si7.14Zr2.86 | 1.7 | |
| 315. | Fe76.55B13.79Si6.9Zr2.76 | 0.5 | |
| 316. | Fe70Ni10P13C7 | 3 | [102] |
| 317. | Fe60Ni20P13C7 | 2.3 | |
| 318. | Fe50Ni30P13C7 | 1.4 | |
| 319. | Fe73.33B16.67Si8.33Zr1.67 | 6.7 | [103] |
| 320. | Fe72.50B16.67Si8.33Zr2.50 | 3 | |
| 321. | Fe71.67B16.67Si8.33Zr3.33 | 1.6 | |
| 322. | Fe70.83B16.67Si8.33Zr4.17 | 1.8 | |
| 323. | Fe77Al3P9B2C9 | 2 | [104] |
| 324. | Fe89Hf7Al3Zr1 | 4.3 | [105] |
| 325. | (Fe0.76B0.24)96Nb4 | 3 | [108] |
| 326. | (Fe0.75Dy0.01B0.24)96Nb4 | 1.9 | |
| 327. | (Fe0.74Dy0.02B0.24)96Nb4 | 2.6 | |
| 328. | (Fe0.73Dy0.03B0.24)96Nb4 | 4.2 | |
| 329. | (Fe0.72Dy0.04B0.24)96Nb4 | 7.7 | |
| 330. | (Fe0.71Dy0.05B0.24)96Nb4 | 9.8 | |
| 331. | (Fe0.70Dy0.06B0.24)96Nb4 | 19.9 | |
| 332. | (Fe0.69Dy0.07B0.24)96Nb4 | 21.6 | |
| 333. | (Fe0.71Gd0.05B0.24)96Nb4 | 3.86 | [110] |
| 334. | (Fe0.71Ho0.05B0.24)96Nb4 | 3.97 | |
| 335. | (Fe0.71Er0.05B0.24)96Nb4 | 3.54 | |
| 336. | (Fe0.71Tm0.05B0.24)96Nb4 | 1.23 | |
| 337. | Fe75P10C10B5 | 8.95 | [113] |
| 338. | Fe72.5Mo2.5P10C10B5 | 4.87 | |
| 339. | Fe70Mo5P10C10B5 | 2.36 | |
| 340. | Fe65Mo10P10C10B5 | 1.05 | |
| 341. | ((Fe0.7Co0.3)71.2B24Y4.8)96Nb4 | 1.1 | [115] |
| 342. | ((Fe0.9Co0.1)71.2B24Y4.8)96Nb4 | 1.7 | |
| 343. | (Fe71.2B24Y4.8)96Nb4 | 1 | |
| 344. | Fe64Co7Zr6Nd3B20 | 4.2 | [117] |
| 345. | Fe56Co7Ni7Zr8W2B20 | 5.7 | [118] |
| 346. | Fe56Co7Ni7Zr8Mo2B20 | 4.9 | |
| 347. | Fe56Co7Ni7Zr8Cr2B20 | 1.9 | |
| 348. | Fe56Co7Ni7Zr8V2B20 | 4.4 | |
| 349. | Fe56Co7Ni7Zr8Hf2B20 | 2.2 | |
| 350. | Fe56Co7Ni7Zr8Ti2B20 | 1.9 | |
| 351. | (Fe0.68Dy0.07B0.2Si0.05)96Nb4 | 36.3 | [120] |
| 352. | (Fe0.69Dy0.06B0.2Si0.05)96Nb4 | 14.1 | |
| 353. | (Fe0.7Dy0.05B0.2Si0.05)96Nb4 | 5.3 | |
| 354. | (Fe0.71Dy0.04B0.2Si0.05)96Nb4 | 4.3 | |
| 355. | (Fe0.72Dy0.03B0.2Si0.05)96Nb4 | 3.9 | |
| 356. | (Fe0.73Dy0.02B0.2Si0.05)96Nb4 | 3.7 | |
| 357. | (Fe0.74Dy0.01B0.2Si0.05)96Nb4 | 3.8 | |
| 358. | Fe83B9C3Si4P1 | 6.2 | [124] |
| 359. | Fe83B10C2Si4P1 | 4.5 | |
| 360. | Fe76Mo4(P0.45C0.2B0.2Si0.15)20 | 2.7 | [125] |
| 361. | (Fe0.9Co0.1)76Mo4(P0.45C0.2B0.2Si0.15)20 | 1.6 | |
| 362. | (Fe0.7Co0.3)76Mo4(P0.45C0.2B0.2Si0.15)20 | 3 | |
| 363. | (Fe0.8Co0.2)76Mo4(P0.45C0.2B0.2Si0.15)20 | 2.1 | |
| 364. | Fe77Al3P8B3C9 | | |
| 365. | Fe80P8B3C9 | 14.2 | |

| 366. | Fe80P9B2C8Si1 | 7 | |
|---|---|---|---|
| 367. | Fe80P9B2C7Si2 | 13.9 | |
| 368. | Fe80P9B2C6Si3 | 10.3 | |
| 369. | Fe80P9B2C9 | 8.8 | |
| 370. | Fe80P10B1C9 | 7.4 | |
| 371. | Fe84B8P4Si2Mo2 | 10.8 | [129] |
| 372. | Fe84B8P3.75Si1.75Mo2Cu0.5 | 11.8 | |
| 373. | Fe85B8P3.5Si1.5Mo1Cu1 | 6.7 | |
| 374. | Fe84B8P3.5Si1.5Mo2Cu1 | 7.9 | |
| 375. | Fe76P12C10.8Si1.2 | 5.1 | [131] |
| 376. | Fe78P11C9.9Si1.1 | 4.3 | |
| 377. | Fe80P10C9Si1 | 3.1 | |
| 378. | Fe82P9C8.1Si0.9 | 4.2 | |
| 379. | Fe76Mo4P12C4B4 | 1.8 | [132] |
| 380. | Fe66Ni10Mo4P12C4B4 | 1.5 | |
| 381. | Fe56Ni20Mo4P12C4B4 | 1 | |
| 382. | Fe56Ni20Mo4P11C4B4Si1 | 1.9 | |
| 383. | Fe56Ni20Mo4P10C4B4Si2 | 3.5 | |

**Table S3** 311 Fe-based MG compositions and their $D_c$ values collected from experimental measurements in literature.

| | Amorphous Alloy | $D_c$ (mm) | Reference |
|---|---|---|---|
| 1. | Fe79P10C4B4Si3 | 1 | [1] |
| 2. | Fe78Mo1P10C4B4Si3 | 1.5 | |
| 3. | Fe77Mo2P10C4B4Si3 | 2.5 | |
| 4. | Fe76Mo3P10C4B4Si3 | 3.5 | |
| 5. | Fe75Mo4P10C4B4Si3 | 4 | |
| 6. | Fe74Mo5P10C4B4Si3 | 3 | |
| 7. | Fe76C7.0Si3.3B5P8.7 | 1 | [2] |
| 8. | Fe75C7.0Si3.3B5P8.7Mo1 | 3 | |
| 9. | Fe73C7.0Si3.3B5P8.7Mo3 | 5 | |
| 10. | Fe71C7.0Si3.3B5P8.7Mo5 | 3 | |
| 11. | (Fe0.76Si0.096B0.084P0.06)100Mo0 | 2.5 | [3] |
| 12. | (Fe0.76Si0.096B0.084P0.06)98Mo2 | 3.5 | |
| 13. | (Fe0.76Si0.096B0.084P0.06)96Mo4 | 3 | |
| 14. | (Fe0.76Si0.096B0.084P0.06)94Mo6 | 1.5 | |
| 15. | Fe75B15Si10 | 0.25 | [4] |
| 16. | (Fe0.75B0.15Si0.10)99Zr1 | 0.75 | |
| 17. | (Fe75B15Si10)99Nb1 | 0.5 | |
| 18. | (Fe75B15Si10)98Nb2 | 1 | |
| 19. | (Fe75B15Si10)96Nb4 | 1.5 | |
| 20. | Fe77Ga3P9.5C4B4Si2.5 | 2.5 | [5] |
| 21. | Fe78Ga2P9.5C4B4Si2.5 | 2 | |
| 22. | (Fe0.74Tb0.01B0.2Si0.05)96Nb4 | 1 | [6] |
| 23. | (Fe0.73Tb0.02B0.2Si0.05)96Nb4 | 2.5 | |
| 24. | (Fe0.72Tb0.03B0.2Si0.05)96Nb4 | 3 | |

| 25. | (Fe0.71Tb0.04B0.2Si0.05)96Nb4 | 3.5 | |
|---|---|---|---|
| 26. | (Fe0.70Tb0.05B0.2Si0.05)96Nb4 | 3.5 | |
| 27. | (Fe0.69Tb0.06B0.2Si0.05)96Nb4 | 3 | |
| 28. | (Fe0.68Tb0.07B0.2Si0.05)96Nb4 | 2 | |
| 29. | [(Fe0.9Co0.1)0.75B0.2Si0.05]96Nb4 | 2 | [7] |
| 30. | [(Fe0.8Co0.2)0.75B0.2Si0.05]96Nb4 | 2.5 | |
| 31. | [(Fe0.7Co0.3)0.75B0.2Si0.05]96Nb4 | 3.5 | |
| 32. | [(Fe0.6Co0.4)0.75B0.2Si0.05]96Nb4 | 4 | |
| 33. | [(Fe0.5Co0.5)0.75B0.2Si0.05]96Nb4 | 5 | |
| 34. | Fe76Mo3.5P10C4B4Si2.5 | 4.5 | [8] |
| 35. | Fe71Ni5Mo3.5P10C4B4Si2.5 | 5.5 | |
| 36. | Fe66Ni10Mo3.5P10C4B4Si2.5 | 4 | |
| 37. | Fe61Ni15Mo3.5P10C4B4Si2.5 | 2.5 | |
| 38. | Fe56Ni20Mo3.5P10C4B4Si2.5 | 1 | |
| 39. | Fe46Ni30Mo3.5P10C4B4Si2.5 | | |
| 40. | (Fe0.75B0.15Si0.1)96Nb4 | 1.5 | [9] |
| 41. | [(Fe0.8Co0.1Ni0.1)0.75B0.2Si0.05]96Nb4 | 2.5 | |
| 42. | [(Fe0.6Co0.1Ni0.3)0.75B0.2Si0.05]96Nb4 | 3 | |
| 43. | [(Fe0.6Co0.2Ni0.2)0.75B0.2Si0.05]96Nb4 | 4 | |
| 44. | [(Fe0.6Co0.3Ni0.1)0.75B0.2Si0.05]96Nb4 | 4 | |
| 45. | Fe76Mo2Ga2P10C4B4Si2 | 2 | [10] |
| 46. | Fe74Mo4Ga2P10C4B4Si2 | 1.5 | |
| 47. | Fe75Mo2Ga3P10C4B4Si2 | 2.5 | |
| 48. | Fe73Mo4Ga3P10C4B4Si2 | 2 | |
| 49. | (Fe0.75B0.15Si0.10)99Nb1 | 0.5 | [11] |
| 50. | (Fe0.75B0.15Si0.10)98Nb2 | 1 | |
| 51. | (Fe0.75B0.15Si0.10)96Nb4 | 1.5 | |
| 52. | (Fe0.775B0.125Si0.10)98Nb2 | 0.5 | |
| 53. | Fe76Si9B10P5 | 2.5 | [12] |
| 54. | Fe73Al5Ga2P11C5Si4 | 1 | |
| 55. | Fe72Al5Ga2P10C6B4Si1 | 2 | |
| 56. | Fe30Co30Ni15Si8B17 | 1.2 | |
| 57. | Fe74Nb6Y3B17 | 2 | |
| 58. | Fe74Cr2Mo2Ga2P10C4B4Si2 | 2.5 | [13] |
| 59. | Fe72Cr4Mo2Ga2P10C4B4Si2 | 2.5 | |
| 60. | Fe70Cr6Mo2Ga2P10C4B4Si2 | 3 | |
| 61. | Fe73Al5Ga2P11C5B4 | 1 | [14] |
| 62. | Fe72Al5Ga2P11C6B4 | 1 | |
| 63. | Fe72Al5Ga2P11C5B4Si1 | 2 | |
| 64. | Fe75Mo5P10C7.5B2.5 | 2 | [19] |
| 65. | Fe80P9C8B2Si1 | 1.5 | [23] |
| 66. | Fe79Sn1P9C8B2Si1 | 3 | |
| 67. | Fe78Sn2P9C8B2Si1 | 3.5 | |

| 68. | Fe77Sn3P9C8B2Si1 | 1.5 | |
|---|---|---|---|
| 69. | Fe76P5(Si0.3B0.5C0.2)19 | 2 | [25] |
| 70. | Fe80P11C9 | 1.5 | [26] |
| 71. | Fe77.3C5.9Si3.3B4.8P8.7 | 2 | [27] |
| 72. | Fe77Si8B10P5 | 2 | [28] |
| 73. | Fe78Si7B10P5 | 1.5 | |
| 74. | Fe79Si6B10P5 | 1.5 | |
| 75. | Fe80Si5B10P5 | 1 | |
| 76. | Fe81Si4B10P5 | 1 | |
| 77. | Fe76Si8P9C7 | 1 | [30] |
| 78. | Fe80P13C7 | 2 | [31] |
| 79. | (Fe95Co5)80P13C7 | 2.1 | |
| 80. | (Fe90Co10)80P13C7 | 2.3 | |
| 81. | (Fe85Co15)80P13C7 | 2.2 | |
| 82. | (Fe90Co10)82P11C7 | 1.2 | |
| 83. | (Fe90Co10)82P8C7B3 | 1.3 | |
| 84. | (Fe90Co10)82P6C7B3Si2 | 1 | |
| 85. | Fe81P8.5C5.5B2Si3 | 1 | [33] |
| 86. | Fe83C1B11Si2P3 | 0.081 | [35] |
| 87. | Fe72B20Si4Nb4 | 1.5 | [38] |
| 88. | Fe71.8Cu0.2B20Si4Nb4 | 1.5 | |
| 89. | Fe71.6Cu0.4B20Si4Nb4 | 1.5 | |
| 90. | Fe71.4Cu0.6B20Si4Nb4 | 1.5 | |
| 91. | Fe71.2Cu0.8B20Si4Nb4 | 1 | |
| 92. | Fe71Cu1B20Si4Nb4 | 1 | |
| 93. | Fe72B20Si4Nb3.9Cu0.1 | 1.5 | [39] |
| 94. | Fe72B20Si4Nb3.8Cu0.2 | 1.5 | |
| 95. | Fe72B20Si4Nb3.7Cu0.3 | 1.5 | |
| 96. | Fe67.7C7.0Si3.3B5.5P8.7Cr2.3Mo2.5Al2.0Co1 | 6 | [41] |
| 97. | Fe68.7C7.0Si3.3B5.5P8.7Cr2.3Mo2.5Al2.0 | 5 | |
| 98. | Fe65.7C7.0Si3.3B5.5P8.7Cr2.3Mo2.5Al2.0Co3 | 5 | |
| 99. | Fe63.7C7.0Si3.3B5.5P8.7Cr2.3Mo2.5Al2.0Co5 | 5 | |
| 100. | Fe61.7C7.0Si3.3B5.5P8.7Cr2.3Mo2.5Al2.0Co7 | 4 | |
| 101. | Fe58.7C7.0Si3.3B5.5P8.7Cr2.3Mo2.5Al2.0Co10 | 3 | |
| 102. | Fe68.2C7Si3.3B5.5P8.7Cr2.3Al2Co3 | 4 | [42] |
| 103. | Fe71.2C7Si3.3B5.5P8.7Cr2.3Al2 | 3 | |
| 104. | Fe73.5C7Si3.3B5.5P8.7Nb2.0 | 3 | [43] |
| 105. | Fe75.5C7.0Si3.3B5.5P8.7 | 1 | |
| 106. | Fe75Cr5P9B4C7 | 2 | [44] |
| 107. | (Fe0.8P0.09C0.09B0.02)99.3Cu0.7 | 1.5 | [45] |
| 108. | (Fe36Co36B19.2Si4.8Nb4)99.5Cu0.5 | 2 | |
| 109. | (Fe40Ni40P14B6)96Ga4 | 3 | |
| 110. | Fe75.3C7.0Si3.3B5.0P8.7Cu0.7 | 1 | |

| 111. | Fe75.7C7.0Si3.3B5.0P8.7Cu0.3 | 3 | |
|---|---|---|---|
| 112. | Fe72Y6B22 | 2 | [46] |
| 113. | (Fe0.39Ni0.39B0.16P0.06)97Nb3 | 1.5 | |
| 114. | (Fe0.39Ni0.39B0.16P0.06)97.5Nb2.5 | 2.5 | |
| 115. | (Fe0.39Ni0.39B0.16P0.06)98Nb2 | 2 | |
| 116. | (Fe0.39Ni0.39B0.16P0.06)98.5Nb1.5 | 1 | |
| 117. | Fe77B18Ti1Zr4 | 1 | |
| 118. | Fe77B18Ti1.5Zr3.5 | 1 | |
| 119. | Fe77B18Ti2Zr3 | 1 | |
| 120. | Fe77B18Ti2.5Zr2.5 | 1 | |
| 121. | Fe77B18Ti3Zr2 | 1.5 | |
| 122. | Fe77B18Ti3.5Zr1.5 | 1 | |
| 123. | Fe77B18Ti4Zr1 | 1 | |
| 124. | Fe73Ga4P11C5B4Si3 | 3 | [51] |
| 125. | (Fe0.8Co0.2)73Ga4P11C5B4Si3 | 5 | |
| 126. | (Fe0.85Co0.15)77Ga2P10C5B3.5Si2.5 | 3 | |
| 127. | Fe76Si9.6B9.6P4.8 | 2.5 | [52] |
| 128. | (Fe0.76Si0.096B0.096P0.048)98Cr2 | 2.5 | |
| 129. | (Fe0.76Si0.096B0.096P0.048)96Cr4 | 3 | |
| 130. | (Fe0.76Si0.096B0.096P0.048)94Cr6 | 2.5 | |
| 131. | (Fe0.72B0.24Nb0.04)95.5Y4.5 | 7 | [53] |
| 132. | Fe71Nb6B23 | 1 | [56] |
| 133. | Fe70Ho1Nb6B23 | 1 | |
| 134. | Fe69Ho2Nb6B23 | 1.5 | |
| 135. | Fe68Ho3Nb6B23 | 2 | |
| 136. | Fe67Ho4Nb6B23 | 2.5 | |
| 137. | Fe66Ho5Nb6B23 | 3 | |
| 138. | Fe65Ho6Nb6B23 | 1.5 | |
| 139. | Fe70Er1Nb6B23 | 1.5 | [57] |
| 140. | Fe68Er3Nb6B23 | 2.5 | |
| 141. | Fe66Er5Nb6B23 | 4 | |
| 142. | Fe64Er7Nb6B23 | 2 | |
| 143. | Fe78P13C9 | 1.5 | [58] |
| 144. | Fe79.9Cu0.1P13C7 | 2.5 | [59] |
| 145. | Fe79.7Cu0.3P13C7 | 2 | |
| 146. | Fe79.4Cu0.6P13C7 | 1.5 | |
| 147. | Fe83B17 | 0.02 | [61] |
| 148. | Fe83B15Si2 | 0.028 | |
| 149. | Fe83B14Si2C1 | 0.034 | |
| 150. | Fe83B12Si2P3 | 0.044 | |
| 151. | Fe83B11Si2P3C1 | 0.076 | |
| 152. | Fe75Co5P13C7 | 2.3 | [65] |
| 153. | Fe70Co10P13C7 | 2.5 | |

| 154. | Fe65Co15P13C7 | 2 | |
|---|---|---|---|
| 155. | Fe60Co20P13C7 | 1.8 | |
| 156. | (Fe0.9Co0.1)76Si9B10P5 | 3 | [67] |
| 157. | (Fe0.75Si0.1B0.15)96Nb4 | 1.5 | |
| 158. | Fe73Al15Ga2P11C5B4 | 1 | |
| 159. | Fe72Al15Ga2P10C6B4Si1 | 2 | |
| 160. | Fe75Ga5P12C4B4 | 1 | [68] |
| 161. | Fe74Mo4Ga2P12C4B4 | 2 | |
| 162. | Fe78Ga2P12C4B4 | 2.5 | |
| 163. | Fe74Mo6P10C7.5B2.5 | 3 | [71] |
| 164. | Fe62.9Ni11.1Mo6P10C7.5B2.5 | 3 | |
| 165. | Fe61Co10Zr5W4B20 | 1 | [73] |
| 166. | [(Fe0.9Ni0.1)0.75B0.2Si0.05]96Nb4 | 1.5 | |
| 167. | [(Fe0.8Ni0.2)0.75B0.2Si0.05]96Nb4 | 2.5 | |
| 168. | [(Fe0.7Ni0.3)0.75B0.2Si0.05]96Nb4 | 3 | |
| 169. | [(Fe0.6Ni0.4)0.75B0.2Si0.05]96Nb4 | 4 | |
| 170. | [(Fe0.6Co0.4)0.75B0.2Si0.05]0.96Nb0.04 | 4 | [75] |
| 171. | {[(Fe0.6Co0.4)0.75B0.2Si0.05]0.96Nb0.04}99Cr1 | 4 | |
| 172. | {[(Fe0.6Co0.4)0.75B0.2Si0.05]0.96Nb0.04}98Cr2 | 4 | |
| 173. | {[(Fe0.6Co0.4)0.75B0.2Si0.05]0.96Nb0.04}97Cr3 | 3.5 | |
| 174. | {[(Fe0.6Co0.4)0.75B0.2Si0.05]0.96Nb0.04}96Cr4 | 3 | |
| 175. | (Fe0.75Si0.1B0.15)99Zr1 | 0.75 | [76] |
| 176. | Fe76.0C7.0Si3.3B5.0P8.7 | 1 | |
| 177. | Fe75.0C7.0Si3.3B5.0P8.7Ga1.0 | 3 | |
| 178. | Fe74.0C7.0Si3.3B5.0P8.7Ga2.0 | 3 | |
| 179. | Fe72.0C7.0Si3.3B5.0P8.7Ga4.0 | 2 | |
| 180. | Fe73.33B16.67Si8.33Hf1.67 | 1 | |
| 181. | Fe72.5B16.67Si8.33Hf2.5 | 2.5 | |
| 182. | Fe71.67B16.67Si8.33Hf3.33 | 2 | |
| 183. | Fe70.83B16.67Si8.33Hf4.17 | 1.5 | |
| 184. | Fe70B16.67Si8.33Hf5 | 1 | |
| 185. | Fe70Zr6Nb4B20 | 1 | |
| 186. | Fe70Zr4Nb4Ti2B20 | 1.5 | |
| 187. | Fe53Co18Mo9P14B6 | 3 | [85] |
| 188. | Fe71Mo9P14B6 | 2 | |
| 189. | Fe72B17Si5Mo6 | 1 | [90] |
| 190. | Fe72B17Si5(Mo0.5Nb0.5)6 | 1.5 | |
| 191. | Fe66Dy5Nb6B23 | 2 | [91] |
| 192. | Fe66Tm5Nb6B23 | 4.5 | |
| 193. | Fe70.2Ni7.8B14.2Si2.75P2.75Nb2.3 | 1 | |
| 194. | Fe62.4Ni15.6B14.2Si2.75P2.75Nb2.3 | 1 | |
| 195. | Fe54.6Ni23.4B14.2Si2.75P2.75Nb2.3 | 1.5 | |
| 196. | Fe46.5Ni31.2B14.2Si2.75P2.75Nb2.3 | 2 | |
| 197. | Fe80P10C7B3 | 1.6 | [94] |
| 198. | Fe80P8C7B5 | 1.3 | |
| 199. | Fe80P13C4B3 | 1.3 | |
| 200. | Fe80P8C9B3 | 1.4 | |
| 201. | Fe80P12C8 | 1.5 | [95] |
| 202. | Fe80P11C8B1 | 1.8 | |
| 203. | Fe80P10C8B2 | 2 | |
| 204. | Fe80P9C8B3 | 1.5 | |
| 205. | Fe80P8C8B4 | 1 | |
| 206. | Fe80P9C9B2 | 1.8 | [97] |
| 207. | Fe76Ga1P8.7C7Si3.3B5 | 3 | |
| 208. | Fe66Co10Mo4P9C4B4Si3 | 6 | |
| 209. | Fe76Mo2P10C7.5B2.5Si2 | 3 | |
| 210. | Fe78Mo1P9C6.5B3.5Si2 | 2 | |
| 211. | Fe77Al3P9C9B2 | 3 | |
| 212. | Fe67.5Mo7.5P10C10B5 | 3 | [98] |
| 213. | Fe62.5Co5Mo7.5P10C10B5 | 4 | |
| 214. | Fe60Ni7.5Mo7.5P10C10B5 | 4 | |
| 215. | Fe60Co5Ni2.5Mo7.5P10C10B5 | 3 | |
| 216. | Fe68Dy6B22Nb4 | 3 | [99] |
| 217. | (Fe0.9Co0.1)68Dy6B22Nb4 | 4 | |
| 218. | (Fe0.8Co0.2)68Dy6B22Nb4 | 4 | |
| 219. | (Fe0.7Co0.3)68Dy6B22Nb4 | 3 | |
| 220. | (Fe0.6Co0.4)68Dy6B22Nb4 | 2 | |
| 221. | (Fe0.72Mo0.04B0.24)94Dy6 | 2 | |
| 222. | [(Fe0.8Co0.2)72Mo4B24]94Dy6 | 3 | |
| 223. | Fe72.8B16Si8Zr3.2 | 2 | [101] |
| 224. | Fe73.85B15.38Si7.69Zr3.08 | 2 | |
| 225. | Fe74.81B14.8Si7.41Zr2.96 | 2 | |
| 226. | Fe75.71B14.29Si7.14Zr2.86 | 1.5 | |
| 227. | Fe76.55B13.79Si6.9Zr2.76 | 1 | |
| 228. | Fe73.33B16.67Si8.33Zr1.67 | 2 | [103] |
| 229. | Fe72.50B16.67Si8.33Zr2.50 | 2.5 | |
| 230. | Fe71.67B16.67Si8.33Zr3.33 | 2.5 | |
| 231. | Fe70.83B16.67Si8.33Zr4.17 | 2 | |
| 232. | Fe77Al3P9B2C9 | 3 | [104] |
| 233. | Fe75Ni5P13C7 | 2.5 | [106] |
| 234. | Fe65Ni15P13C7 | 2.4 | |
| 235. | Fe55Ni25P13C7 | 2 | |
| 236. | Fe60Co20P14B6 | 1.2 | [107] |
| 237. | Fe50Co30P14B6 | 1.6 | |
| 238. | (Fe0.75Dy0.01B0.24)96Nb4 | 0.5 | |
| 239. | (Fe0.74Dy0.02B0.24)96Nb4 | 2.5 | |

| 240. | (Fe0.73Dy0.03B0.24)96Nb4 | 3.5 | |
|---|---|---|---|
| 241. | (Fe0.72Dy0.04B0.24)96Nb4 | 4.5 | |
| 242. | (Fe0.71Dy0.05B0.24)96Nb4 | 5.5 | |
| 243. | (Fe0.70Dy0.06B0.24)96Nb4 | 4.5 | |
| 244. | (Fe0.69Dy0.07B0.24)96Nb4 | 3 | |
| 245. | (Fe0.71Gd0.05B0.24)96Nb4 | 3.5 | [110] |
| 246. | (Fe0.71Tb0.05B0.24)96Nb4 | 3.5 | |
| 247. | (Fe0.71Ho0.05B0.24)96Nb4 | 5 | |
| 248. | (Fe0.71Er0.05B0.24)96Nb4 | 5.5 | |
| 249. | (Fe0.71Tm0.05B0.24)96Nb4 | 6.5 | |
| 250. | Fe60Ni20P14B6 | 1 | [111] |
| 251. | Fe50Ni30P14B6 | 2 | |
| 252. | Fe72.5Mo2.5P10C10B5 | 1.5 | |
| 253. | Fe70Mo5P10C10B5 | 3 | |
| 254. | Fe62Ni10Y6B22 | 1 | [114] |
| 255. | Fe67Ni5Y6B22 | 1 | |
| 256. | Fe42Co30Y6B22 | 2 | |
| 257. | Fe52Co20Y6B22 | 2 | |
| 258. | Fe62Co10Y6B22 | 2 | |
| 259. | Fe72Y2Ta4B22 | 1 | [116] |
| 260. | Fe72Y1Nb5B22 | 1 | |
| 261. | Fe64Co7Zr6Nd3B20 | 1 | [117] |
| 262. | (Fe0.68Dy0.07B0.2Si0.05)96Nb4 | 2 | [120] |
| 263. | (Fe0.69Dy0.06B0.2Si0.05)96Nb4 | 3 | |
| 264. | (Fe0.7Dy0.05B0.2Si0.05)96Nb4 | 3 | |
| 265. | (Fe0.71Dy0.04B0.2Si0.05)96Nb4 | 4 | |
| 266. | (Fe0.72Dy0.03B0.2Si0.05)96Nb4 | 2.5 | |
| 267. | (Fe0.73Dy0.02B0.2Si0.05)96Nb4 | 2 | |
| 268. | (Fe0.74Dy0.01B0.2Si0.05)96Nb4 | 1 | |
| 269. | Fe69Nb6B17Y3Co5 | 3 | [121] |
| 270. | Fe81Mo1Si3P7.5C5.5B2 | 1 | [122] |
| 271. | Fe78Mo1Si2P9C6.5B3.5 | 2 | |
| 272. | (Fe0.8P0.09C0.09B0.02)99.5Cu0.5 | 1.8 | |
| 273. | (Fe0.8P0.09C0.09B0.02)99.7Cu0.3 | 2 | |
| 274. | (Fe0.8P0.09C0.09B0.02)99.9Cu0.1 | 1.8 | |
| 275. | Fe76Mo4(P0.45C0.2B0.2Si0.15)20 | 4 | [125] |
| 276. | (Fe0.9Co0.1)76Mo4(P0.45C0.2B0.2Si0.15)20 | 6 | |
| 277. | (Fe0.7Co0.3)76Mo4(P0.45C0.2B0.2Si0.15)20 | 3 | |
| 278. | (Fe0.8Co0.2)76Mo4(P0.45C0.2B0.2Si0.15)20 | 5 | |
| 279. | Fe67Co7Mo4Si2P10C7.5B2.5 | 3 | [126] |
| 280. | Fe70.3Ni3.7Mo6P10C7.5B2.5 | 5 | |
| 281. | Fe69Ni5Mo6P10C7.5B2.5 | 5 | |
| 282. | Fe66.6Ni7.4Mo6P10C7.5B2.5 | 4 | |

| 283. | Fe70.3Ni3.7Mo5Si1P10C7.5B2.5 | 4 | |
|---|---|---|---|
| 284. | Fe69Ni5Mo5Si1P10C7.5B2.5 | 4 | |
| 285. | Fe62.9Co7.4Ni3.7Mo6P10C7.5B2.5 | 5 | |
| 286. | Fe60Co7Ni7Mo6P10C7.5B2.5 | 3 | |
| 287. | Fe62Co7Ni7Mo4P10C7.5B2.5 | 3 | |
| 288. | Fe74Mo5Nb1P10C7.5B2.5 | 3 | |
| 289. | Fe74Mo5Cr1P10C7.5B2.5 | 3 | |
| 290. | Fe66.6Co7.4Mo6P10C7.5B2.5 | 4 | |
| 291. | Fe77Al3P6B5C9 | 1.5 | |
| 292. | Fe77Al3P7B4C9 | 2 | |
| 293. | Fe77Al3P8B3C9 | 3 | |
| 294. | Fe80P8B3C9 | 1.5 | |
| 295. | Fe80P9B2C8Si1 | 2 | |
| 296. | Fe80P9B2C7Si2 | 1.8 | |
| 297. | Fe80P9B2C6Si3 | 1.5 | |
| 298. | Fe80P9B2C9 | 1.8 | |
| 299. | Fe80P10B1C9 | 1.5 | |
| 300. | Fe77Al3P10B1C9 | 3 | |
| 301. | Fe77Al3P11C9 | 2.5 | |
| 302. | Fe74Mo5P10C7.5B2.5Si1 | 4 | [127] |
| 303. | Fe76P12C10.8Si1.2 | 1.5 | [131] |
| 304. | Fe78P11C9.9Si1.1 | 1.5 | |
| 305. | Fe80P10C9Si1 | | |
| 306. | Fe82P9C8.1Si0.9 | 1 | |
| 307. | Fe76Mo4P12C4B4 | 2 | [132] |
| 308. | Fe66Ni10Mo4P12C4B4 | 1.5 | |
| 309. | Fe56Ni20Mo4P12C4B4 | 1.5 | |
| 310. | Fe56Ni20Mo4P11C4B4Si1 | 2 | |
| 311. | Fe56Ni20Mo4P10C4B4Si2 | 1 | |

**Tabel S4.** Optimized hyperparameters for traditional ML baselines (SVM, KNN, RF, XGBoost) on $B_s$, $\ln(H_c)$, and $D_c$ using 5-fold cross-validation.

| Dataset | Model | Best Parameters |
|---|---|---|
| $B_s$ (T) | SVR | Kernel = 'rbf', Gamma = 0.1, C = 120 |
| $\ln(H_c)$ (A/m) | SVR | Kernel = 'rbf', Gamma = 0.1, C = 240 |
| $D_c$ (mm) | SVR | Kernel = 'linear', Gamma = 0.01, C = 240 |
| $B_s$ (T) | KNN | Weights = 'distance', N_neighbors = 5, leaf_size = 30 |
| $\ln(H_c)$ (A/m) | KNN | Weights = 'distance', N_neighbors = 6, leaf_size = 50 |
| $D_c$ (mm) | KNN | Weights = 'distance', N_neighbors = 7, leaf_size = 40 |
| $B_s$ (T) | XGBoost | Reg_lambda = 0.09, Reg_alpha = 0.06, N_estimators = 200, Max_depth = 2, Learning_rate = 0.4, Gamma = 0.003 |
| $\ln(H_c)$ (A/m) | XGBoost | Reg_lambda = 0.05, Reg_alpha = 0.06, N_estimators = 100, Max_depth = 4, Learning_rate = 0.1, Gamma = 0.01 |
| $D_c$ (mm) | XGBoost | Reg_lambda = 0.07, Reg_alpha = 0.09, N_estimators = 200, Max_depth = 2, Learning_rate = 0.2, Gamma = 0.003 |
| $B_s$ (T) | RF | N_estimators = 150, Min_samples_split = 2, Min_samples_leaf = 1, Max_features = 'log2', Max_depth = 28 |
| $\ln(H_c)$ (A/m) | RF | N_estimators = 50, Min_samples_split = 8, Min_samples_leaf = 1, Max_features = 'sqrt', Max_depth = 16 |
| $D_c$ (mm) | RF | N_estimators = 100, Min_samples_split = 2, Min_samples_leaf = 1, Max_features = 'sqrt', max_depth = 20 |

**Table 5.** 95 Fe-based alloy candidates with $B_s$>1.75T, ln($H_c$)<1.5A/m, and $D_c$>1mm predicted by the NSMTWAE framework.

| No. | Predicted Fe-based amorphous alloy composition | $B_s$ (T) | ln($H_c$) (A/m) | $D_c$ (mm) |
|---|---|---|---|---|
| 1 | Fe65.76B13.43Si0.95Co19.43Nb0.33 | 1.98 | 0.75 | 1.13 |
| 2 | Fe68.47B13.11Si1.20Co17.03 | 1.97 | 1.20 | 1.38 |
| 3 | Fe66.34B12.48Si1.78Co18.87Nb0.43 | 1.97 | 0.53 | 1.06 |
| 4 | Fe69.11B11.38Si0.87Co18.36Nb0.18 | 1.97 | 1.04 | 1.27 |
| 5 | Fe67.66B11.62Si1.59Co18.74Nb0.29 | 1.96 | 0.49 | 1.16 |
| 6 | Fe73.24B11.29Si1.18Co14.13 | 1.96 | 0.74 | 1.27 |
| 7 | Fe69.98B13.08Si1.47Co15.30 | 1.96 | 1.47 | 1.44 |
| 8 | Fe67.54B13.56Si1.23Co17.46Nb0.11 | 1.96 | 1.16 | 1.44 |
| 9 | Fe71.26B11.35Si1.22Co15.98Nb0.10 | 1.95 | 0.70 | 1.34 |
| 10 | Fe67.13B12.79Si2.09Co17.44Nb0.43 | 1.94 | 0.47 | 1.27 |
| 11 | Fe41.92B18.88Si0.83Co37.22Nb0.83Y0.13 | 1.93 | 0.32 | 1.17 |
| 12 | Fe41.76B19.04Si0.73Co37.29Nb0.84Y0.15 | 1.93 | 0.32 | 1.17 |
| 13 | Fe40.20B19.75Si0.19Co38.42Nb0.95Y0.27 | 1.93 | 0.19 | 1.42 |
| 14 | Fe70.62B14.08Si1.12Co14.03 | 1.93 | 1.16 | 1.45 |
| 15 | Fe74.41B11.53Si1.01Co12.84 | 1.92 | 0.87 | 1.56 |
| 16 | Fe74.52B12.00Si1.13Co12.14 | 1.92 | 0.90 | 1.57 |
| 17 | Fe39.71B18.37Si1.64Co36.28Nb3.69 | 1.92 | 0.12 | 1.16 |
| 18 | Fe75.68B12.45Si1.58Co10.07Cu0.10 | 1.92 | 0.53 | 1.46 |
| 19 | Fe75.16B12.56Si1.19Co10.93 | 1.92 | 0.83 | 1.47 |
| 20 | Fe72.89B11.78Si1.07Co13.99 | 1.92 | 1.21 | 1.62 |
| 21 | Fe78.57B12.15Si1.55Co7.49Cu0.11 | 1.91 | 0.48 | 1.48 |
| 22 | Fe42.81B18.30Si0.13Co37.48Nb0.64Y0.35 | 1.91 | 0.47 | 1.63 |
| 23 | Fe42.65B18.87Si0.12Co36.98Nb0.78Y0.35 | 1.90 | 0.35 | 1.68 |
| 24 | Fe67.99B22.41Si5.71Co0.29Nb1.35Ni1.81Cu0.13Hf0.21 | 1.89 | 1.08 | 2.15 |
| 25 | Fe82.95B13.94Si0.20P0.15Nb0.26Ni2.13Hf0.19 | 1.89 | 1.41 | 2.17 |
| 26 | Fe65.66B21.78Si7.98Co0.73Nb2.54Ni0.90Cu0.21Hf0.12 | 1.89 | 0.94 | 1.84 |
| 27 | Fe77.80B17.19Si0.19Co0.20Nb0.34Ni3.69Hf0.43 | 1.88 | 1.37 | 2.38 |
| 28 | Fe41.33B19.23Si0.10Co37.88Nb0.58Y0.61 | 1.87 | 0.15 | 1.61 |
| 29 | Fe70.45B20.26Si5.68Co0.25Nb1.34Ni1.58Cu0.13Hf0.24 | 1.87 | 1.01 | 2.12 |
| 30 | Fe45.23B18.19Si0.13Co35.57Nb0.30Y0.32 | 1.87 | 0.83 | 1.89 |
| 31 | Fe40.93B19.35Si0.26Co37.15Nb1.70Y0.37 | 1.87 | 0.10 | 1.74 |
| 32 | Fe43.53B18.48Si0.12Co36.84Nb0.35Y0.37 | 1.87 | 0.51 | 1.89 |
| 33 | Fe83.61B14.26Si0.17P0.15Nb0.31Ni1.19Hf0.14 | 1.87 | 1.43 | 2.60 |
| 34 | Fe38.58B19.30Si0.53Co37.49Nb3.70Y0.18 | 1.87 | -0.06 | 1.65 |
| 35 | Fe45.82B18.42Si0.12Co34.76Nb0.27Y0.35 | 1.87 | 0.68 | 1.89 |
| 36 | Fe43.50B19.26Co35.35Nb0.44Dy0.10Y1.02 | 1.87 | 0.30 | 1.86 |
| 37 | Fe43.43B18.59Si0.12Co36.77Nb0.38Y0.39 | 1.87 | 0.48 | 1.89 |
| 38 | Fe45.23B18.32Si0.12Co35.40Nb0.34Y0.34 | 1.86 | 0.59 | 1.90 |
| 39 | Fe84.38B13.69Si0.14P0.17Nb0.27Ni1.06Hf0.13 | 1.86 | 1.44 | 2.84 |
| 40 | Fe84.57B13.52Si0.12P0.19Nb0.26Ni1.05Hf0.12 | 1.86 | 1.44 | 2.80 |
| 41 | Fe63.27B21.01Si8.17Co2.23Nb4.73Ni0.32Cu0.14 | 1.85 | 0.98 | 2.45 |
| 42 | Fe65.70B22.93Si6.53Co1.08Nb2.73Ni0.76Hf0.10 | 1.85 | 1.14 | 2.80 |
| 43 | Fe44.51B19.66Co33.71Nb0.43Dy0.11Y1.26 | 1.84 | 0.26 | 1.87 |
| 44 | Fe43.14B19.64Co35.06Nb0.56Dy0.11Y1.20 | 1.84 | 0.11 | 1.91 |
| 45 | Fe60.90B21.35Si7.96Co4.09Nb4.96Ni0.48Cu0.12 | 1.84 | 0.94 | 2.61 |
| 46 | Fe38.82B19.10Si0.64Co37.20Nb3.79Y0.24 | 1.84 | -0.05 | 1.68 |
| 47 | Fe38.81B20.29Si0.24Co36.42Nb3.61Y0.33 | 1.84 | -0.06 | 1.67 |
| 48 | Fe45.97B20.28Si9.00Co19.52Nb4.20Ni0.63Cr0.21 | 1.83 | -0.20 | 1.59 |
| 49 | Fe45.60B19.59Si8.52Co21.28Nb4.15Ni0.37Cr0.29 | 1.82 | -0.33 | 1.56 |
| 50 | Fe47.07B19.29Co32.12Nb0.42Dy0.18Y0.63 | 1.82 | 0.35 | 1.98 |
| 51 | Fe64.64B22.24Si6.72Co2.50Nb3.18Ni0.50 | 1.82 | 1.07 | 3.08 |
| 52 | Fe63.84B23.05Si6.68Co2.28Nb3.21Ni0.72Hf0.10 | 1.82 | 1.08 | 3.54 |
| 53 | Fe43.94B19.14Si0.15Co34.91Nb1.21Y0.41 | 1.82 | 0.08 | 1.88 |
| 54 | Fe51.30B19.58Si10.44Co13.80Nb3.83Ni0.50Cr0.32 | 1.81 | -0.38 | 1.40 |
| 55 | Fe57.96B22.29Si7.49Co6.16Nb5.03Ni0.82 | 1.81 | 0.96 | 3.39 |
| 56 | Fe58.03B21.33Si7.91Co6.72Nb5.14Ni0.62 | 1.81 | 0.83 | 2.98 |
| 57 | Fe41.28B20.15Si0.38Co33.53Nb3.99Y0.44 | 1.81 | -0.15 | 1.80 |
| 58 | Fe58.94B21.83Si7.65Co5.74Nb5.03Ni0.57 | 1.81 | 0.85 | 3.08 |
| 59 | Fe62.18B22.29Si7.00Co3.37Nb4.27Ni0.68 | 1.81 | 1.05 | 3.80 |
| 60 | Fe65.85B22.55Si6.69Co2.22Nb2.05Ni0.36Hf0.11 | 1.81 | 0.78 | 2.08 |
| 61 | Fe46.71B20.07Co30.94Nb0.65Dy0.21Y1.08 | 1.81 | 0.04 | 1.92 |
| 62 | Fe59.00B20.58Si7.28Co7.61Nb5.05Ni0.29 | 1.80 | 0.73 | 2.68 |
| 63 | Fe61.98B22.54Si6.94Co3.44Nb4.13Ni0.76 | 1.80 | 1.07 | 3.98 |
| 64 | Fe61.80B22.44Si6.98Co3.59Nb4.22Ni0.74 | 1.80 | 1.06 | 4.00 |
| 65 | Fe50.95B19.03Si10.33Co14.93Nb3.71Ni0.44Cr0.37 | 1.79 | -0.40 | 1.40 |
| 66 | Fe59.82B22.61Si7.63Co3.81Nb5.49Ni0.43 | 1.79 | 0.85 | 4.35 |
| 67 | Fe48.70B19.20Co30.37Nb0.76Dy0.24Y0.48 | 1.79 | 0.17 | 1.96 |
| 68 | Fe48.39B20.00Co29.63Nb0.76Dy0.18Y0.77 | 1.79 | 0.01 | 1.98 |
| 69 | Fe59.42B22.16Si7.61Co4.69Nb5.42Ni0.47 | 1.79 | 0.79 | 4.02 |
| 70 | Fe59.69B22.72Si8.26Co3.23Nb5.55Ni0.30 | 1.79 | 0.68 | 3.57 |
| 71 | Fe59.68B22.19Si7.64Co4.41Nb5.44Ni0.43 | 1.79 | 0.79 | 4.06 |
| 72 | Fe58.38B22.66Si7.52Co5.17Nb5.42Ni0.63 | 1.78 | 0.95 | 4.45 |
| 73 | Fe61.28B21.80Si8.12Co2.90Nb5.48Ni0.21 | 1.78 | 0.66 | 3.37 |
| 74 | Fe61.30B21.79Si8.05Co2.97Nb5.48Ni0.20 | 1.78 | 0.65 | 3.39 |
| 75 | Fe45.65B19.02Si7.20Co23.31Nb4.21Ni0.21Cr0.23 | 1.78 | -0.17 | 1.71 |
| 76 | Fe59.05B22.34Si7.84Co4.68Nb5.48Ni0.38 | 1.78 | 0.66 | 4.01 |
| 77 | Fe64.03B22.59Si6.54Co3.44Nb2.82Ni0.33 | 1.78 | 0.52 | 2.75 |
| 78 | Fe45.34B21.61Si6.89Co20.98Nb4.29Ni0.63 | 1.78 | 0.41 | 2.28 |
| 79 | Fe59.67B22.58Si7.50Co4.18Nb5.44Ni0.43 | 1.77 | 0.95 | 4.63 |
| 80 | Fe60.73B21.87Si7.92Co3.54Nb5.44Ni0.28 | 1.77 | 0.64 | 3.86 |
| 81 | Fe54.25B21.37Si6.02Co12.74Nb5.34 | 1.77 | 0.16 | 2.21 |

| 82 | Fe42.87B16.64Si6.44Co29.61Nb3.90Ni0.11Cr0.27 | 1.77 | -0.39 | 1.63 |
|---|---|---|---|---|
| 83 | Fe43.97B17.22Si6.84Co27.38Nb4.00Ni0.12Cr0.31 | 1.77 | -0.42 | 1.63 |
| 84 | Fe61.15B22.36Si7.38Co3.17Nb5.35Ni0.40 | 1.77 | 0.81 | 4.54 |
| 85 | Fe45.91B18.52Si7.09Co23.66Nb4.24Ni0.17Cr0.25 | 1.77 | -0.29 | 1.64 |
| 86 | Fe60.28B22.50Si7.66Co3.57Nb5.46Ni0.33 | 1.76 | 0.68 | 4.45 |
| 87 | Fe54.69B21.44Si7.64Co10.20Nb5.28Ni0.50 | 1.76 | 0.63 | 3.39 |
| 88 | Fe47.59B20.03Co29.94Nb0.88Dy0.23Y1.04 | 1.76 | -0.01 | 1.99 |
| 89 | Fe65.29B24.62Si0.40Co7.81Nb1.56 | 1.76 | 0.47 | 2.35 |
| 90 | Fe54.12B22.01Si6.10Co12.02Nb5.47 | 1.76 | 0.23 | 2.37 |
| 91 | Fe63.69B20.68Si6.71Co5.51Nb3.08Ni0.16 | 1.76 | 0.53 | 2.89 |
| 92 | Fe62.64B21.56Si6.44Co5.68Nb3.27Ni0.21 | 1.76 | 0.56 | 3.11 |
| 93 | Fe59.52B21.56Si6.85Co7.00Nb4.59Ni0.30 | 1.75 | 0.57 | 3.99 |
| 94 | Fe54.13B21.66Si2.42Co16.28Nb5.25Y0.12 | 1.75 | 0.14 | 2.43 |
| 95 | Fe53.59B21.05Si6.98Co12.58Nb5.10Ni0.46 | 1.75 | 0.79 | 4.91 |